\documentclass{bmcart}

\usepackage{hyperref}
\usepackage[utf8]{inputenc} 
\usepackage{graphicx}
\usepackage{amsmath}
\usepackage{amssymb}
\usepackage{esint}
\usepackage{dsfont}
\usepackage{subfig}
\usepackage{verbatim}
\usepackage{color}
\usepackage{breakcites}

\startlocaldefs
\endlocaldefs

\begin{document}

\begin{frontmatter}

\begin{fmbox}
\dochead{Research}


\title{A generative model for age and income distribution}


\author[
   addressref={aff1},                   
   email={fatih.ozhamaratli.19@ucl.ac.uk}   
]{\inits{FO}\fnm{Fatih} \snm{Ozhamaratli}}
\author[
   addressref={aff2},
   email={oik22@cam.ac.uk}
]{\inits{OK}\fnm{Oleg} \snm{Kitov}}
\author[
   addressref={aff1},
   email={p.barucca@ucl.ac.uk}
]{\inits{PB}\fnm{Paolo} \snm{Barucca}}


\address[id=aff1]{
  \orgname{University College London}, 
  \street{Gower Street},                     %
  \postcode{WC1E 6BT}                                
  \city{London},                              
  \cny{UK}                                    
}
\address[id=aff2]{%
  \orgname{University of Cambridge, The Old Schools},
  \street{Trinity Ln},
  \postcode{CB2 1TN}
  \city{Cambridge},
  \cny{UK}
}


\begin{artnotes}
\end{artnotes}

\end{fmbox}


\begin{abstractbox}

\begin{abstract} 
Each individual in society experiences an evolution of their income during their lifetime. 
Macroscopically, this dynamics creates a statistical relationship between age and income for each society. 
In this study, we investigate income distribution and its relationship with age and identify a stable joint distribution function for age and income within the United Kingdom and the United States. 
We demonstrate a flexible calibration methodology using panel and population surveys and capture the characteristic differences between the UK and the US populations. 
The model here presented can be utilised for forecasting income and planning pensions.
\end{abstract}




\end{abstractbox}
%

\end{frontmatter}



\section{Introduction}
A universal element of societies is the emergence of hierarchical organisation structures within professions. 
People develop work experience through time and manage to obtain jobs of increasing responsibility and increasing level of income with time. 
Hence, it is a natural property of income distribution to be correlated with work experience and age; nevertheless, most income models do not study the relationship between income and age, and consequently between income distribution and demographic changes. 
This paper introduces a model of income, dependent on age-specific model parameters and random shocks. 
The model contributes to the understanding of the relationship between age and income and its dynamics.

Our aim is to compare the estimated parameters in the UK and the US age and income distribution to find out similar characteristics of age and income across states, as well as the contrasting differences.
A simple age and income model is fundamental for the development of a sustainable pension system. 
The model focuses on the age and income relationship and further factors, such as occupational levels, are not considered.  
The model is estimated via panel survey data from the UK and population surveys from the USA. 
The data from panel surveys track the same individuals for the duration of the survey, and the population survey is repeated with different people each wave. 
The results reflect a clear income-age relationship in the UK and US, a clear structure of the joint distribution characterised by rapidly increasing income at younger ages, followed by income levels stabilising near mean income but spreading till retirement. 
At this point, the income decreases and concentrates around mean retirement income. 
The paper demonstrates a flexible methodology to estimate parameters from population surveys, as well as panel surveys. 
The paper provides a simple generative model to evolve age-income population for simulation and forecasting purposes, which can constitute the foundation for future studies of financially sustainable pension systems by providing a benchmark for capturing age and income relationship.
The purpose is to have a baseline model simple enough for isolating age and income relationship of income dynamics. 
Such a model will serve for investigating the properties of a sustainable and balanced pension system. 
The mean and standard deviation statistics from the panel and population surveys on Fig. \ref{fig:uk_labour_real_stat}, Fig. \ref{fig:uk_labour_sim_stat}, Fig.\ref{fig:jdf_uk_usa_income_15_100} from observed panel data and simulation results reflect a clear relationship between age and income. 
More complex models, which investigate additional factors, and profile heterogeneity of income dynamics are out of the scope of our work.
\newline

Previous research on income have been conducted, and the research focused on investigating and explaining wage dynamics. Champernowne explicitly introduces a first-order Markov process to model the time-evolution of wages\cite{champernowne1953model}. Following Markov process path, the validity of the first-order Markov assumption is tested by Shorrocks\cite{shorrocks1976inequality}. Following research introduces a second-order Markov process, yet neither of these works links individual wage dynamics to time-evolution of the distribution of wages  \cite{shorrocks1978income}.
A different approach focused on poverty, which deals with modelling individual data using linear regression and transitioning to poverty (probit model)\cite{lillard1978dynamic}. 
A more comprehensive model incorporating various factors is developed to estimate  transition probability in wage quintiles conditioned on various regressors, including education, experience and age \cite{buchinsky1999wage}, and study both intra- and inter-group inequality. 
The persistence of the low pay state and on factors affecting the low pay probability in a generalized regression model is expressed. 
For modelling low income transitions the previous research use British Panel Data for the $'90$s, focus on the transition probability and state dependence for the poverty status\cite{cappellari2004modelling}. They define poverty transition equation, a coarse-grained dynamics.
Focus on inequality and upward mobility between quintiles considering gender effects are investigated \cite{kopczuk2010earnings}. 
The previous models in literature either incorporate numerous external variables, distribution characteristics and functions, such as innovation constants 
or limiting their scope to the investigation of dependence on a single variable\cite{firpo2009unconditional}\cite{firpo2011occupational}.
A more recent article by
Guvenen investigated a model for which focal variables are the human capital consisting of education, work experience, and idiosyncratic shocks \cite{guvenenlabour}, following research modelled male income for studying the impact of labour income taxation policy on inequality \cite{guvenentaxation}
The referred life-cycle model's distribution characteristics of the pre-tax income arise from the differences in the individual's ability to learn new things and idiosyncratic shocks.
Previous research tried to capture the income dynamics with Markov Models, linear autoregressive models, or by relying on econometric toolset such as covariance matrices. 
We investigate a generative model with an empirical distribution for sustainable age and income relationship in a population; we achieve this via an income evolution model with an age-dependent parameter, estimated from previous population and panel surveys.

In contrast to previous research, our study introduces a dynamic model that describes the income-dependent only on age and previous income. This paper investigates the stationary property of the income distribution dependent on age.
We provide a model in which the mean and variance of income given age are preserved at any time point. 

\section{Methods}

We introduce a simple model which focuses on age and income relationship and differs from recent literature by not incorporating other variables such as occupational level, level of education and skill coefficients. 
The model is stationary, i.e. the mean and variance of income given age are preserved in time.
The model is utilised to represent observed panel data for gaining empirical insights regarding age dependant, income dynamics and mobility.
The calibrated model can be utilised as a simple generative model to evolve an age-income population for simulation purposes and it provides a theoretical background for studies focusing on ageing and pension income of the population.  
We initially assume the following model, by which $\mu(.)$ and $\sigma(.)$ represent a function of age, income, and individual-specific additional parameters $\theta_{i}$ or $\lambda_{i}$, for the sake of generality. $\mu(.)$ is a function capturing mean income characteristics, and $\sigma(.)$ captures the variational characteristics of the income.
We consider the following individual income stochastic process for an economic agent $i$ characterised at each time step $t$ by a given age $a_i$ and income $y_i$:
\begin{equation}\label{eq:incomeAgeDyn}
y_{i(t+1)}= \mu(a_{i(t+1)},\,y_{it}|\mathbf{\theta}_{i}) + \sigma(a_{i(t+1)},\,y_{it}|\mathbf{\lambda}_{i})\eta_{it}
\end{equation}
The characterising insights on Fig.\ref{fig:jdf_uk_usa_income_15_100} from the panel data lead to the assumption that the probabilistic step at time $t$ depends only on the age and income of the preceding step. 

\subsection{Defining Income and Age Dynamics}
Earnings of individual $i$ at the time step $t$ is denoted as $Y_{i,t}$ and its logarithm is $y_{i,t}$.
The parameters that describe the income process are: age-dependent persistence parameter $q_{a}$, age-dependent mean $\mu_{a}$ and age-dependent standard deviation $\sigma_{a}$. 
The income shock process consists of independent random shock $\eta_{t}^{i}$ which is normally distributed with mean zero and variance 1, and it is applied to $\sigma_{a}$, the model can be defined as follows: 

\begin{equation}
y^{i}_{a+1,t+1} = q_{a} y^{i}_{a,t} + \mu_{a} + \sigma_{a}\eta^{i}_{t}
\label{eq:agent_income_evo}
\end{equation}

Averaging income $y$, for individuals who are $a$ years old, gives $\bar{y}_{a}$ ,which denotes the average income for age group $a$ across all individuals $i$ and periods $t$.
Assuming that the age-dependent income profiles are  stationary, we can average incomes $y^{i}_{a,t}$ across individuals and time to get the following equation:

\begin{equation}
\bar{y}_{a+1}=q_{a}\bar{y}_{a} + \mu_{a}
\end{equation}
\noindent where $\bar{y}_{a}$ denotes the average income for age group $a$, taken across all individuals $i$ and periods $t$. The following equation can find the estimator for $\mu_{a}$:
\begin{equation}
\mu_{a}= \bar{y}_{a+1} - q_{a} \bar{y}_{a}
\label{eq:agent_income_evo_average}
\end{equation}
The income data from different waves are inflation adjusted to isolate effects of economic growth.
\subsection{Data}
The British Household Panel Survey (BHPS)\cite{bhps} from the UK and The Current Population Survey (IPUMS CPS)\cite{ipumscps} from the USA are used for estimating the model parameters of the model (\ref{eq:agent_income_evo}) and comparing the results of simulated data and surveys.
The BHPS is a Panel Survey conducted between 1991-2008. 
For our model we focus on labour income data, which captures wage, salary or self-employment income. 
To investigate population characteristics, we also incorporate other income sources and call it "Total Income", which additionally captures the transfers, pensions, grants, aids, state-benefits, dividends, capital income and rents. 
BHPS provides individuals specific longitudinal weights for ensuring the representativeness of the population.
Two types of weights are provided with BHPS. 
The first wave is weighted for adjusting population marginals at the households and post-stratified to the population age by sex marginals.
Consecutive waves are re-weighted to take into account sample attrition, variables such as address change, household region, age, sex, race, employment status, income total and composition, educational qualifications\cite{bhpsmanual}. 
Panel Survey is conducted via questionnaires with tracked individuals of the initial sample. 
There is an extension to the sample population in 1999.

For the USA, IPUMS CPS is used, which is annually conducted with different samples each year. 
In contrast with BHPS, the Labour Income does not include self-employed income, and the weights are cross-sectional.
Income distribution, age distribution and income-dependent age distribution from the surveys are utilised for parameter estimation and further analysis. 
${q_{a}}$, ${\mu}_{a}$, ${\sigma}_{a}$ are the key parameters estimated according to the proposed model. 
Following investigation and interpretation of the estimated parameters, these parameters are used to simulate the population's income transitions. 
The simulation is initialised using the panel data from the wave 1, and the income evolution function \ref{eq:agent_income_evo} of the model is applied transitively in an iterative approach to the data for simulating successive waves.   
The simulated data is plotted, interpreted and finally compared with the observed data.

\begin{figure}[htp]%
         \centering
         \subfloat[UK Pyramid Wave 1]{\includegraphics[width=0.52\linewidth]{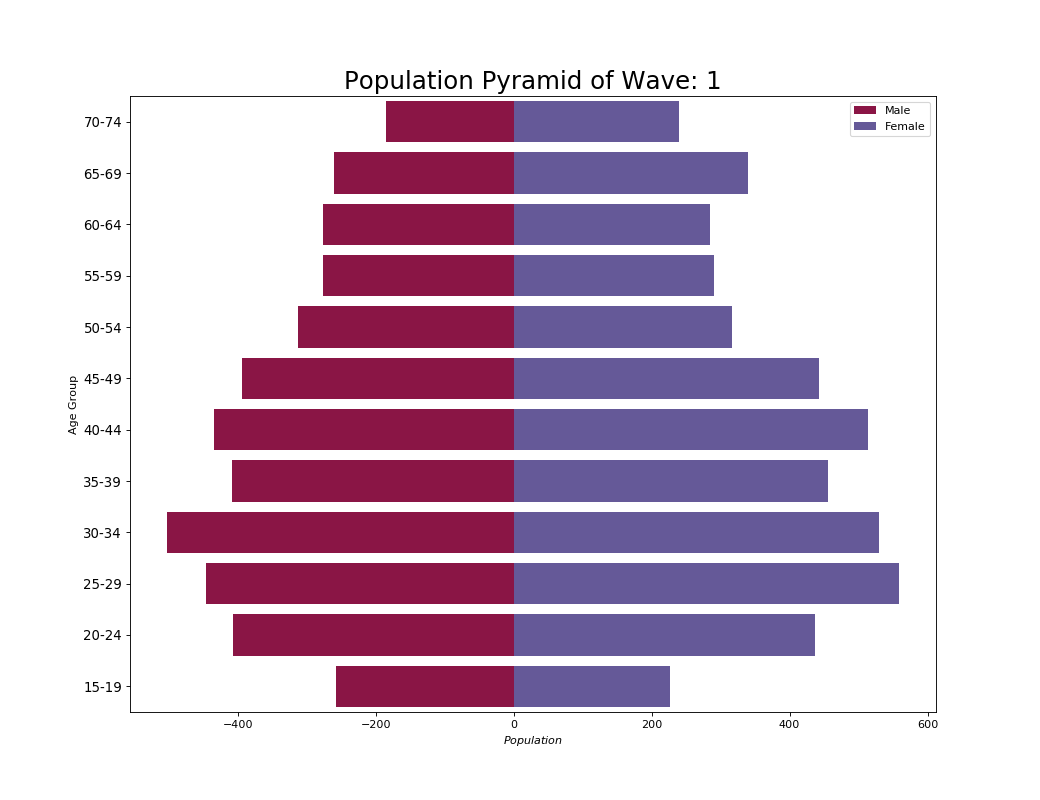}}\qquad
         \subfloat[UK Pyramid Wave 18]{\includegraphics[width=0.52\linewidth]{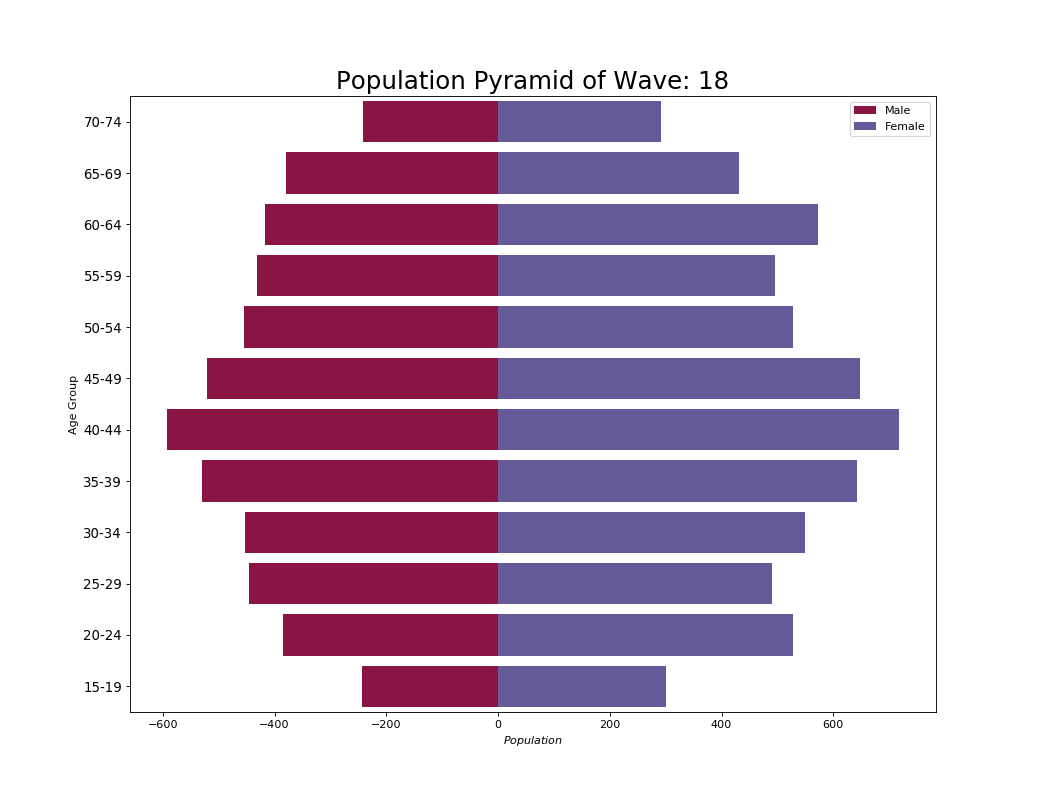}}\\
         \subfloat[USA Pyramid Wave 1]{\includegraphics[width=0.52\linewidth]{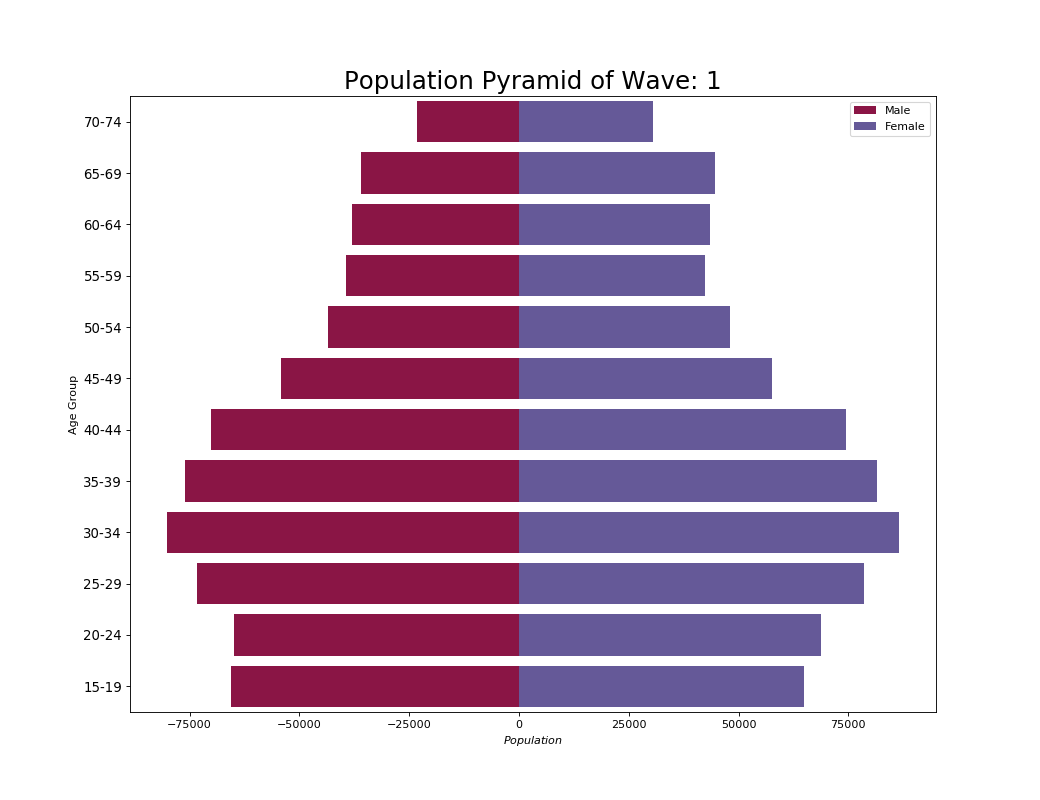}}\qquad
         \subfloat[USA Pyramid Wave 18]{\includegraphics[width=0.52\linewidth]{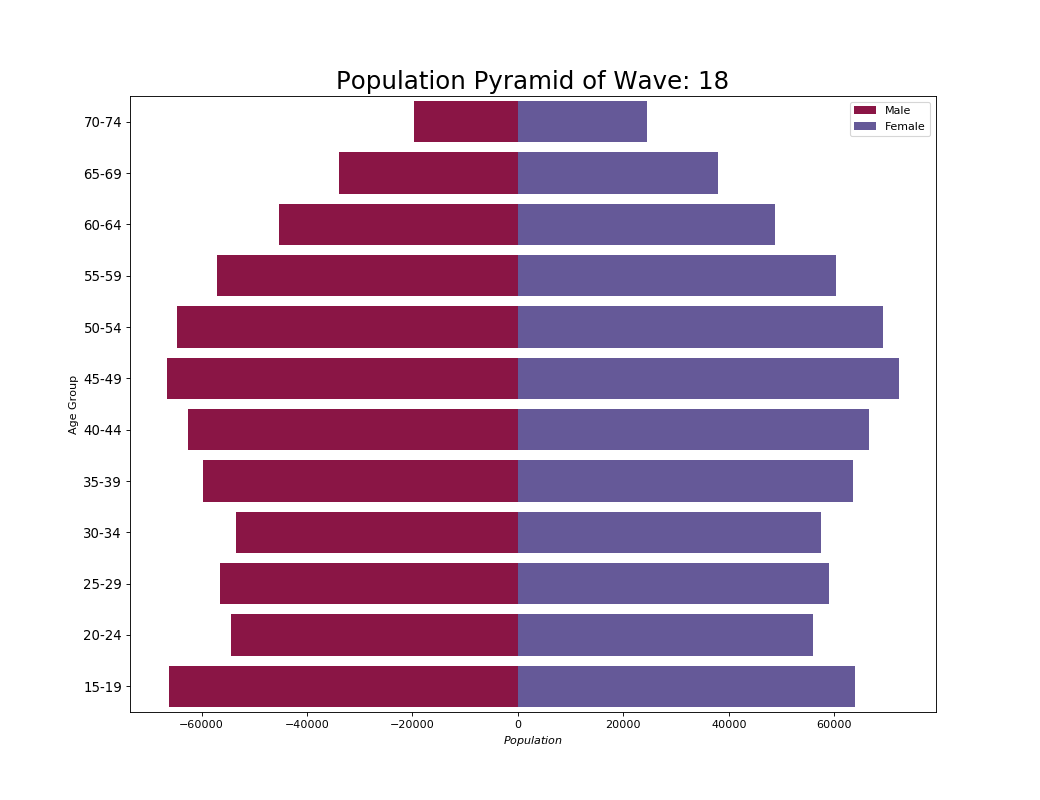}}\\
         \caption{Population Pyramid for the UK \& USA  Income between Ages 15-100}
         \label{fig:uk_usa_pyramid}
\end{figure}
\begin{figure}[htp]%
         \centering
         \subfloat[UK Labour JDF]{\includegraphics[width=0.45\linewidth]{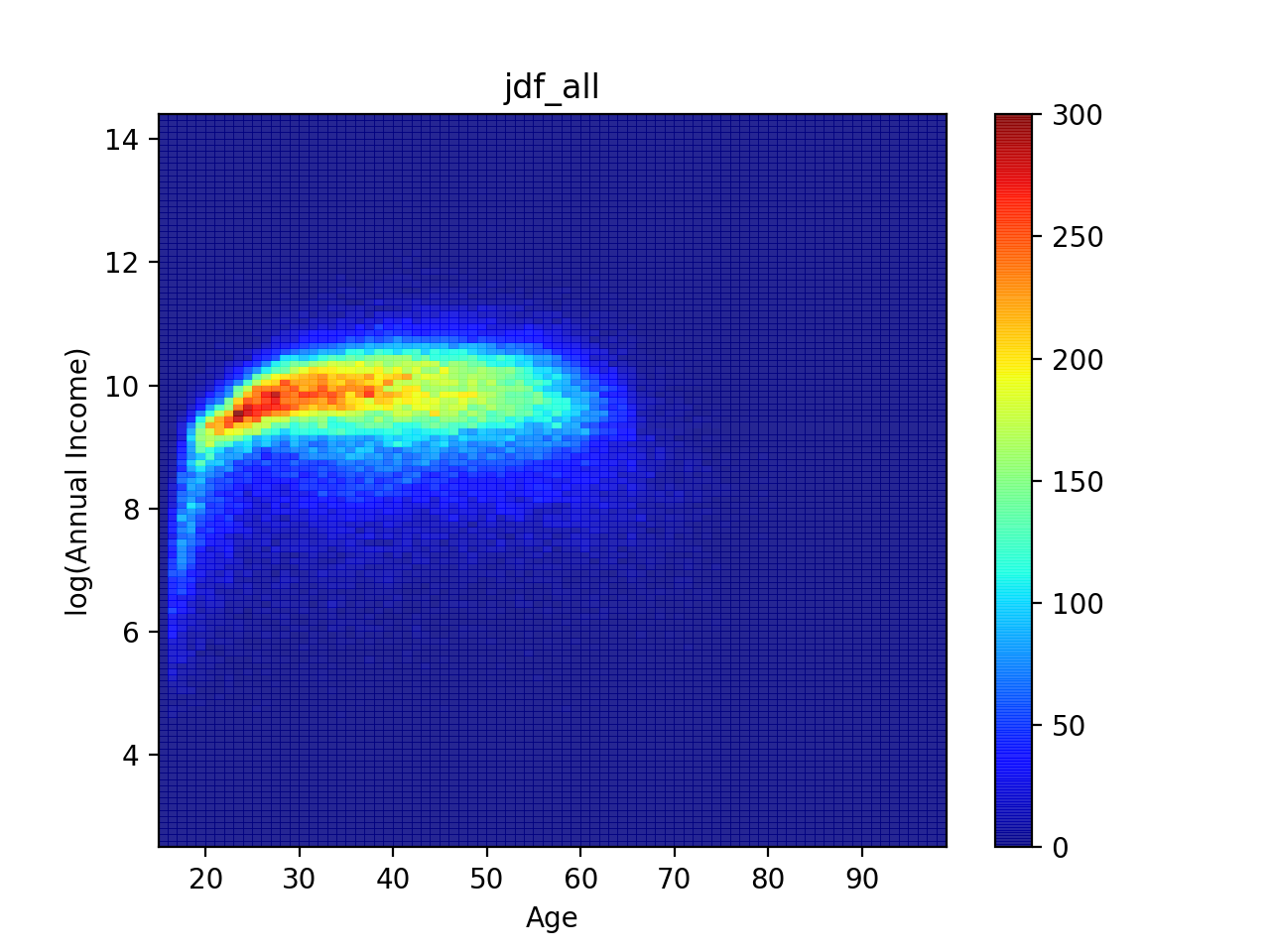}}\qquad
         \subfloat[UK Total JDF]{\includegraphics[width=0.45\linewidth]{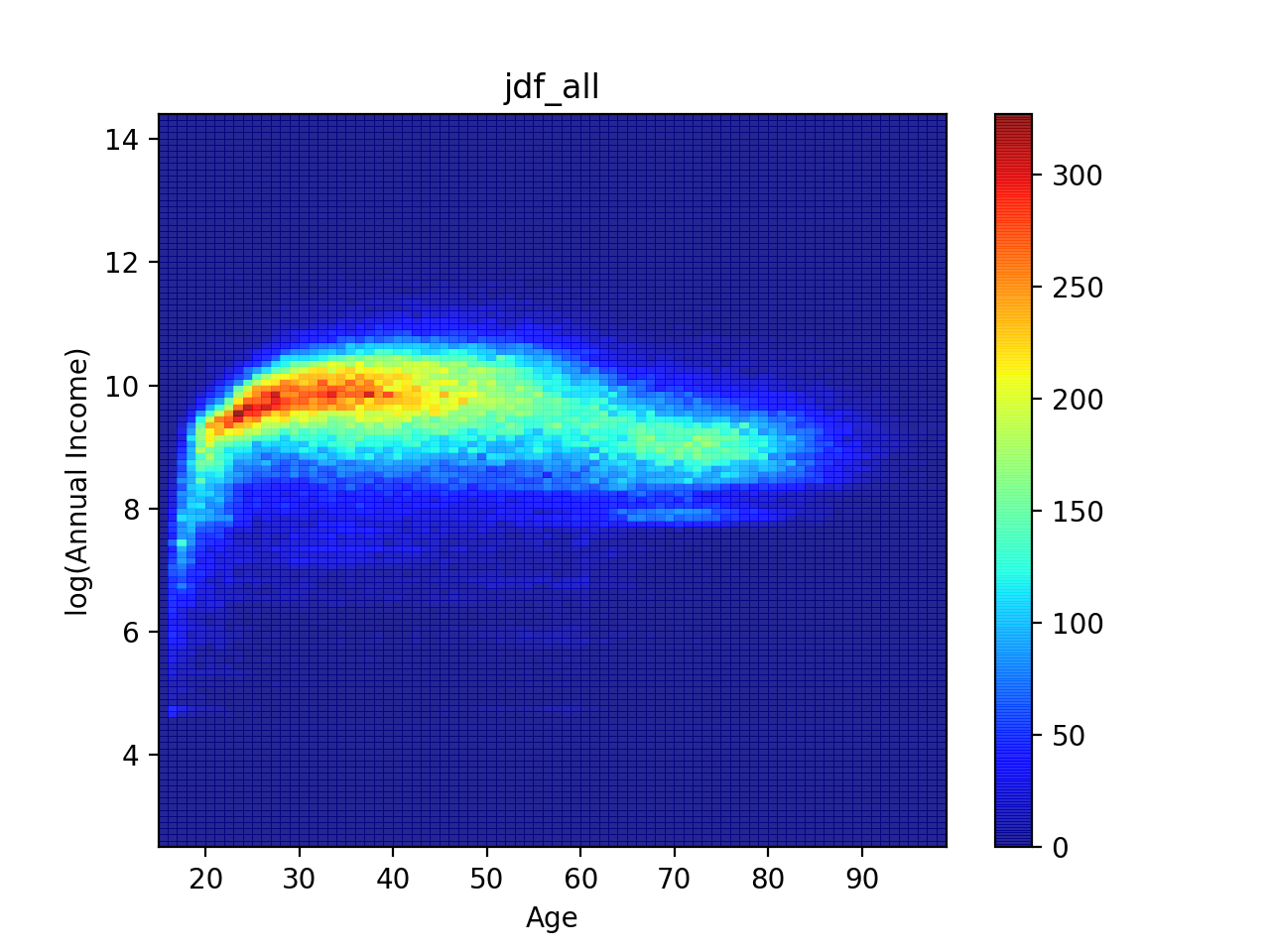}}\\
         \subfloat[USA Labour JDF]{\includegraphics[width=0.45\linewidth]{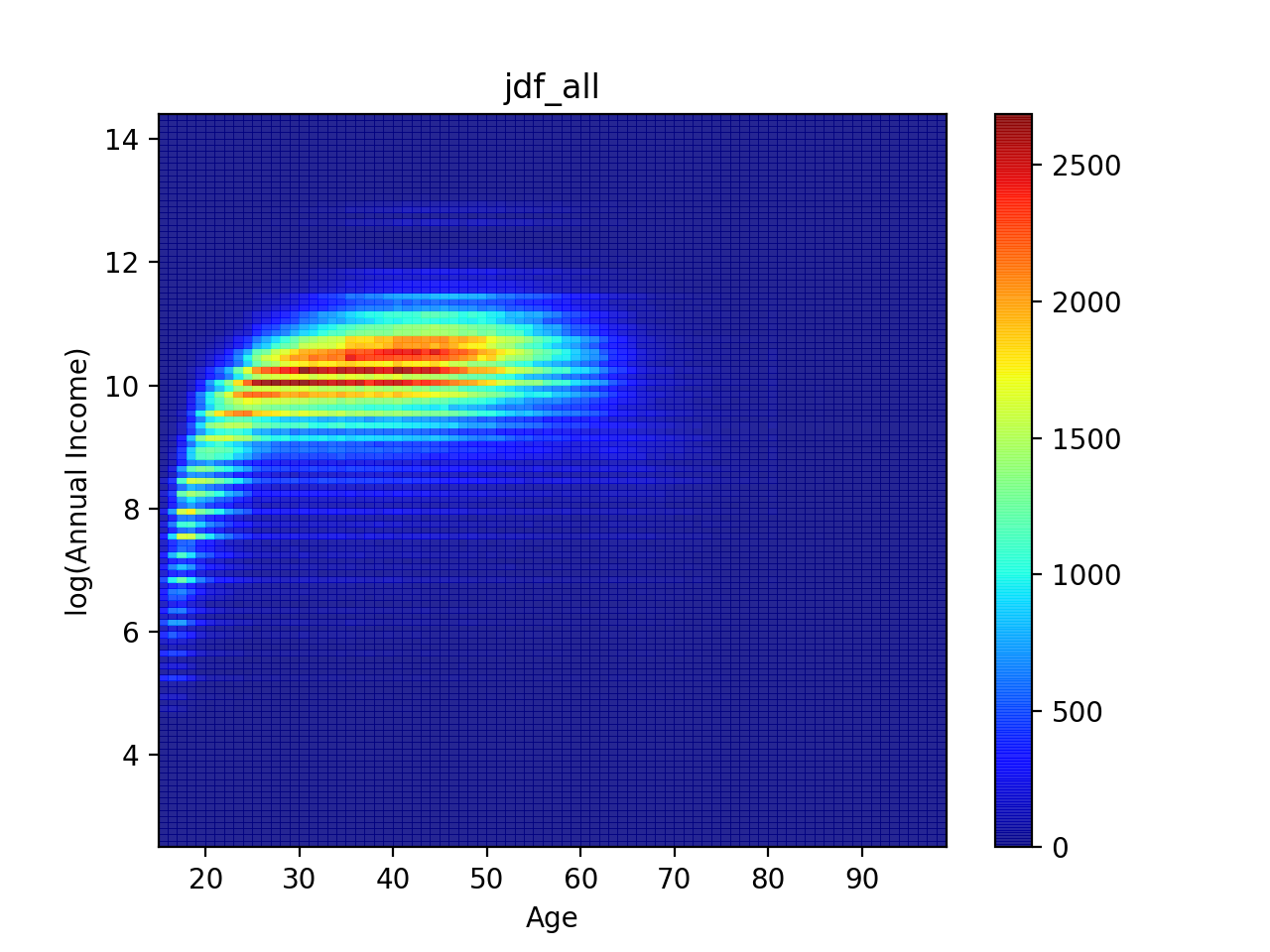}}\qquad
         \subfloat[USA Total JDF]{\includegraphics[width=0.45\linewidth]{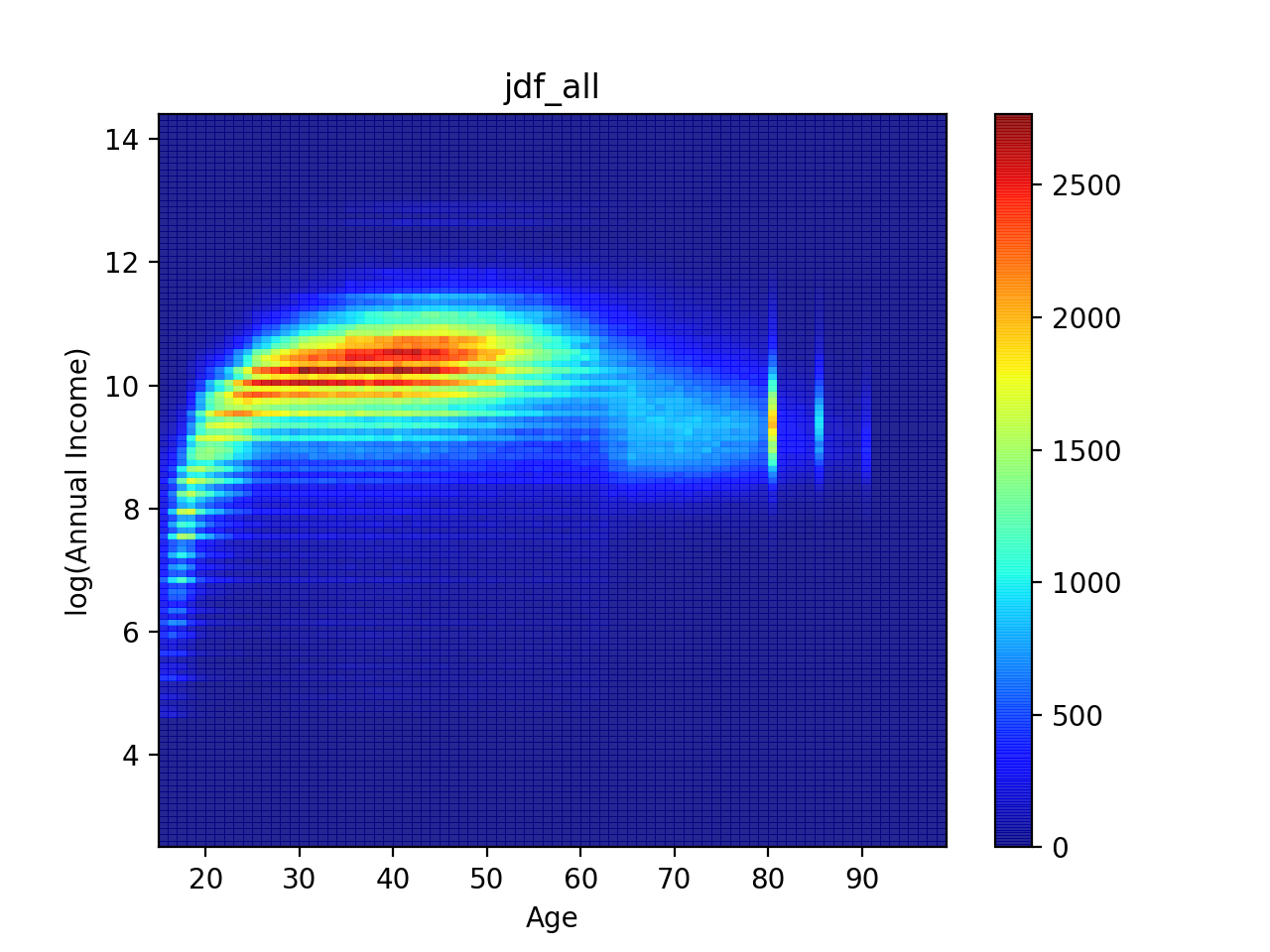}}\qquad
         \caption{All Waves JDF for UK and USA  Income between Ages 15-100}
         \label{fig:jdf_uk_usa_income_15_100}
\end{figure}

Fig. \ref{fig:uk_usa_pyramid} reflects the Population Pyramid in the UK and USA, and how the shape evolved over the 18 years considered. The UK population sample from BHPS has a relatively balanced population with a slight weight towards younger cohorts initially in 1991, which denotes Wave 1. The UK population gradually got older, and the population pyramid reflects mass's shift towards older generations, this shift happened gradually over the years. The US population from CPS reflects a young population in Wave 1 with a notable skew towards younger cohorts, after 17 years the US population loses this property towards younger cohorts and gets significantly older. 
Both the UK and US population get older and reflect a trend towards an ageing population, which will significantly impact the pension system.

The shape of the population pyramid and its evolution with time from the panel survey reflect an ageing community \cite{onspopulation}. 
JDFs of Total and Labour Income in the UK and USA are presented in Fig.\ref{fig:jdf_uk_usa_income_15_100}. 
There is a sharp increase between the ages 15-20, which can be interpreted as the beginning of the work-life, transitioning from part-time work to full-time work, and graduation from higher or vocational education. 
The most significant difference of the UK Total JDF for the UK Labour JDF is the tail section corresponding to the retired population, which denotes the significant percentage of  individuals older than 55. The tail section is relatively concentrated, which can be explained by the state pension benefit levels and mandatory social security system.
The US population reflects a surprisingly sparse older cohort for the Total Income data, and the most significant difference to the UK is the relatively lower income levels compared to the wage income. 
A higher variance spread to a wider band, which might be caused by a non-standard retirement system not supported by strong state pension benefit and mandatory pension schemes during employment.

The comparison for the model simulation and observed data shows common characteristics, as the joint distribution of age and income in logarithmic scale is presented in Fig.\ref{fig:jdf_uk_usa_income_15_100} : an initial sharp rise between the entrance to the graph on 16 years old, the amount of 16 years old includes pocket money, allowance and part-time or internship jobs. There is a steep increase in mean and variance between the initial income and income at the age of 20. The increase is sharper for the mean in comparison to the variance. The population's mass has similar characteristics with near 23k GBP annual income, and for ages between 20-45. 
Between ages 65-75, there is a significant decrease in income and after 75, the income converges to a certain mean. 
The data and the models provide an essential tool to tackle problems related to an ageing population and shocks introduced by technological and political changes. 

In the following sections we will focus on Labour Income and employed labour population. 
Total Income covers all of the income streams including Transfer Income such as pocket-money, Labour Income, Capital Income, and Pension Income; these different streams might be governed by varying dynamics non-uniform across the type of income; so we decided to focus on Labour Income, which involves the broadest section of the population; with most significant impact. 
The only other primary source of income in terms of gross value is the Capital Income, which might be significantly affected by other factors such as inter-generational shifts, market conditions and global financial state. 
In order to focus on labour income dynamics, the other income sources are left out of our modeling. 

\section{Data Processing and Calibration}

BHPS provides a vast amount of socio-economical data for each individual and household participating in the study. 
The columns of income, age data, the individual's statistical weight -representative of the British population and overall survey- with the individual's intra-wave unique identifier mostly suffice for this paper's scope.
PID, Wxage12, wFIYR, Wxrwght fields of BHPS are used for each wave.

The Income variable xfiyr is each individual's annual income, including labour income, benefits, pensions, transfer income, and investment income. Participants were asked according to annual income in the reference year from September in the year prior to the interview until September in which interviewing begins \cite{bhpsmanual}.  
The income figures are adjusted for inflation, as part of pre-processing. 
During the dataset preparation, a floor wage is determined to exclude in labour income, which denotes to excluding part-time and short-employment income. 
The income data is inflation-adjusted and transformed into log-domain.

IPUMS harmonizes the CPS and provides IPSUM CPS micro-data. The IPSUM CPS includes a large spectrum of topics such as demographics and employment, as well as supplemental studies such as the Annual Social and Economic Supplement (ASEC).
Each individual can be identified by "CPSIDP", "INCTOT" and "INCWAGE" correspond to the total income and wage income, and "ASECWT" denotes the weights derived from ASEC Supplement. The data set is topcoded, and specific codes are used for labelling missing and incorrect data. The ages over 80, 90 and 99 are top-coded till 2004, and after 2004, the top-coding bins are determined as 80,85,90 and 99 by the panel data collectors\cite{ipumscps}.
Although this dataset contains high-income individuals, there is top-coding applied, so individuals with very high income are not included.

\subsection{Fitting Distributions}

Estimating the income evolution function parameters is the most critical part of the research, and the decision depends on various factors such as the type of data, bias, and assumptions. 
Various techniques are investigated, leading to different results, with each having unique strengths and weaknesses. 
\\
The first method investigated is Generalized Method of Moments (GMM), which presumes that the first three moments of the income evolution functions provide the necessary information for approximating the underlying generative process. 
The equations of the first three moments of the income evolution equation can be solved for the parameters  $q_{a}$, ${\sigma}_{a}$, ${\mu}_{a}$. 
Both of the BHPS from the UK and the IPUMS CPS from the USA can be used for estimation with Generalized Method of Moments (GMM) with first three moments.
\\
The second method utilises the micro-data from the longitudinal surveys, which tracks the individual for consecutive years. The parameters are approximated to fit the income evolution function using least squares minimisation for the individuals participating in the studies for consecutive years. 
The BHPS from the UK is a suitable micro-data consisting of a panel survey, and the survey tracks income of the same individuals over the years. 

\subsection{Estimation for Generalized Method of Moments (GMM)}
The three moments of the age income evolution function are utilised to find a polynomial; afterwards, the equation is solved for $q_{a}$, ${\sigma}_{a}$ and ${\mu}_{a}$; at this point, a observed solution for parameters is found, but the relationship captures only the dynamics of the first three moments. 
Calculations can be found in appendices. 
The statistical variables such as $\bar{y}_{a}$ are found for each wave and than averaged across waves for finding a one set of stationary variables, which can be used to estimate  $q_{a}$, ${\sigma}_{a}$, ${\mu}_{a}$.
The details and derivations for the GMM estimation technique can be found in appendix.
\subsection{Estimation of Least Squares for Micro Data}
The Least Square Method requires that an individual's income for two consecutive years be existent in the dataset, this restriction is fulfilled by the BHPS, a panel survey, but the CPS IPUMS population survey does not satisfy this condition. 
The income data from two consecutive years per agent is used to estimate age-specific parameters, which characterise the income evolution function at Eq.\ref{eq:agent_income_evo}. LSM tries to estimate parameters by fitting the data to the income evolution function. 

\section{The generative model}
The model can also be used for simulation and forecast, tracking income trajectories of the individuals, providing a bench table for observing the stylized facts and complex properties of the income dynamics.
Following the estimation of model parameters, the model is bootstrapped with data from wave 1 for initialising the simulation. 
Each individual from wave 1 is initialised as an agent in our model. According to Age Income Evolution Dynamics Eq. \ref{eq:agent_income_evo} , the income of these agents is transitively updated at each consecutive wave update. 
$ \eta_{i t}$ provides the random feed, which introduces variability for the income evolution of the agents. 
At each wave update, a new generation of agents consisting of 25 years old agents from the initial wave is injected.
Following each wave, distributions corresponding to the state of the simulated population are calculated. 
A full calibration of the model is shown in the Supplementary Material. 

\section{Parameters}
The optimized performance of these three methods are compared and discussed in the following sections. 

No boundaries are explicitly imposed by LSM estimation.

\begin{figure}[htp]%
         \centering
         \subfloat[$q_{a}$ with GMM]{\includegraphics[width=0.40\linewidth]{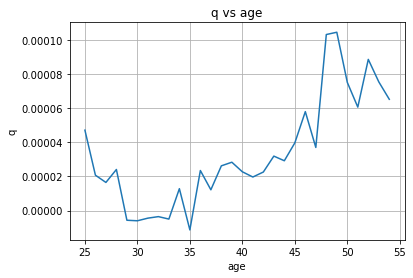}}\qquad
         \subfloat[$\sigma_{a}$ with GMM]{\includegraphics[width=0.40\linewidth]{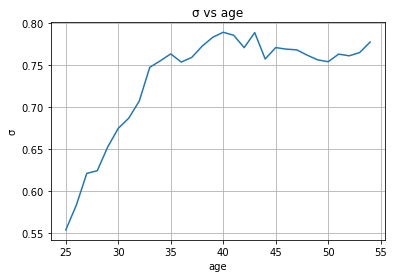}}\qquad
         \subfloat[$\mu_{a}$ with GMM]{\includegraphics[width=0.40\linewidth]{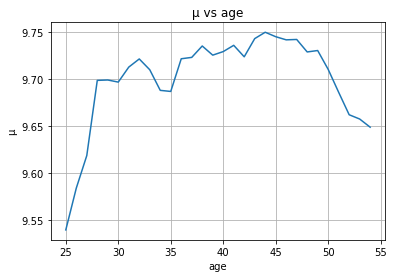}}\\
         \subfloat[$q_{a}$ with LSM]{\includegraphics[width=0.40\linewidth]{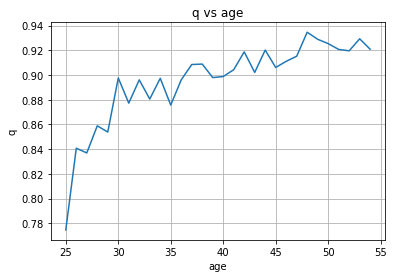}}\qquad
         \subfloat[$\sigma_{a}$ with LSM]{\includegraphics[width=0.40\linewidth]{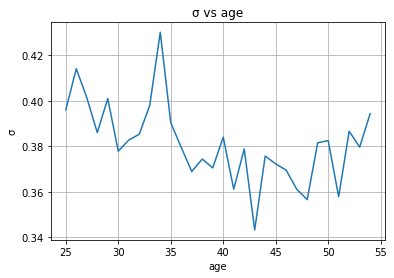}}\qquad
         \subfloat[$\mu_{a}$ with LSM]{\includegraphics[width=0.40\linewidth]{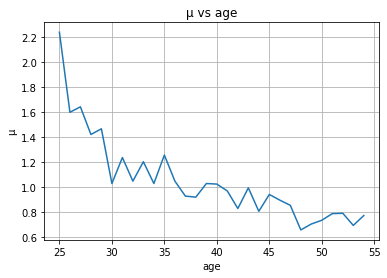}}\\
         \caption{$q_{a}$, $\sigma_{a} $ and $ \mu_{a} $ Plots for UK Labour Income}
         \label{fig:qms_uk_labour}
\end{figure}
The $\mu$, $\sigma$ and $q$ variables are independent of each other, but the estimation process or data itself can introduce a slight dependence. 
The GMM estimation technique results in minimal $q$ values near 0, so the estimated parameters approximately resemble an auto-regressive model. However, despite near 0 negligible $q$ values, the $q$ plot has a distinct shape with an increasing trend with a small decrease  between 25 and 30, has very different characteristics depending on the estimation method. 
The GMM estimation method results in minimal $q$ and the $\mu$ reflects the characteristics of $\bar{y}$, which is in compliance with this estimation method's nature.  
The $\mu$ value increases at first and then plateaus and slightly decreases near retirement. 
On the contrary LSM estimation mainly characterises the income with an increasing $q$ parameter, so the $\mu$ parameter has limited effect and reflects a decreasing trend. $\sigma$ values reflect a distinct trend of initially decreasing values with a spike around the age of 34 followed by a stable decrease and noisy plateau with a minor increase towards 55. 
The LSM with bootstrap is the most accurate estimation method and reflects the characteristics of the model clearly.

\begin{figure}[htp]%
         \centering
         \subfloat[Simulation Statistics with GMM Estimation]{\includegraphics[width=0.45\linewidth]{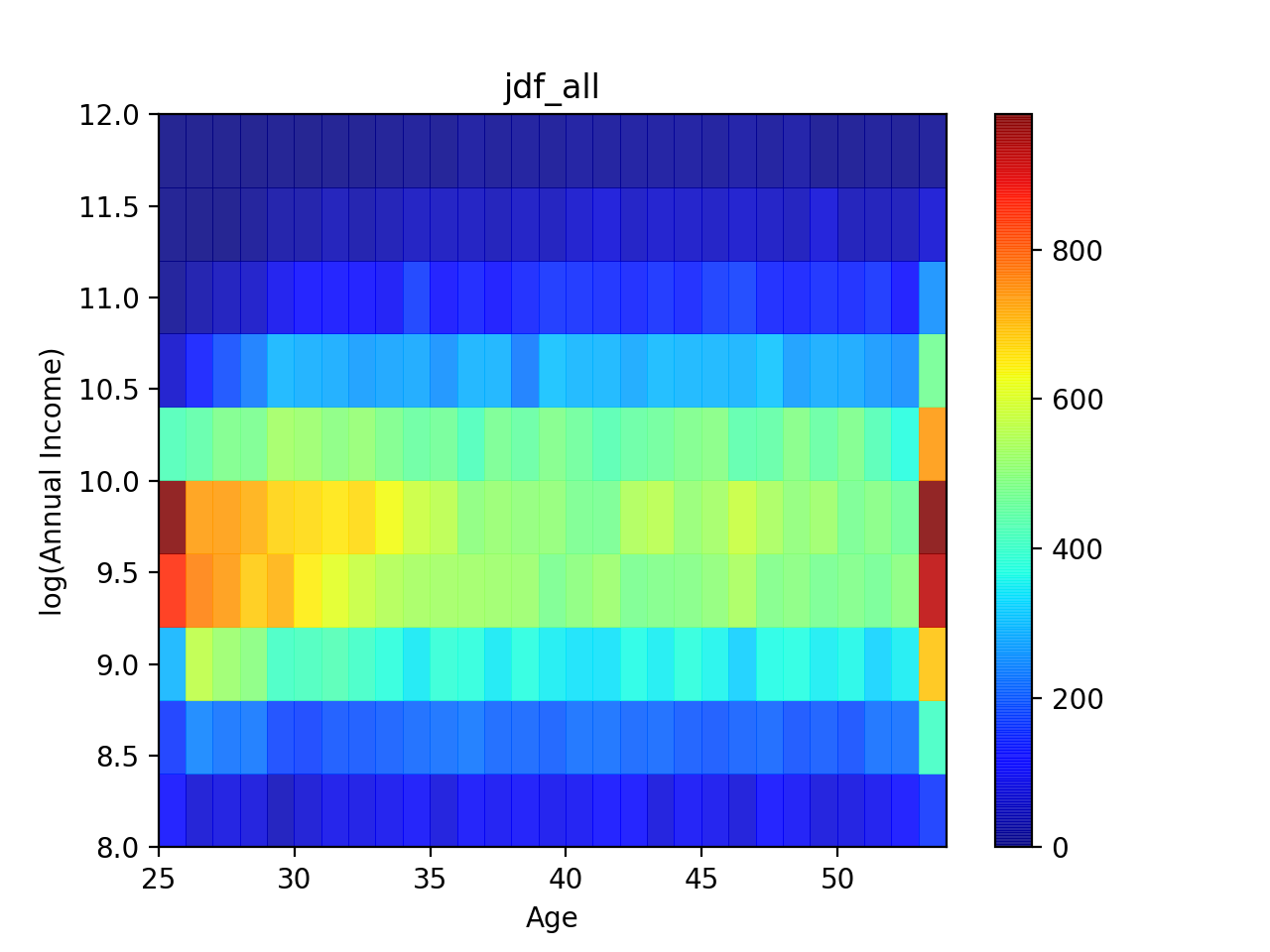}}\qquad
         \subfloat[Simulation Statistics with LSM]{\includegraphics[width=0.45\linewidth]{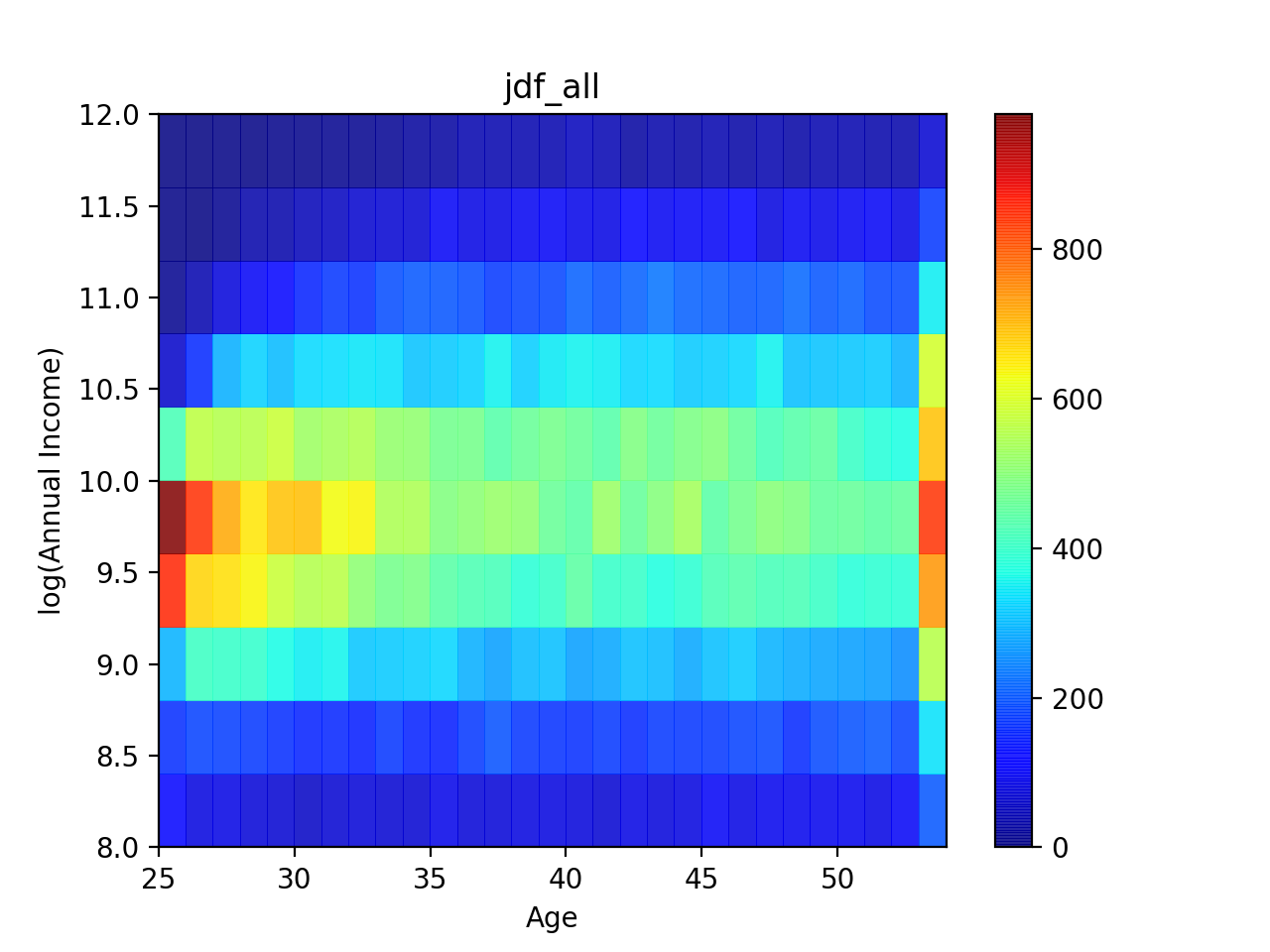}}\\
         \caption{UK Labour Data Simulation All Waves JDF}
         \label{fig:uk_labour_sim_jdf}
\end{figure}

\textbf{GMM}
\newline
GMM estimation technique approximates the $\mu_{a}$ values to be consistently around 10 and the $q_{a}$ values are around 0 with an initial sine-like wave followed by a steady increase. 
The $\sigma_{a}$ values are around 0.8 and have a positive trend. 
$q_{a}$ values display a positive trend as well. 
The  $ \bar{y}_{a} $ and $ std(y)_{a} $ plots of the simulation is similar to the observed data, but the standard deviation plot is particularly noisy. 
The JDF of the simulation on Fig. \ref{fig:uk_labour_sim_jdf} is sparse, consistent; but not highly concentrated around mean. 
Both of these methods depends on assumptions about the dynamics of the income evolution function. 
The GMM method assumes that the first three moments of the equation are enough for estimating the parameters because they provide a solvable system. 
However individual characteristics in an age group such as different income levels and clusters within are lost during the moment estimation.
\newline

\begin{figure}[htp]%
         \centering
         \subfloat[$q_{a}$ with GMM]{\includegraphics[width=0.40\linewidth]{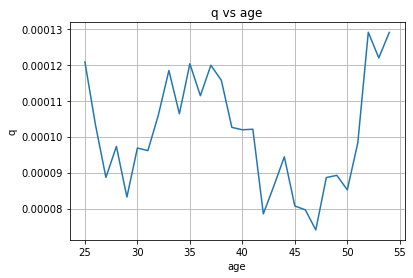}}\qquad
         \subfloat[$\sigma_{a}$ with GMM]{\includegraphics[width=0.40\linewidth]{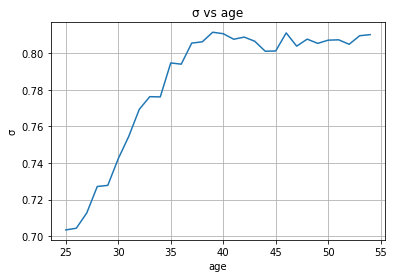}}\qquad
         \subfloat[$\mu_{a}$ with GMM]{\includegraphics[width=0.40\linewidth]{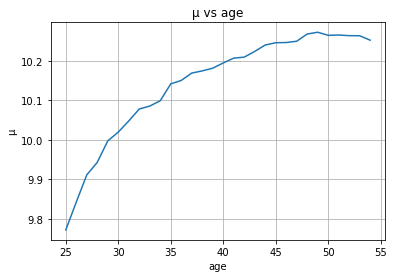}}\\
         \caption{$q_{a}$, $\sigma_{a} $ and $ \mu_{a} $ Plots for USA Labour Income}
         \label{fig:qms_usa_labour}
\end{figure}
\begin{figure}[htp]%
         \centering
         \subfloat[Observed Statistics]{\includegraphics[width=0.90\linewidth]{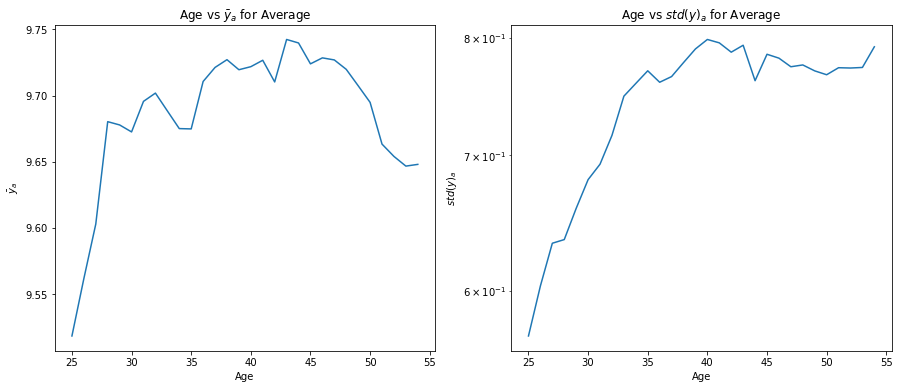}}\\         
         \caption{UK Labour Data Observed Statistics}
         \label{fig:uk_labour_real_stat}
\end{figure}
\begin{figure}[htp]%
         \centering
         \subfloat[Simulation Statistics with GMM Estimation]{\includegraphics[width=0.90\linewidth]{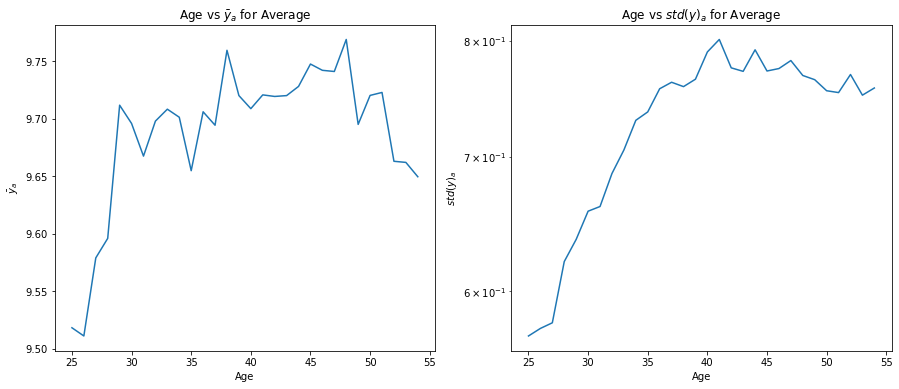}}\\
         \subfloat[Simulation Statistics with LSM Estimation ]{\includegraphics[width=0.90\linewidth]{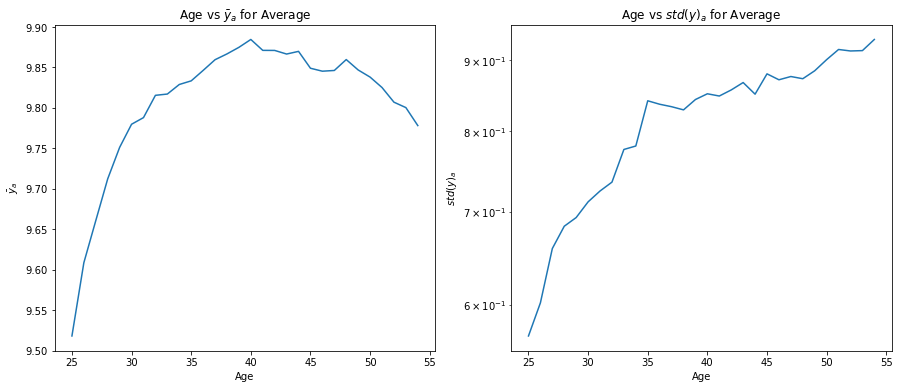}}\\
         \caption{UK Labour Data Simulation Statistics}
         \label{fig:uk_labour_sim_stat}
\end{figure}

\textbf{LSM by individual transitions}
\newline
\\
To use LSM to approximate the parameters, one needs the individual income transitions in consecutive years, thus identifying the same individual in consecutive cohorts is necessary, the panel studies such as BHPS satisfy this condition. 
The age-dependent income evolution function is fitted with individual income transitions of consecutive years. 
The JDF of simulation has concentrated heat regions around the mean, and the trend is decreasing unlike the observed data.  
Imposing boundaries to the parameter space results in better parameters, which results in consistent parameter plots and the  $ \bar{y}_{a} $ plot of the simulation reflect similar shape with the observed data Fig. \ref{fig:uk_labour_real_stat} and Fig. \ref{fig:uk_labour_sim_stat}. 
The JDF of the simulation on Fig.\ref{fig:uk_labour_wave_jdfs} is able to reflect the dispersion among various clusters better because unlike the other methods heavily depending on the statistics such as mean and standard deviation of the entire age group, the LSM utilises individual-level microdata. 

\begin{figure}[htp]%
         \centering
         \subfloat[$q_{a}$ with LSM]{\includegraphics[width=0.40\linewidth]{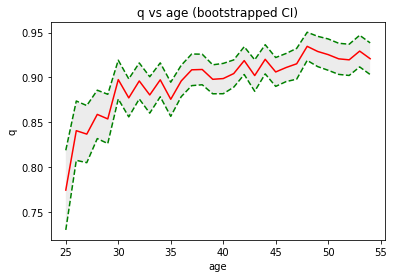}}\qquad
         \subfloat[$\sigma_{a}$ with LSM]{\includegraphics[width=0.40\linewidth]{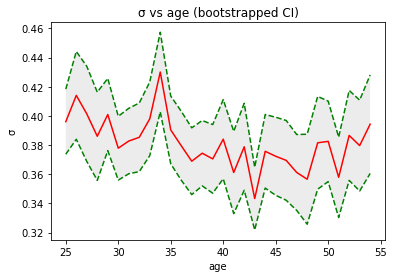}}\qquad
         \subfloat[$\mu_{a}$ with LSM]{\includegraphics[width=0.40\linewidth]{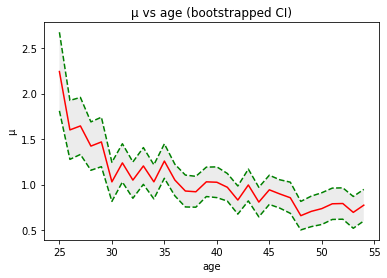}}
         \caption{$q_{a}$, $\sigma_{a} $ and $ \mu_{a} $ Confidence Interval for UK Data LSM Estimation}
         \label{fig:qms_ci_uk_lsm}
\end{figure}

The 95\% confidence interval with 2000 bootstrap samples for the estimated parameters from UK microdata by LSM can be found on Fig. \ref{fig:qms_ci_uk_lsm} 

It is evident from the plots of $\bar{y}_{a}$ and $std(y_{a})$ for the observed and simulated data that the model can capture the characteristics of the income conditional on age distribution, and the characteristic stationary property of this model can be observed on Fig. \ref{fig:uk_labour_real_stat} and Fig. \ref{fig:uk_labour_sim_stat}.

A close investigation of Fig. \ref{fig:uk_labour_real_stat} and Fig. \ref{fig:uk_labour_sim_stat} on UK Labour Income Data suggests that the GMM is most successful for reflecting the outcomes with similar mean and standard deviation characteristics of all waves after simulation with 18 waves that were simulated with the parameters $q_{a}$, ${\sigma}_{a}$, ${\mu}_{a}$estimated by the GMM. But LSM reflects the individual trajectories, and JDF more accurately.


\begin{figure}[htp]%
         \centering
         \subfloat[Observed Statistics]{\includegraphics[width=0.90\linewidth]{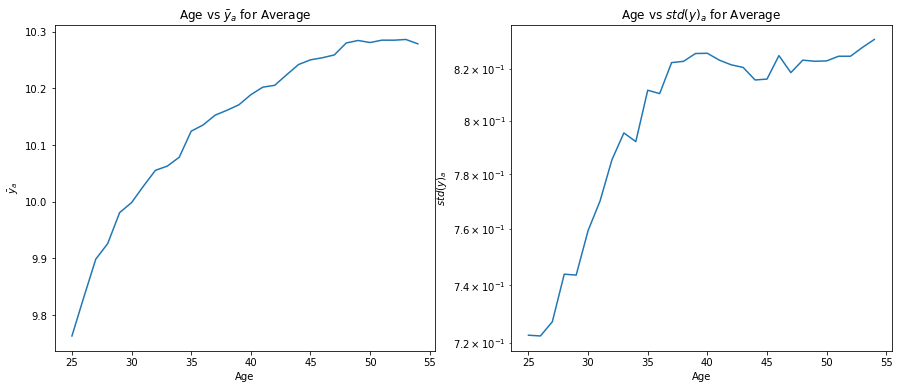}}\qquad
  \subfloat[Simulation Statistics with GMM]{\includegraphics[width=0.90\linewidth]{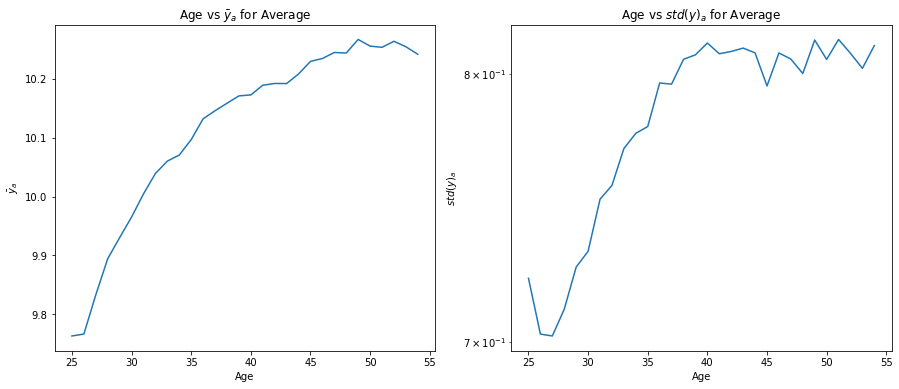}}\\
         \caption{USA Labour Data Observed and Simulation Statistics}
         \label{fig:usa_labour_real_sim_stat}
\end{figure}

The results showing the performance of GMM method is in Fig.\ref{fig:usa_labour_real_sim_stat}.

A general analysis of the comparison of joint distribution of age and inflation-adjusted income results in the following plots for weighted observed data and simulated data in Fig.\ref{fig:uk_labour_real_jdf} and Fig.\ref{fig:uk_labour_sim_jdf}:

\begin{figure}[htp]%
         \centering
         \subfloat[Observed Statistics]{\includegraphics[width=0.55\linewidth]{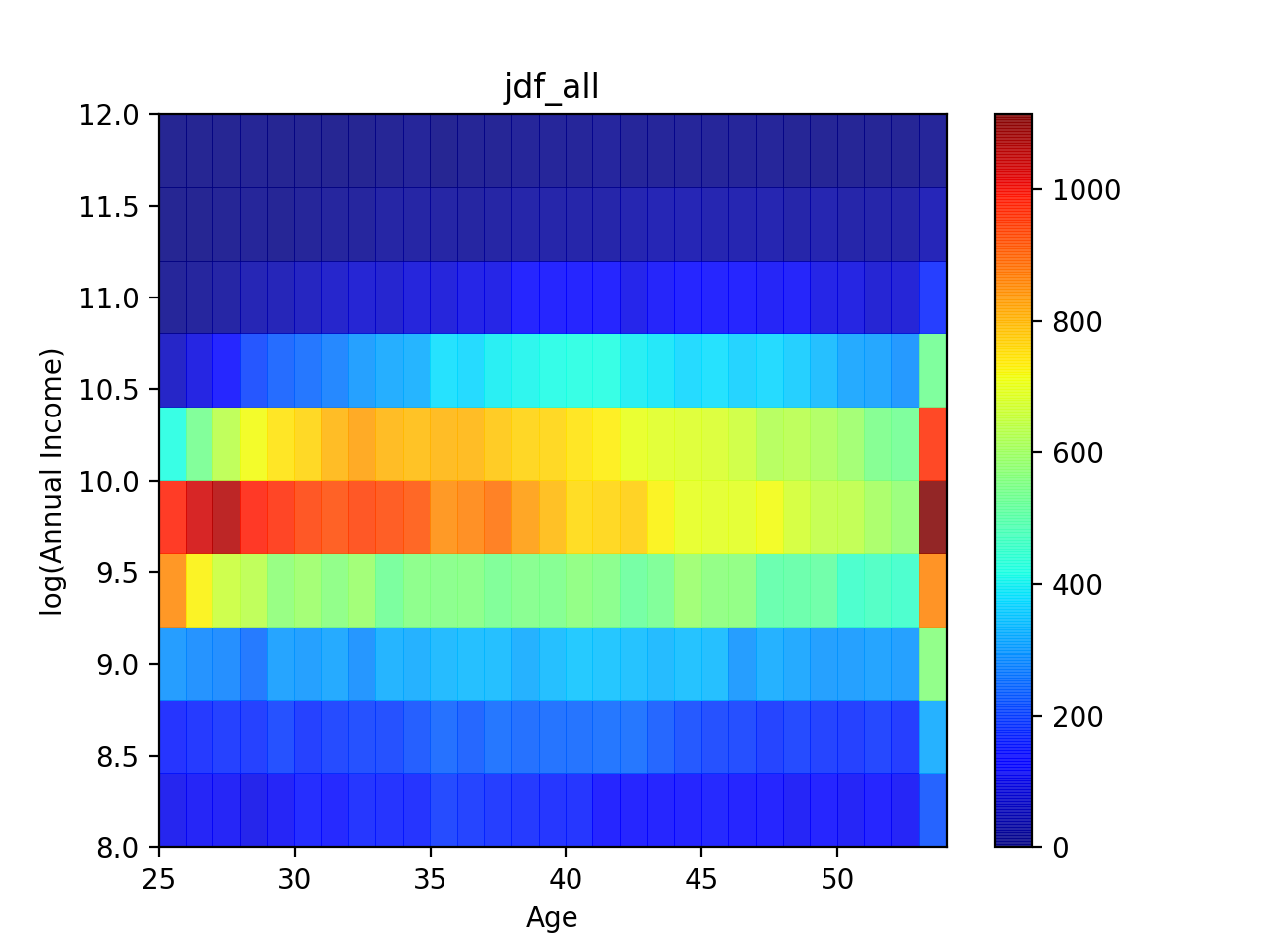}}\\         
         \caption{UK Labour Data Observed All Waves JDF}
         \label{fig:uk_labour_real_jdf}
\end{figure}

JDF of the simulated UK Labour data is in parallel with the expectations for GMM Estimation method, consistent and stable, resembling a similar shape but not concentrated for the heat regions with intense concentrations on Fig. $\ref{fig:uk_labour_real_stat} $ and Fig. $\ref{fig:uk_labour_sim_stat} $. 

The main differences between the observed and simulated JDFs are concentration of the mass of the population between 23 and 50.

\subsection{Wave-Specific analysis}

\begin{figure}[htp]%
         \centering
         \subfloat[1992 Observed Data]{\includegraphics[width=0.32\linewidth]{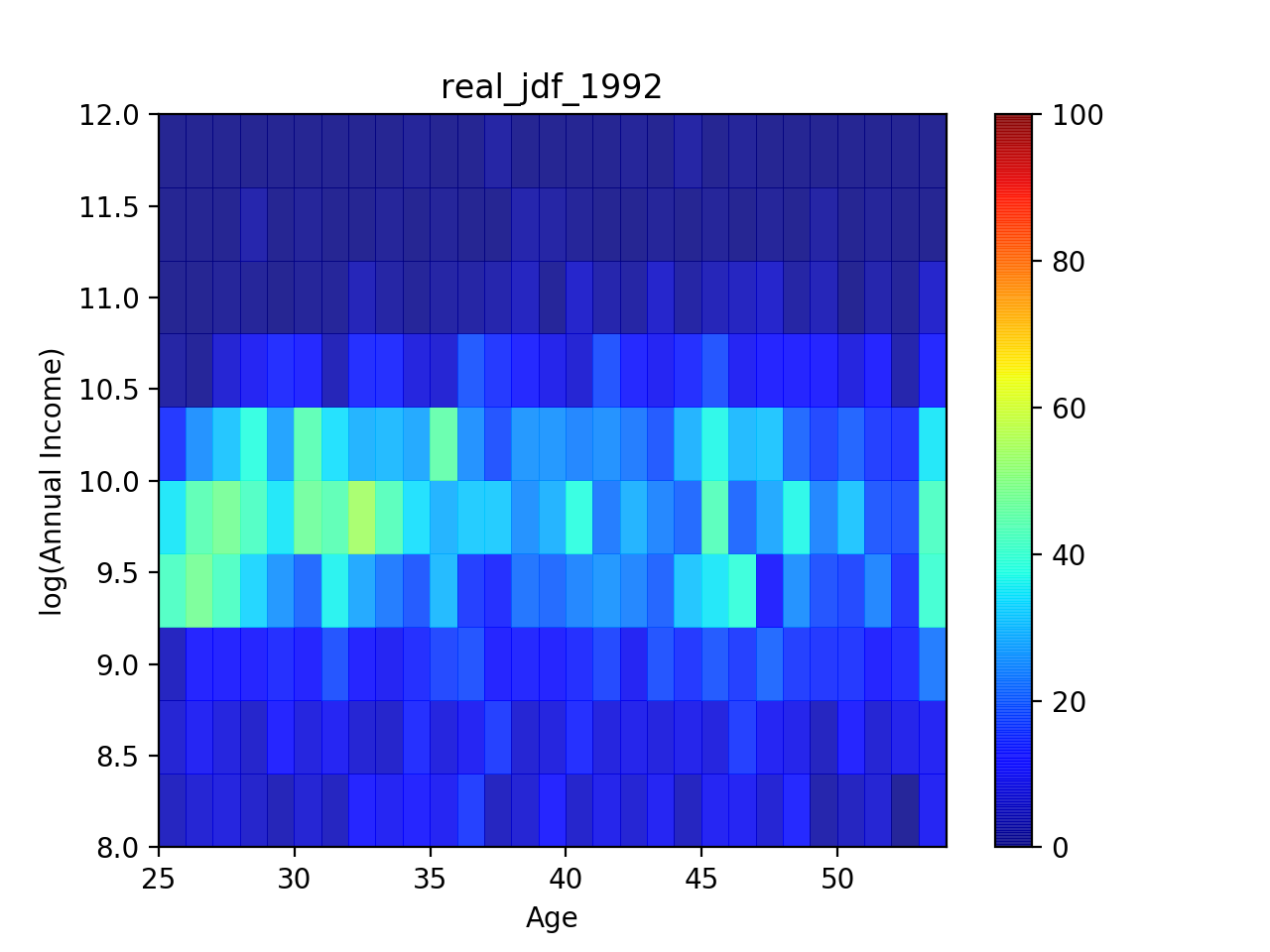}}\qquad
         \subfloat[1995 Observed Data]{\includegraphics[width=0.32\linewidth]{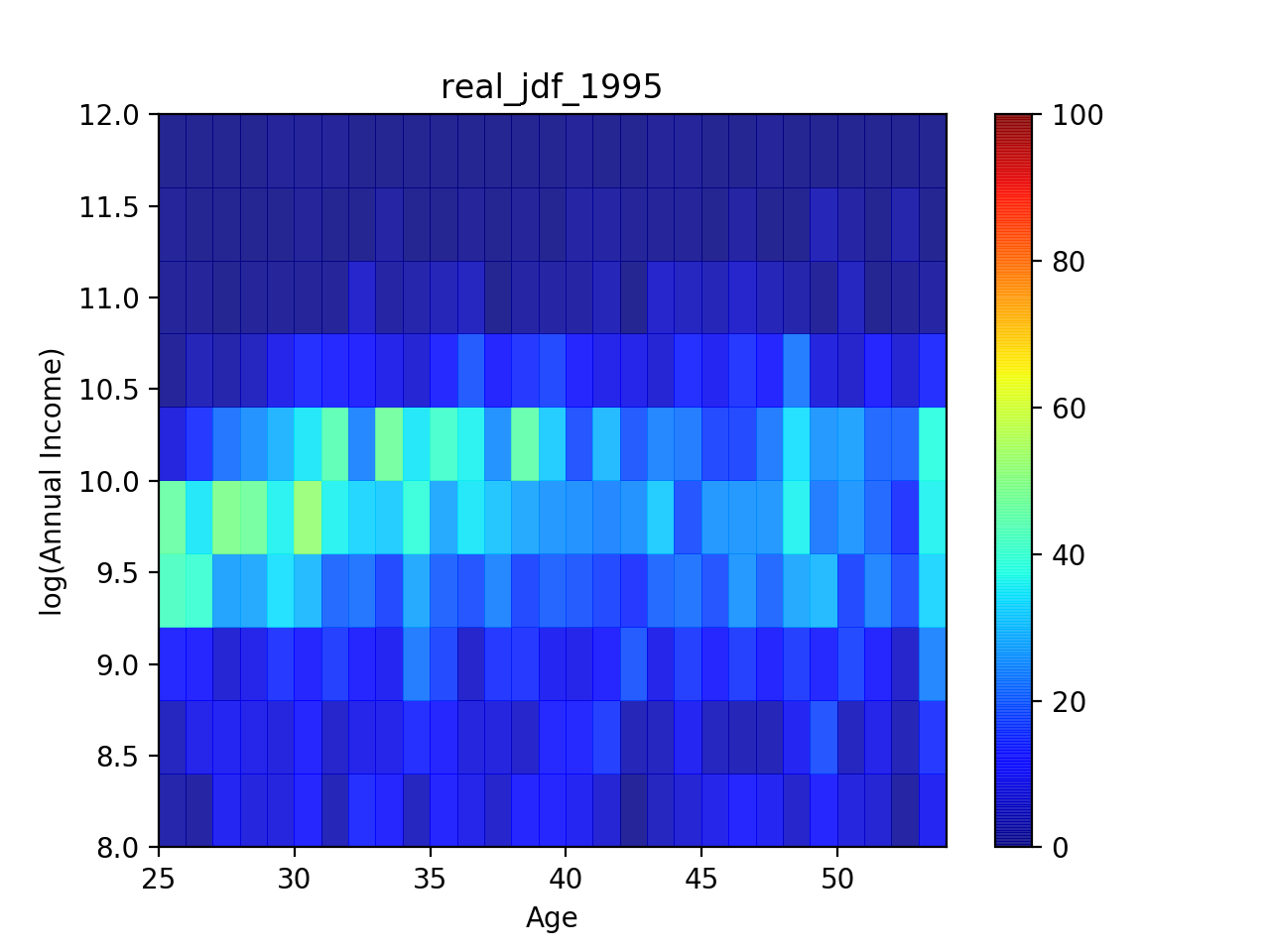}}\qquad
         \subfloat[2005 Observed Data]{\includegraphics[width=0.32\linewidth]{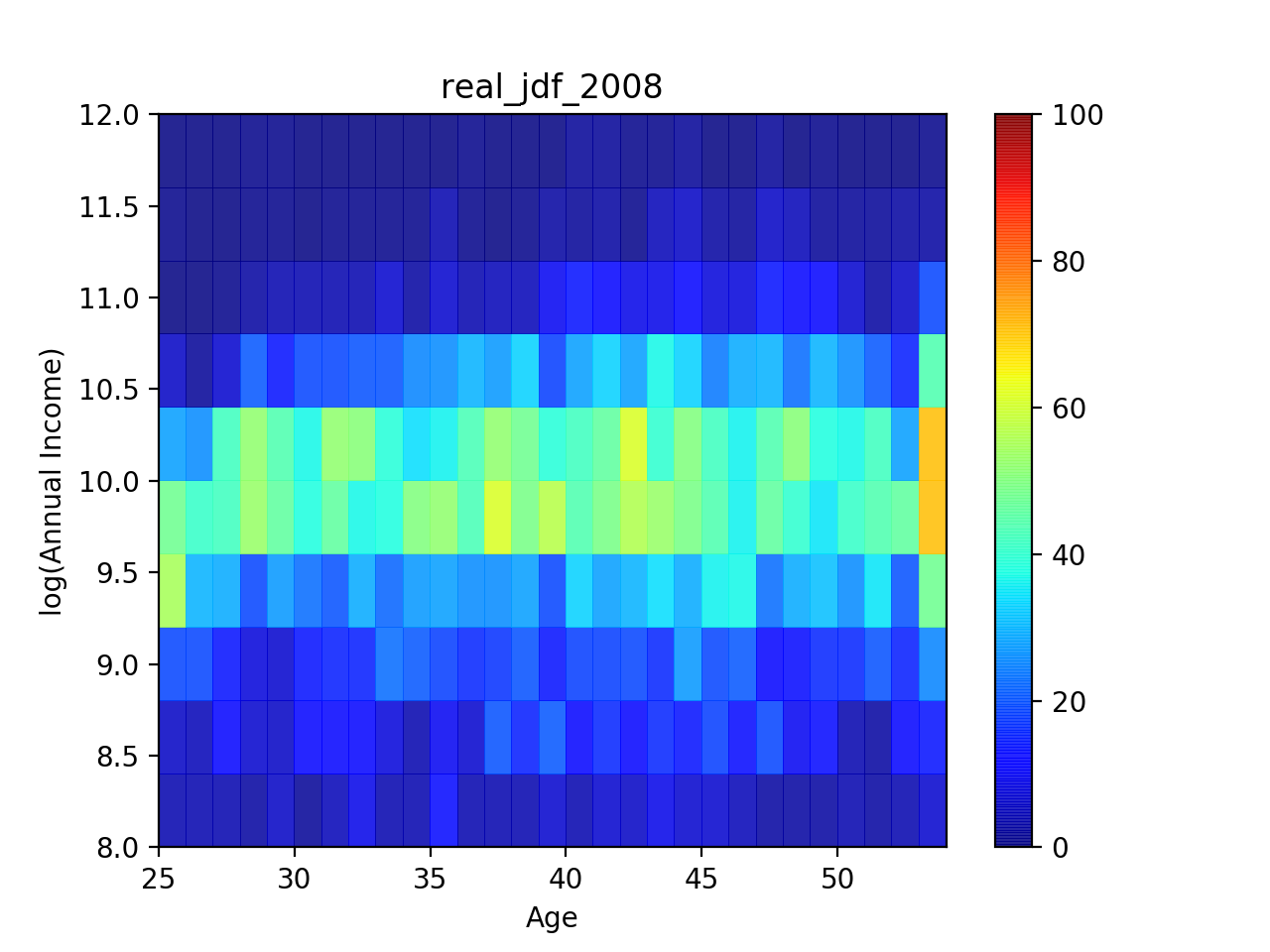}}\\
         \subfloat[1992 with GMM Estimation]{\includegraphics[width=0.32\linewidth]{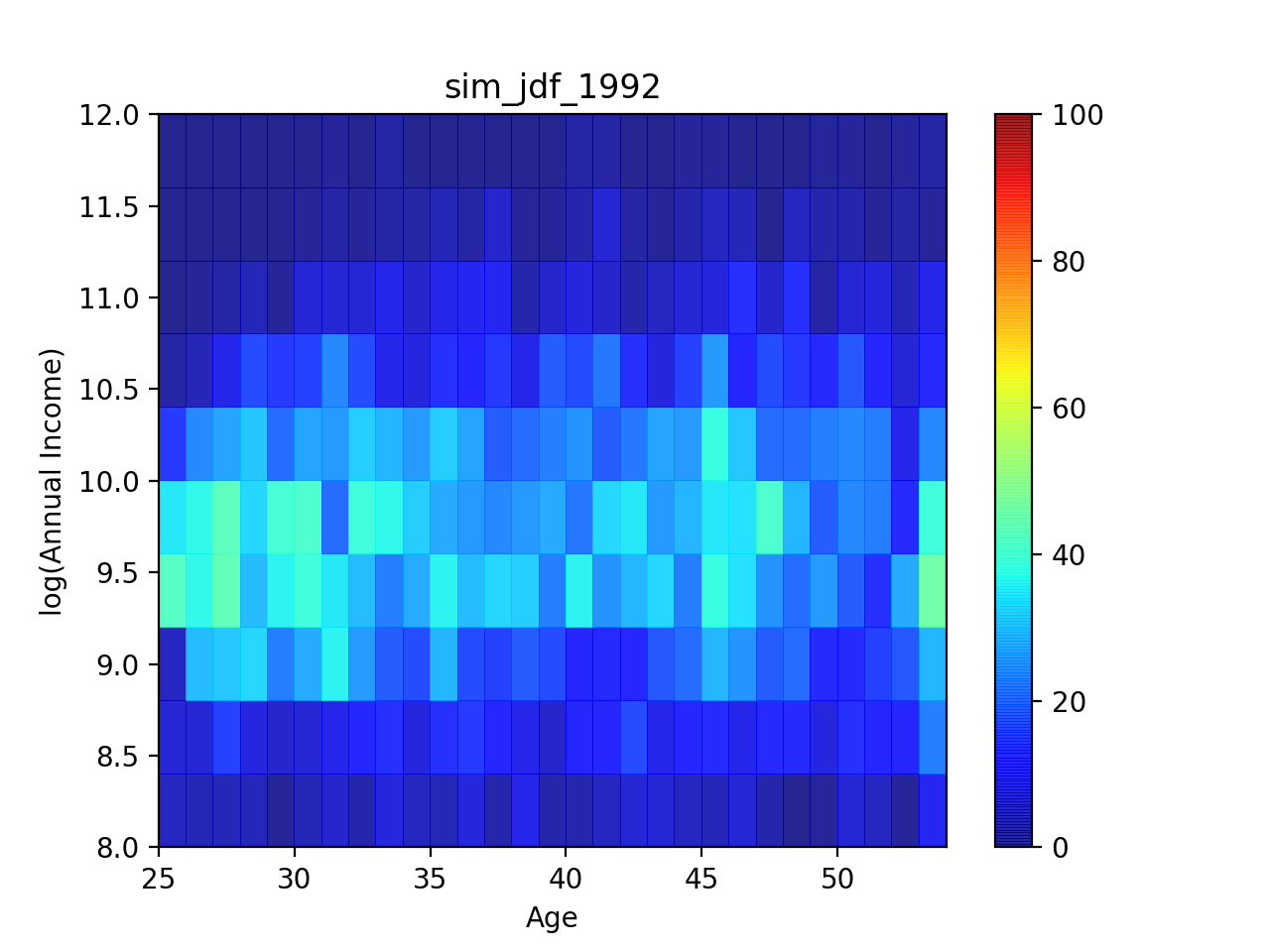}}\qquad
         \subfloat[1995 with GMM Estimation]{\includegraphics[width=0.32\linewidth]{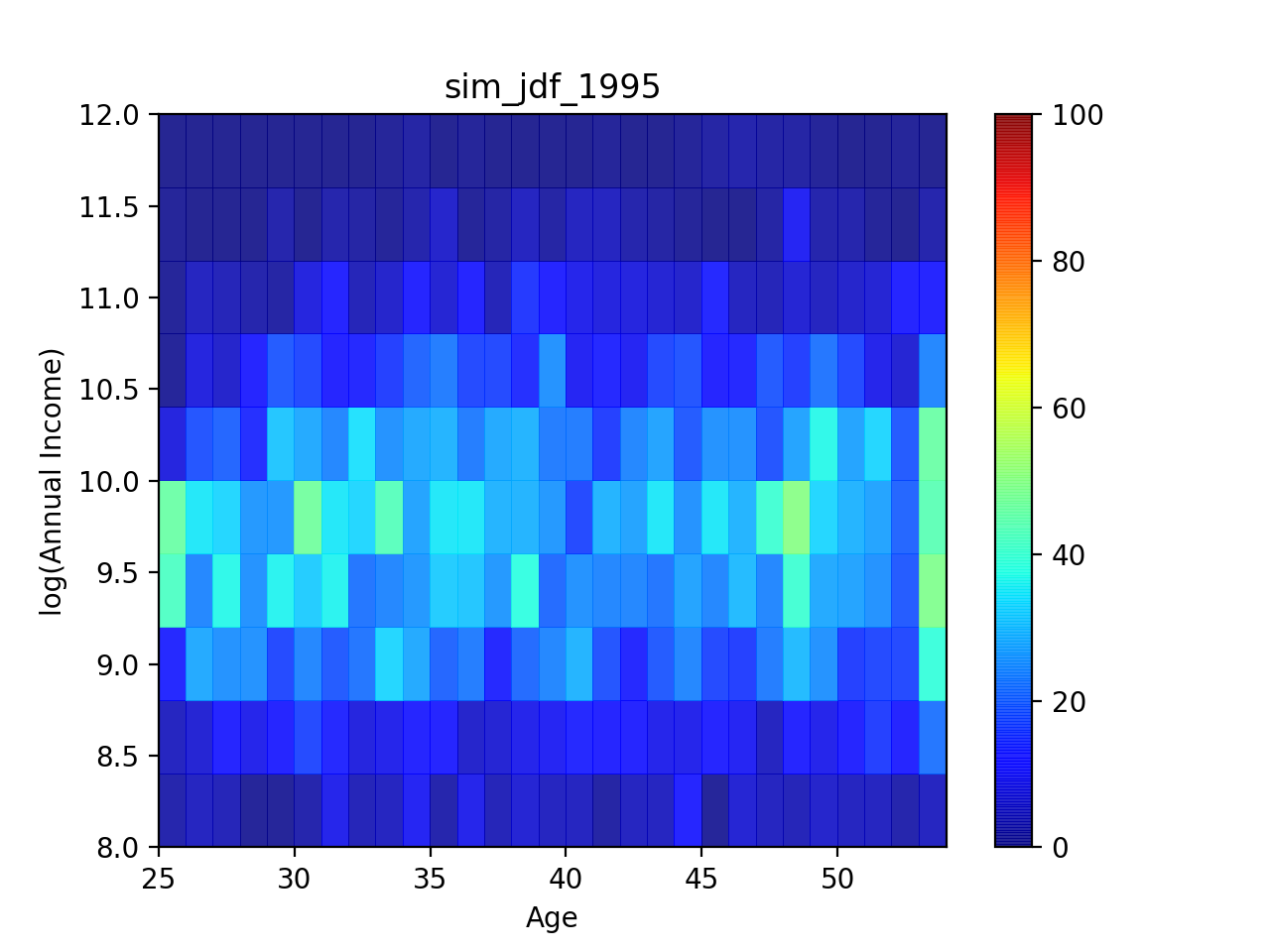}}\qquad
         \subfloat[2005 with GMM Estimation]{\includegraphics[width=0.32\linewidth]{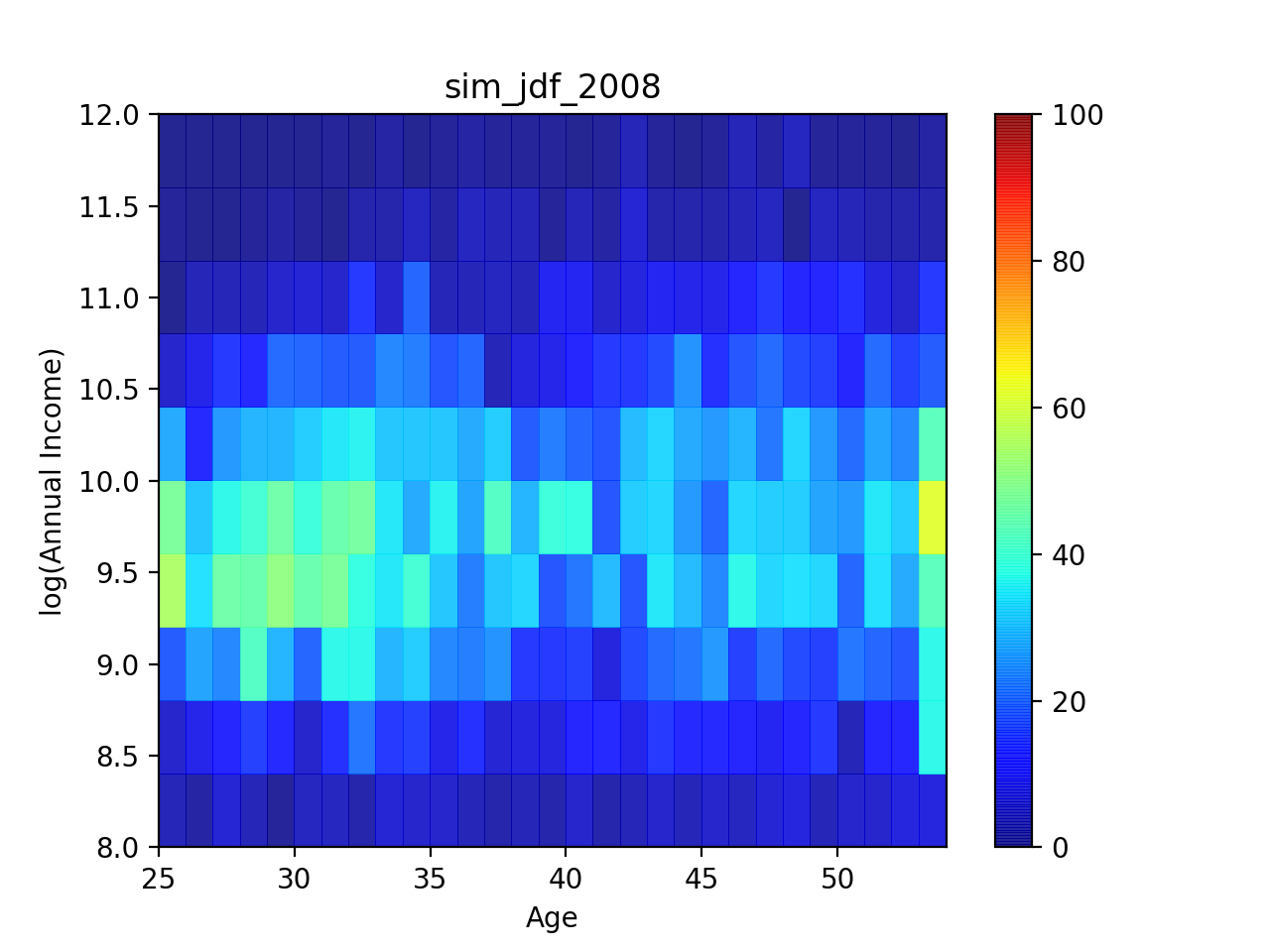}}\\
         \subfloat[1992 with LSM Estimation]{\includegraphics[width=0.32\linewidth]{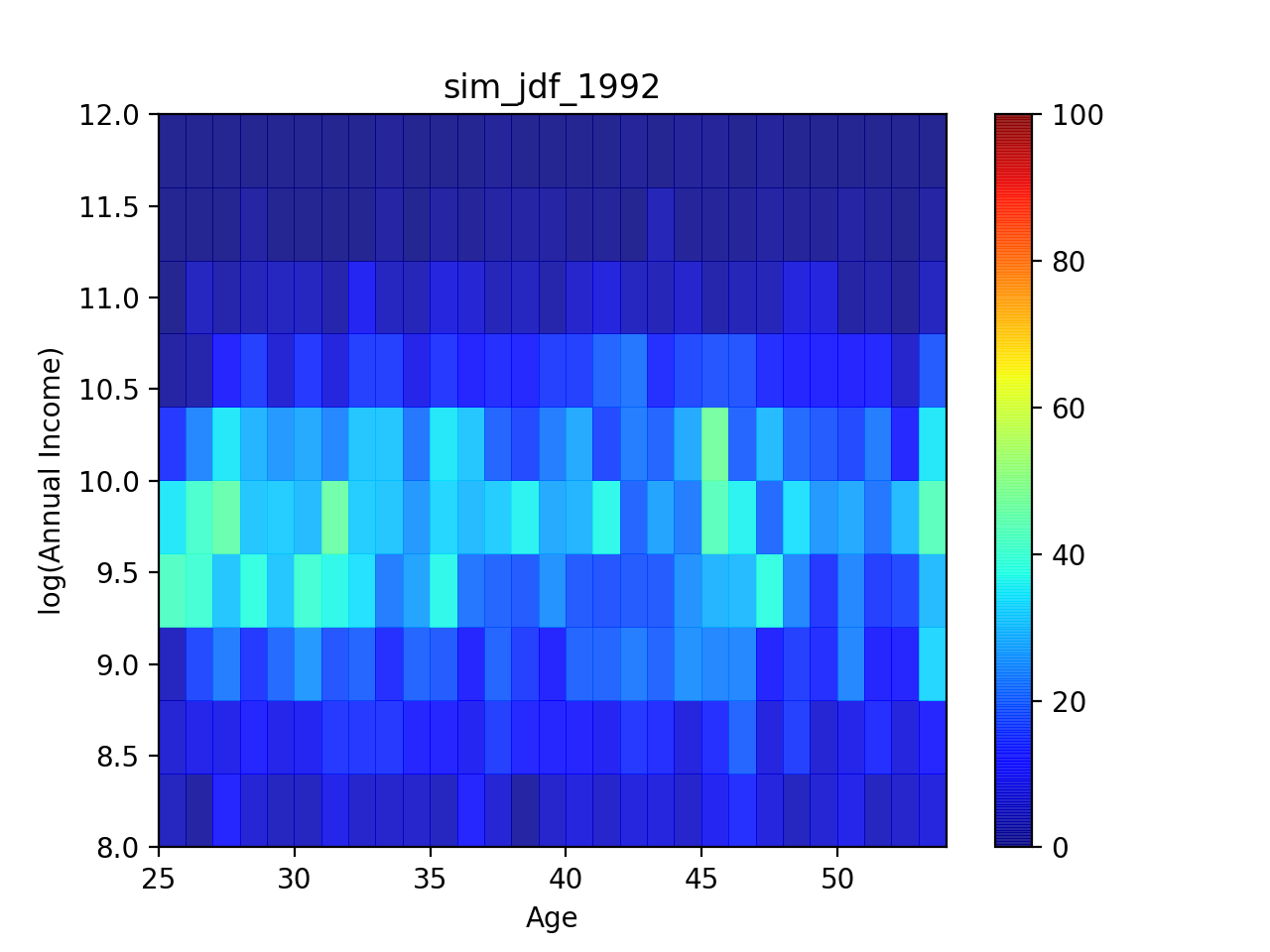}}\qquad
         \subfloat[1995 with LSM Estimation]{\includegraphics[width=0.32\linewidth]{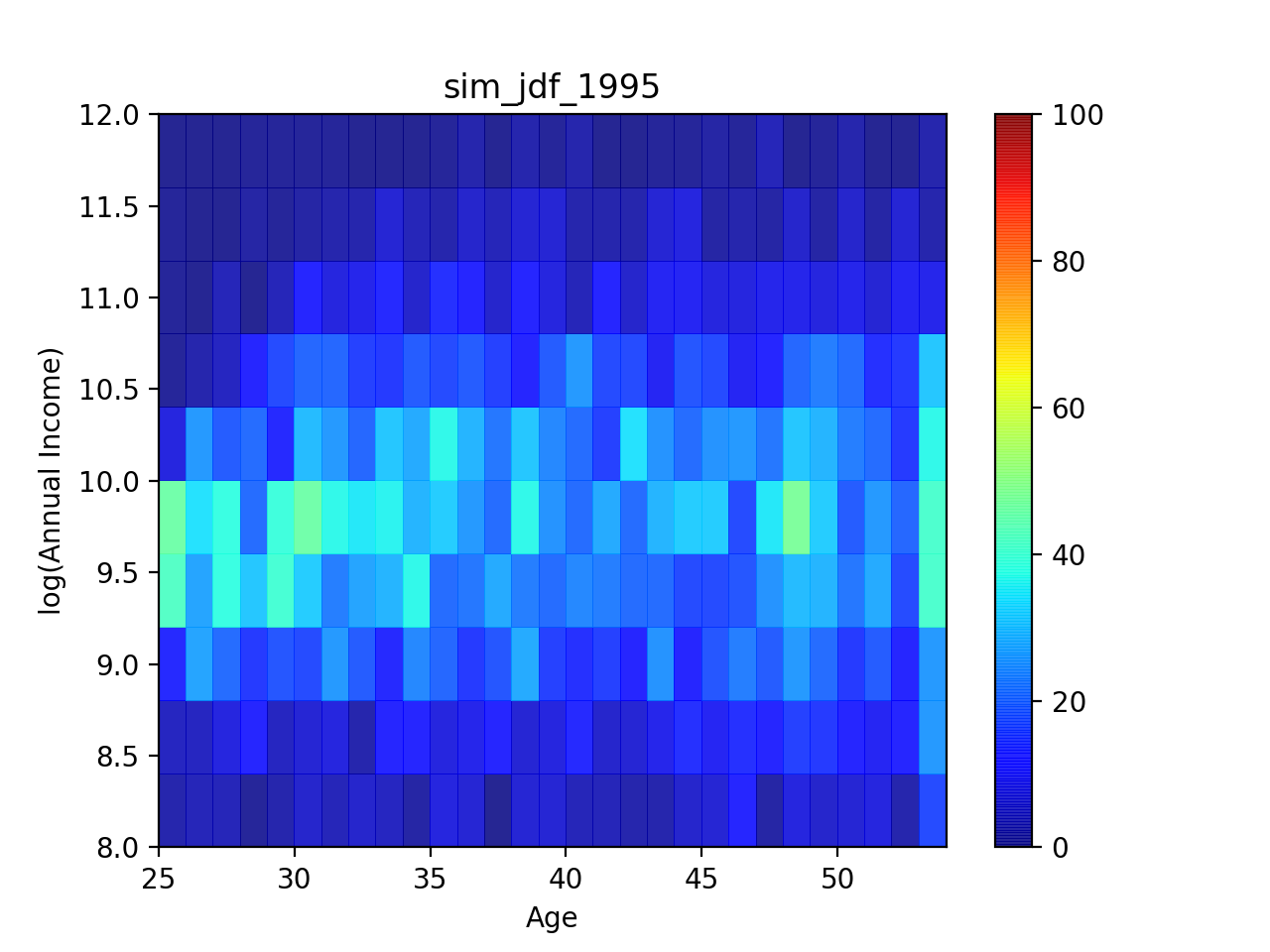}}\qquad
         \subfloat[2005 with LSM Estimation]{\includegraphics[width=0.32\linewidth]{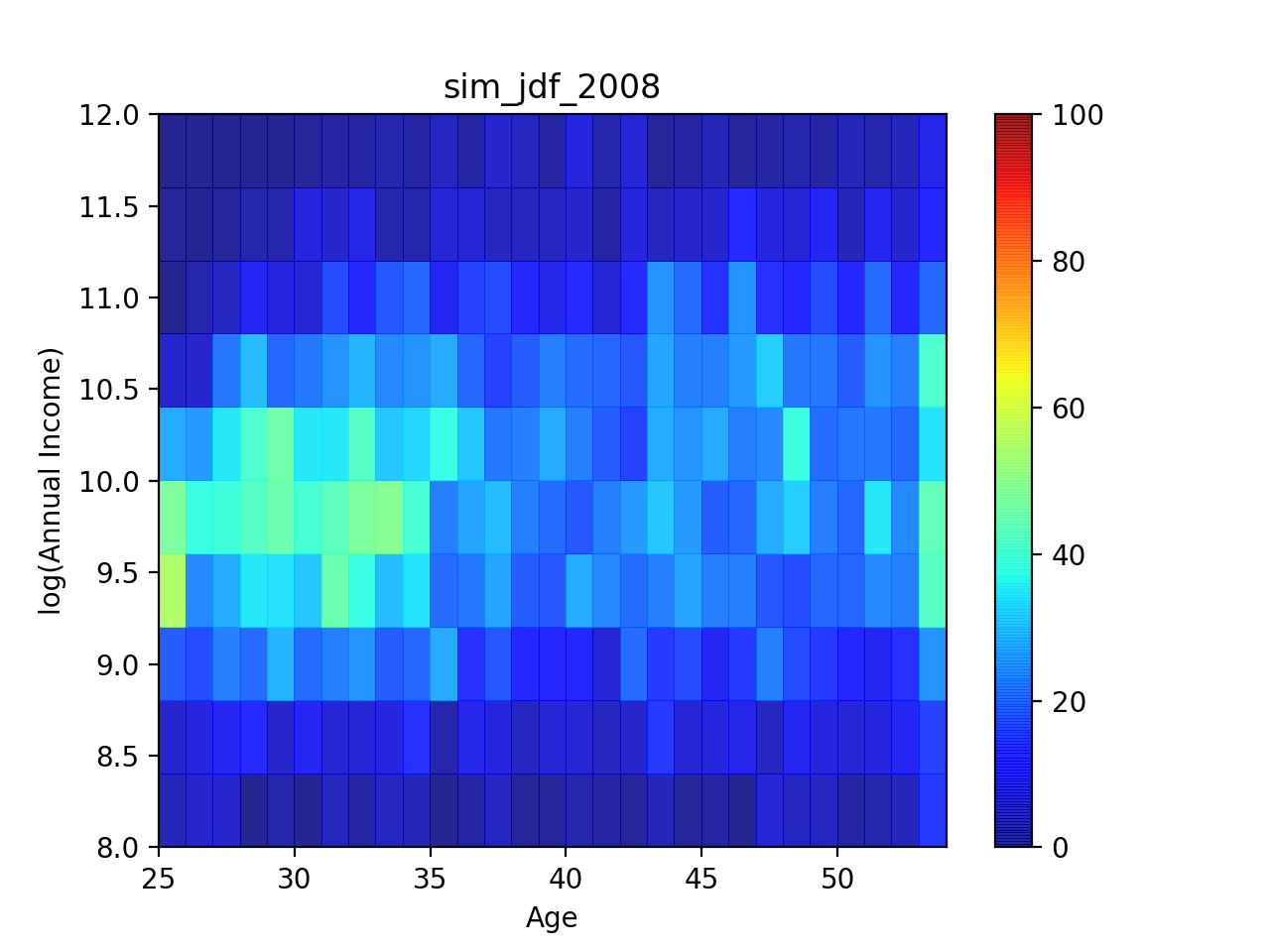}}\\
         \caption{JDF Plots of Simulation for UK Labour Income}
         \label{fig:uk_labour_wave_jdfs}
\end{figure}
The population from wave-1 is used for bootstrapping the simulation and the weights of the individuals are not incorporated to the simulation, because the income evolution Func.(eq. \ref{eq:agent_income_evo}) is the focus of this paper, and the main purpose is not the perfect representativeness of the initial wave. The new agent injection on 1999 by panel survey is reason of difference in the UK simulation and observed JDF plots.
Although the simulation's initial state is bootstrapped as the unweighted dataset, starting from the second wave, the JDF of the simulated population(Fig. \ref{fig:uk_labour_wave_jdfs}) resembles the characteristics of the JDF from the panel survey with the weighted population, which reflects that the model is successfully capturing income evolution dynamics.
\clearpage
\subsection{A Simple Pension System}

\begin{figure}[htp]%
         \centering
         \subfloat[Inflow \& Outflow]{\includegraphics[width=0.55\linewidth]{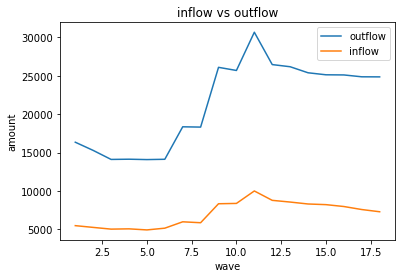}}\\         
         \caption{UK Inflow Outflow Plot of our Simple Pension  System}
         \label{fig:uk_labour_inflow_outflow}
\end{figure}

A financially sustainable pension system can be characterised by the balance between inflow and outflow of funds. The specifics and stability of pension system is out of the scope of this paper, and needs case specific detailed modelling. 
For a general demonstration, we assume simple inflow and outflow dynamics(Eq.\ref{eq:outflow} and Eq.$\ref{eq:inflow}$), which are derived to represent statistical properties of the savings and consumption. 
Figure \ref{fig:uk_labour_inflow_outflow} reflects the imbalance between inflow and outflow, which results in a deficit.

Pension is assumed to be annually \textsterling 16368, in light of the median net income before housing costs for all pensioners from DWP Pensioners Income Series in 2008/2009 \cite{dwppensionseries}. Constant alpha for pension saving rate is selected to be 0.0775 to 0.2, in light of OECD Pension Report statistics \cite{worldbankpensionreport}.

Outflow $O_{t}$ in a given year $t$ is characterised by constant annual pension amount $p$, and count of people above 65 $c_{a>65}$ is assumed to be pensioner counts.
\begin{equation}
O_{t}=p c_{a>65}
\label{eq:outflow}
\end{equation}
Inflow $I_{t}$ in a given year $t$ is characterised by constant pension contribution rate $\alpha$ and total labour income of individuals $y_{a}$

\begin{equation}
I_{t}=\alpha \sum_{i} (y_{a\leq65})
\label{eq:inflow}
\end{equation}

The amounts are adjusted for inflation and reflect the 2009 levels. 
The inflow and outflow plots on Fig. \ref{fig:uk_labour_inflow_outflow} from our simplified generalisation of the pension system reflect a deficit.

\section{Discussion}
The income evolution eq. \ref{eq:agent_income_evo} of the proposed model consists of the parameters ${q}_{a}$, ${\mu}_{a}$, ${\sigma}_{a}$: 
the persistence coefficient for the respective age group ${q}_{a}$, determines the rate of mobility at a given age. \\
Age-dependent mean income parameter ${\mu}_{a}$ expresses the expected age-specific income evolution mean for the next income and behaves such that if the mean parameter is high the mobility is most likely to have lower  ${q}_{a}$. 
If the mean parameter ${\mu}_{a}$ is lower, the persistence parameter is higher which signals a potential widening of the income gap for the population. \\
${\sigma}_{a}$ captures the variability of the individuals according to conditional distribution and incorporates randomness of the shocks. 

The social safety nets, basic pension incomes and the Defined Benefit Pension plans are financed via the working population; the ever-growing unbalance towards ageing cohorts needs careful forecasting and planning. 
The demographic shift will impact the economy's functioning in general, introducing a heavy burden to welfare states financing the health and pension of the retired population, which will reflect society as taxes and benefit cuts. 
The best course of action is forecasting the changes and planning in advance for the future.

\subsection{Interpretation}
The $q_a$ mobility estimated by GMM reflects that UK population reflects an initially high $q_a$ value in youth, followed by relative decrease, and then a consistent increase. The $q$ values estimated by GMM fluctuate around 0 and minimal. 
The income persistence variable of individuals is not captured by GMM, which does not utilise panel survey's tracked individual income micro-data each year.

The LSM results  in a consistently increasing $q_a$ value by the UK model, with a significant jump between ages 25-30, which corresponds with a $\mu_a$ plot consistently decreasing with a significantly sharper decrease between ages 25-30. $q_a$ and $\mu_a$ corresponding each other in an inverse proportion, especially by significant changes, especially by LSM.
There are various examples of mobility that can be observed from BHPS dataset.
Example of ${q}_{a}$ mobility reflected by LSM in the UK:
\begin{itemize}
    \item One example of ${q}_{a}$ mobility is the upward mobility of age-group between 25 and 30, which is reflected by the increasing ${q}_{a}$ values and sharp increase observed on the joint-distribution plot. This mobility can be due to finishing higher education and internships, in addition few years of experience, which results in a widening of income scissors. This change in mobility is healthy for the economy and does not represent a negative effect. One assumption should be researched further; if either this initial difference in mobility might limit of people with lower income for upward mobility.
\end{itemize}
Example of ${\sigma}_{a}$ mobility:
\begin{itemize}
    \item An example of the ${\sigma}_{a}$ mobility is the age group of 30-35, which is reflected by a locally sharp increase of ${\sigma}_{a}$ values. Such mobility reflects a bidirectional movement of income for individuals, and such a variation might arise from the short-time employment, interruption of employment for education, temporary jobs and most importantly this mobility might be caused by the initial differentiation according to the education of individuals such as higher education or vocational education. This window represents an increase in the variation of the income.
 
\end{itemize}
In general, the shape of the distribution can be explained in three periods; the first period is the introduction to employment and teenagers, which represents income from part-time and temporary jobs at the beginning and start of full-time employment it sharply increases on Fig.\ref{fig:jdf_uk_usa_income_15_100}. 

The age group of 25-55 denotes the main productive era of the economic life, and the income reflects a high dispersion. 
All of the factors and random shocks act together and result in dispersed but a consistent distribution. 
Mobility wise this era provides opportunities for upward mobility and possesses downward mobility risks. 
At the end of this period, income tends to decrease slightly, which reflects a decrease in productivity. 
Another limiting factor is the minimum wage and state benefits, which introduces a lower bound envelope for the mass. 
Income sources and affecting factors of individuals in this era vary greatly, which results in the widest dispersion in the entire life-span. 
Some of the factors are education, social strata, adaptability to innovation, total-work hours per week, experience and expertness, seniority of the jobs and ageism.
The third and final era represents the exit from the workforce and retirement, and temporary or part-time jobs for low-income old individuals. 
The income decreases gradually as the number of individuals exiting workforce increases with time, the income stabilises, and variation decreases significantly.
Income in this era is relatively low, and the source is usually pension benefits, state support or temporary jobs.
This model's outcomes can be used for various purposes; the most apparent fields for drawing consequences and planning are the works on inequality and mobility depending on age.
Characteristics of workforce entrance, work-efficiency of individuals per age, the structure of the society, pension system, income stability, and the taxation system are the most obvious fields.

In the paper, two main estimation techniques are investigated, and the corresponding results from the simulated waves are presented. 
The first estimation method investigated is GMM Estimation. 
The income regions appear smoothed and spread.
The second estimation method investigated is LSM, it utilises the microdata and is suitable for capturing an agent's income evolution. 
The JDFs from the simulated waves have the most similar mean characteristics to the observed data.

The LSM evidently performs better by utilizing longitudinal microdata; the GMM estimation method can be applied to both population and panel surveys, provides feasible distributions but with unrealistic modeling of an agent's individual income trajectory. 

\subsection{Conclusions}

We demonstrated
(1) a clear income-age relationship, which is reflected by the data from BHPS and IPSUM CPS, as well as simulations. 
(2) a clear structure of the joint age-income distribution in both the UK and USA. 
(3) a flexible methodology to estimate parameters from population surveys, as well as panel surveys. 
(4) a simple generative model to evolve the age-income population with real constraints for evaluating general policy scenarios, that is agnostic about occupation levels.

The model can be interpreted as delivering a premise that the information of an individual's experience and education can be encapsulated by income. 
Although in early career, the income dynamics are governed by the initial difference at the level of education and profession; the main dynamics governing income transitioning can be reduced to the relationship between income and age, which collectively encapsulate education and experience. 
These premises can be leveraged for developing simplified models for evaluating  mobility, inequality, welfare state, and pensions.

The proposed model focuses on the evolution of age and income population and the paper successfully demonstrates a simple model that can be calibrated for age and income that can be used as a backbone for forecasting income and planning pensions. 
Understanding the dynamics and having the ability to forecast the age and income population is the key to the design of financially sustainable pension systems.

There are different dimensions for the future work: one of the dimensions is injecting random shocks to the distributions itself, which can be in the form of new population injection or withdrawal, as well as tuning the $\eta_{it}$ with various means for simulating a global or regional shock, such as pandemics or mass migration. 
Stress-testing the age and income distribution for different labour market scenarios could lead to relevant policy implications.
The second dimension for future work is modifying the simulation system to estimate parameters on the fly, and provide a more adaptive and granular version of the simulation system. 
The third dimension for future work is incorporating data encompassing more years and more countries and with a higher resolution in time to investigate the role of multiple economic factors for short, medium and long time horizons. 

\bibliographystyle{bmc-mathphys} 
\bibliography{bibliography}      

\clearpage
\section{Supplementary Material}
\subsection{Model Calibration}
We can define the mean and standard deviation of income at a given age $a$ as following:
\begin{equation}
\left(\bar{y}_{a},std(y_{a})\right)
\end{equation}
\begin{equation}
\langle y_{a,t}^{i}\rangle=\bar{y}_{a}
\label{eq:agent_income_avg}\end{equation}
The standard deviation and mean has the following relation with the squared average of incomes:
\begin{equation}
\langle\left( y_{a,t}^{i}\right)^{2}\rangle-\left(\bar{y}_{a}\right)^{2}=\left(std(y_{a})\right)^{2}
\label{eq:agent_income_meanstd_avg}\end{equation}
$\eta_{it}$ has characteristics of the standard normal distribution:
\begin{equation}
\langle \eta_{it}\rangle=0
\label{eq:eta_mean}\end{equation}\begin{equation}
\langle \eta_{it}^{2}\rangle=1
\label{eq:eta_sq_mean}\end{equation}
\begin{equation}
\langle \eta_{it}^{3}\rangle=0
\label{eq:eta_cb_mean}\end{equation}Squaring both sides of income evolution equation (\ref{eq:agent_income_evo}) results in following distribution:
\begin{equation}
\left(y_{a+1,t+1}^{i}\right)^{2}=\left(q_{a} y_{a, t}^{i}+\mu_{a}+\sigma_{a} \eta_{ t}^{i}\right)^{2}
\label{eq:agent_income_sqr}\end{equation}
Eq.(\ref{eq:agent_income_meanstd_avg})  can be formalized as:
\begin{equation}
\left(\bar{y}_{a}\right)^{2}+\left(std(y_{a})\right)^{2}=\langle\left( y_{a,t}^{i}\right)^{2}\rangle
\label{eq:agent_income_sq_avg}\end{equation}Placing Eq.(\ref{eq:agent_income_sq_avg}) for $a+1$ and  Eq.(\ref{eq:agent_income_sqr}) results in following equation:

\begin{equation}
\left(\bar{y}_{a+1}\right)^{2}+\left(std(y_{a+1})\right)^{2}=\langle\left(q_{a} y_{a, t}^{i}+\mu_{a}+\sigma_{a} \eta_{t}^{i}\right)^{2}\rangle
\end{equation}Expanding the right side of the equation results in:
\begin{equation}
\left(\bar{y}_{a+1}\right)^{2}+\left(std(y_{a+1})\right)^{2}=\langle\left(q_{a} y_{a, t}^{i}\right)^{2}+\left(\mu_{a}+\sigma_{a} \eta_{t}^{i}\right)^{2}+2\left(q_{a} y_{a, t}^{i}\right)\left(\mu_{a}+\sigma_{a} \eta_{t}^{i}\right)\rangle
\end{equation}
\begin{equation}
\left(\bar{y}_{a+1}\right)^{2}+\left(std(y_{a+1})\right)^{2}=\langle\left(q_{a} y_{a, t}^{i}\right)^{2}+\left(\mu_{a}+\sigma_{a} \eta_{t}^{i}\right)^{2}+2\left(q_{a} y_{a, t}^{i}\right)\left(\mu_{a}+\sigma_{a} \eta_{t}^{i}\right)\rangle
\end{equation}
\begin{equation}
=\langle\left(q_{a} y_{a,t}^{i}\right)^{2}+\left(\mu_{a}\right)^{2}+\left(\sigma_{a} \eta_{t}^{i}\right)^{2}+2\left(\mu_{a} \sigma_{a}\eta_{it}\right)+2\left(q_{a} y_{a,t}^{i} \mu_{a}+\left(q_{a} y_{a,t}^{i} \right)\sigma_{a} \eta_{t}^{i}\right)\rangle
\end{equation}Averaging the equation by using Eq.(\ref{eq:agent_income_sq_avg}), Eq.(\ref{eq:eta_mean}) and Eq.(\ref{eq:eta_sq_mean}).
\begin{equation}
\left(\bar{y}_{a+1}\right)^{2}+\left(std(y_{a+1})\right)^{2}=\left(q_{a}\right)^{2}\left(\left(\bar{y}_{a}\right)^{2}+\left(std(y_{a})\right)^{2}\right)+\left(\mu_{a}\right)^{2}+\left(\sigma_{a}\right)^{2}+2 q_{a} \mu_{a} \bar{y}_{a}
\label{eq:squared_meanstd_avg_master}\end{equation}
\subsection{Deriving The Update Equations}
For clarity $\left(\bar{y}_{a}\right)^{2}+\left(std(y_{a})\right)^2$ is expressed as $(\Delta_{a})^{2}$, The number of parameters can be reduced to 2 using the third parameter of  Eq.(\ref{eq:squared_meanstd_avg_master}) by expressing $\mu_{a}$ as $\bar{y}_{a+1}-q_{a} \bar{y}_{a}$ according to Eq.(\ref{eq:agent_income_evo_average}):

\begin{equation}
\left(\Delta_{a+1}\right)^{2}=\left(q_{a}\right)^{2}\left(\Delta_{a}\right)^{2}+\left(\mu_{a}\right)^{2}+\left(\sigma_{a}\right)^{2}+2 q_{a} \mu_{a} \bar{y}_{a}
\end{equation}
\begin{equation}
\left(\Delta_{a+1}\right)^{2}=(q_{a})^{2}\left(\Delta_{a}\right)^{2}+\left(\bar{y}_{a+1}-q_{a} \bar{y}_{a}\right)^{2}+\sigma_{a}^{2}+2 q_{a}\left(\bar{y}_{a+1}-q_{a} \bar{y}_{a}\right) \bar{y}_{a}
\end{equation}
unpacking $\Delta$:
\begin{equation}
\left(\bar{y}_{a+1}\right)^{2}+\left(std(y_{a+1})\right)^{2}=q_{a}^{2}\left((\bar{y}_{a})^{2}+(std(y_{a}))^{2}\right)+(\bar{y}_{a+1})^{2}+\left(q_{a} \bar{y}_{a}\right)^{2}-2\left(\bar{y}_{a+1}q_{a} \bar{y}_{a}\right)+\sigma_{a}^{2}+2 q_{a}\bar{y}_{a}\bar{y}_{a+1}-2 (q_{a})^{2} (\bar{y}_{a}) ^{2}
\end{equation}
expressions at both sides of the equation cancel each other and simplify as follows:
\begin{equation}
\left(std(y_{a+1})\right)^{2} =q_{a}^{2}(std(y_{a}))^{2}+\left(\sigma_{a}\right)^{2}
\label{eq:sigma_q_tilde}
\end{equation}
solving in quadratic equation form:
\begin{equation}
0 =q_{a}^{2}(std(y_{a}))^{2}+\left(\sigma_{a}\right)^{2}-(std(y_{a+1}))^{2}
\end{equation}
for $\left(-(\sigma^{a})^{2}\left(\left(\tilde{\sigma}^{a}\right)^{2}-(\sigma^{a+1})^{2}\right)\right)>0$ and $(\sigma^{a})^{2} > 0$, $\tilde{q}$ values can be solved as follows:
\begin{equation}
\tilde{q}_{1}^{a}=\frac{\sqrt{-(\sigma^{a})^{2}\left(\left(\tilde{\sigma}^{a}\right)^{2}-(\sigma^{a+1})^{2}\right)}}{(\sigma^{a})^{2}}
\label{eq:qtilda_1}
\end{equation}
\begin{equation}
\tilde{q}_{2}^{a}=\frac{-\sqrt{-(\sigma^{a})^{2}\left(\left(\tilde{\sigma}^{a}\right)^{2}-(\sigma^{a+1})^{2}\right)}}{(\sigma^{a})^{2}}
\label{eq:qtilda_2}
\end{equation}
Following equations are used in the method of GMM:

Using unnormalized unstandardized third moment of the Equation \ref{eq:agent_income_evo}
\begin{equation}
E\left[\left(y_{a+1}\right)^{3}\right] = E\left[\left(q_{a}y_{a}+\mu_{a}+\sigma_{a}\eta\right)^3\right]
\end{equation}
Expanding the cube equation
\begin{equation}
\begin{array}{l}
E\left[\left(y_{a+1}\right)^{3}\right] = \\ 
\\
E\biggl[
\left(q_{a}y_{a}\right)^{3} + \left(\mu_{a}\right)^{3}+\left(\sigma_{a}{\eta}^{a}\right)^{3} + \left( 6 q_{a}y_{a}\mu_{a}\sigma_{a}\eta \right)  + 3\left(q_{a}y_{a}\right)^{2}\sigma_{a} \\
+3\left( q_{a}y_{a}\right)^{2}\mu_{a} + 3\left(\mu_{a}\right)^{2}q_{a}y_{a}+3\left(\mu_{a}\right)^{2}\sigma_{a}\eta + 3\left(\sigma_{a}\right)^{2}\mu_{a}+3\left(\sigma_{a}\right)^{2}q_{a}y_{a}
\biggr]
\end{array}
\end{equation}
Using Eq.\ref{eq:eta_cb_mean} , $\left(\sigma_{a}{\eta}^{a}\right)^{3}$, $\eta^{3}$ equals zero

\begin{equation}
\begin{array}{l}
E\left[\left(y_{a+1}\right)^{3}\right] = \\ 
\left(q_{a}\right)^{3}E\left[\left(y_{a}\right)^{3}\right] + \left(\mu_{a}\right)^{3} + 3\left( q_{a}\right)^{2}\mu_{a}E\left[\left(y_{a}\right)^{2} \right]+ 3\left(\mu_{a}\right)^{2}q_{a}E\left[y_{a}\right]+ \\
3\left(\sigma_{a}\right)^{2}\mu_{a}+3\left(\sigma_{a}\right)^{2}q_{a}E\left[y_{a}\right]
\end{array}
\end{equation}
\begin{equation}
\begin{array}{l}
E\left[\left(y_{a+1}\right)^{3}\right] = \\ 
\left(q_{a}\right)^{3}E\left[\left(y_{a}\right)^{3}\right] + \left(\mu_{a}\right)^{3} + 3\mu_{a}\left(\left( q_{a}\right)^{2}E\left[\left(y^{a}\right)^{2}\right] +\left(\sigma_{a}\right)^{2}\right) +3q_{a}E\left[y_{a}\right]\left(\left(\mu_{a}\right)^{2}+\left(\sigma_{a}\right)^{2}\right)
\end{array}
\end{equation}
Expressing $\left(\sigma_{a}\right)^{2}$ from Eq.(\ref{eq:sigma_q_tilde}) in terms of $q_{a}$
\begin{equation}
\begin{array}{l}
E\left[\left(y_{a+1}\right)^{3}\right] = \\ 
\left(q_{a}\right)^{3}E\left[\left(y_{a}\right)^{3}\right] + \left(\mu_{a}\right)^{3} + 3\mu_{a}\left(\left( q_{a}\right)^{2}E\left[\left(y_{a}\right)^{2}\right] +\left(std(y_{a+1})\right)^{2}-\left(q_{a}\right)^{2}\left(std(y_{a})\right)^{2}\right) \\
+3q_{a}E\left[y_{a}\right]\left(\left(\mu_{a}\right)^{2}+\left(std(y_{a+1})\right)^{2}-\left(q_{a}\right)^{2}\left(std(y_{a})\right)^{2}\right)
\end{array}
\end{equation}
Replacing $E\left[y_{a}\right] = \bar{y}_{a}$ and $E\left[\left(y_{a}\right)^{2}\right] = \left(std(y_{a})\right)^2+\left(\bar{y}_{a}\right)^2$ from Eq.\ref{eq:agent_income_meanstd_avg} 
\begin{equation}
\begin{array}{l}
E\left[\left(y_{a+1}\right)^{3}\right] = \\ 
\left(q_{a}\right)^{3}E\left[\left(y_{a}\right)^{3}\right] + \left(\mu_{a}\right)^{3}+3\mu_{a}\left(q_{a}\right)^{2}\left(\bar{y}_{a}\right)^{2}\\
+3\mu_{a}\left(std(y_{a+1})\right)^2+3q_{a}\bar{y}_{a}\left(\mu_{a}\right)^2+3\mu_{a}\bar{y}_{a}\left(std(y_{a+1})\right)^{2}-3\left(q_{a}\right)^{3}\bar{y}_{a}\left(std(y_{a})\right)^{2}
\end{array}
\end{equation}Expressing $\mu_{a}$ from Eq.(\ref{eq:agent_income_evo_average}) in terms of $q_{a}$
\begin{equation}
\begin{array}{l}
E\left[\left(y_{a+1}\right)^{3}\right] = \\ 
\left(q_{a}\right)^{3}E\left[\left(y_{a}\right)^{3}\right] + \left(\bar{y}_{a+1}-q_{a}\bar{y}_{a}\right)^{3}+3\left(\bar{y}_{a+1}-q_{a}\bar{y}_{a}\right)\left(std(y_{a+1})\right)^{2}\\
+3q_{a}\bar{y}_{a}\left(\bar{y}_{a+1}-q_{a}\bar{y}_{a}\right)^{2}+3q_{a}\bar{y}_{a}\left(std(y_{a+1})\right)^{2}-3\left(q_{a}\right)^{3}\bar{y}_{a}\left(std(y_{a})\right)^{2}
\end{array}
\end{equation}
Expressing in the form of cubic polynomial equation of $q_{a}$
\begin{equation}
\begin{array}{l}
0   = \\ \left(q_{a}\right)^{3}\left(  E\left[\left(y_{a}\right)^{3}\right] -  \left( \bar{y}_{a}\right)^{3}-3\bar{y}_{a}\left(std(y_{a})\right)^{2} \right)
\\+ \left(q_{a}\right)^{2}\left(  3\bar{y}_{a+1}\left(\bar{y}_{a}\right)^2 -6\left(\bar{y}_{a}\right)^{2}\bar{y}_{a+1} \right)
\\ +\left(q_{a}\right)\left(  3\left(\mu_{a+1}\right)^{2}\bar{y}_{a} -3\bar{y}_{a}\left(std(y_{a+1})\right)^{2}  +3\bar{y}_{a}\left(\bar{y}_{a+1}\right)^{2}    +3\bar{y}_{a}\left(std(y_{a+1})\right)^{2}\right)
\\ +\left(\bar{y}_{a+1}\right)^{3} + 3\bar{y}_{a+1}\left(std(y_{a+1})\right)^{2} -E\left[\left(y_{a+1}\right)^{3}\right] 
\label{eq:three_moments_cubic_poly}
\end{array}
\end{equation}
This equation can be solved for $q_{a}$ corresponding each age group. Cardano solution for cubic equations guarantees single real root to exist, the other two complex roots that Cardano solution provides are not used.
Both of the $\sigma_{a}$ = $std(y_a)$ and GMM estimation techniques can use the following equations for determining the $\mu_{a}$ and $\sigma_{a}$:
For $(q_{a})_{1}$ and $(q_{a})_{2}$  according to Eq.(\ref{eq:agent_income_evo_average}):
\begin{equation}\mu_{a} =  \bar{y}_{a+1}-q_{a} \bar{y}_{a}
\label{eq:mutilda}
\end{equation}
The $\sigma_{a}^{2}$ can also be expressed in terms of $q_{a}$, using Eq.(\ref{eq:agent_income_evo_average}) :
\begin{equation}
\sigma_{a}^{2}=\left(\Delta_{a+1}\right)^{2}-q_{a}^{2}\left(\Delta_{a}\right)^{2}-\left(\bar{y}_{a+1}-q_{a} \mu^{2}\right)^{2}-2 q_{a}\left(\bar{y}_{a+1}-q_{a} \bar{y}_{a}\right) \bar{y}_{a}
\label{eq:sigma_tilde_sq}
\end{equation}

\subsection{Analysis BHPS - Joint Distribution of Age and Income for Observed and Simulated Data}
The parameters are estimated with LSM.

\begin{figure}[htp]%
         \centering
         \subfloat[Wave 1991 JDF of Observed Data.]{\includegraphics[width=0.33\linewidth]{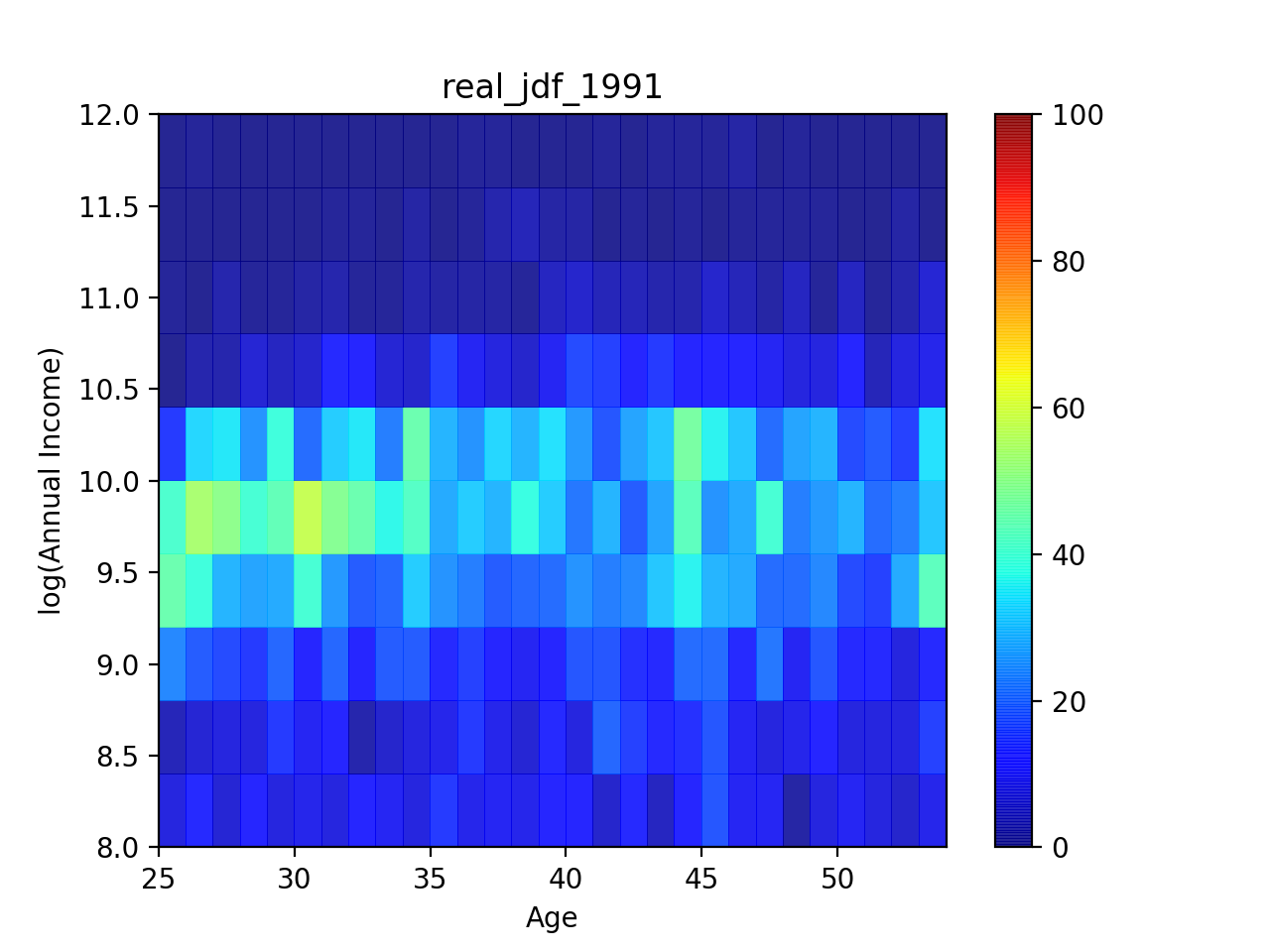}}\qquad
         \subfloat[Wave 1991 JDF of Sim Data]{\includegraphics[width=0.33\linewidth]{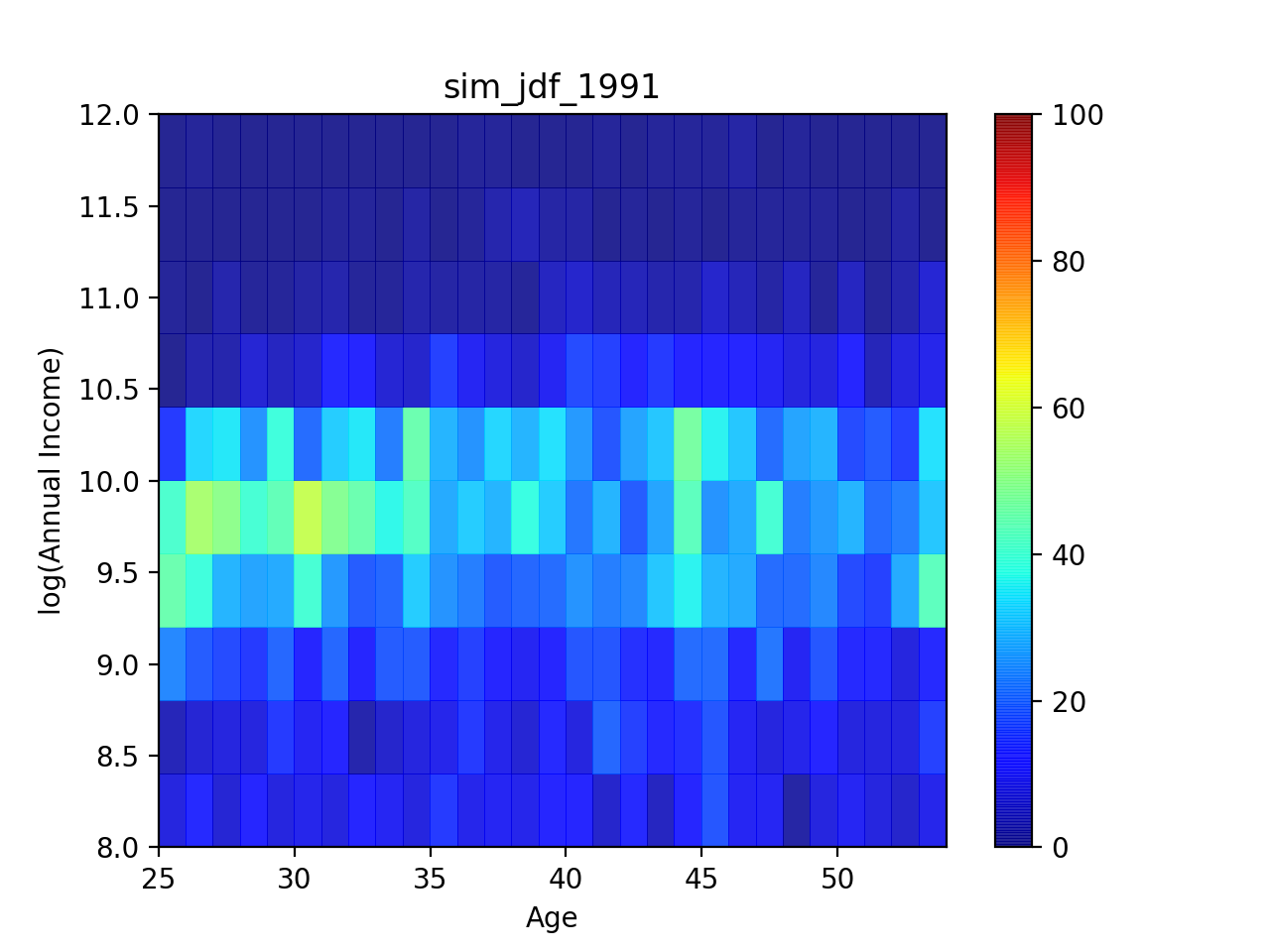}}\\
         \subfloat[Wave 1992 JDF of Observed Data.]{\includegraphics[width=0.33\linewidth]{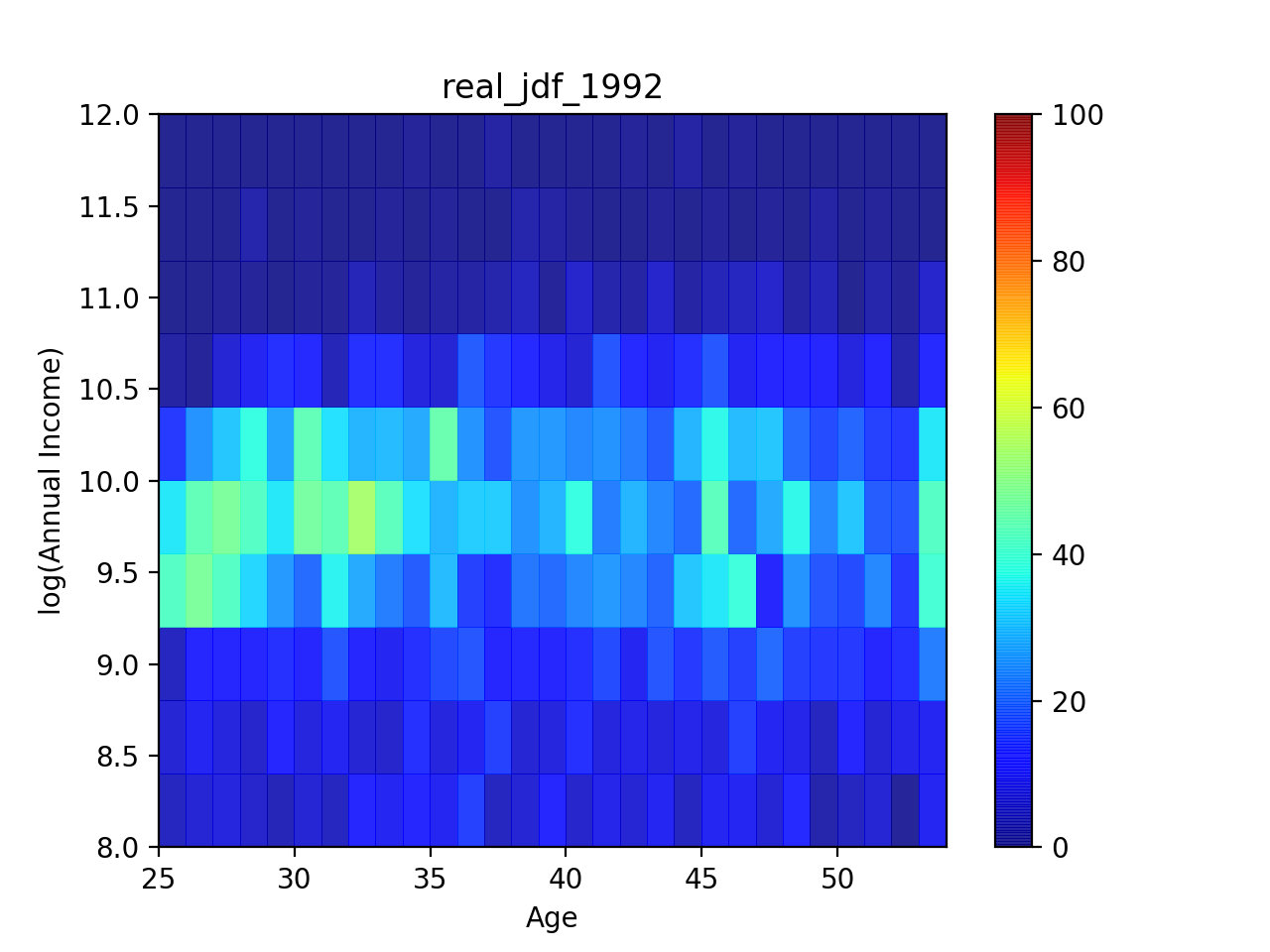}}\qquad
         \subfloat[Wave 1992 JDF of Sim Data]{\includegraphics[width=0.33\linewidth]{figs/uk_lsm_labour_bootstrap/sim_jdf_1992.png}}\\
         \subfloat[Wave 1993 JDF of Observed Data.]{\includegraphics[width=0.33\linewidth]{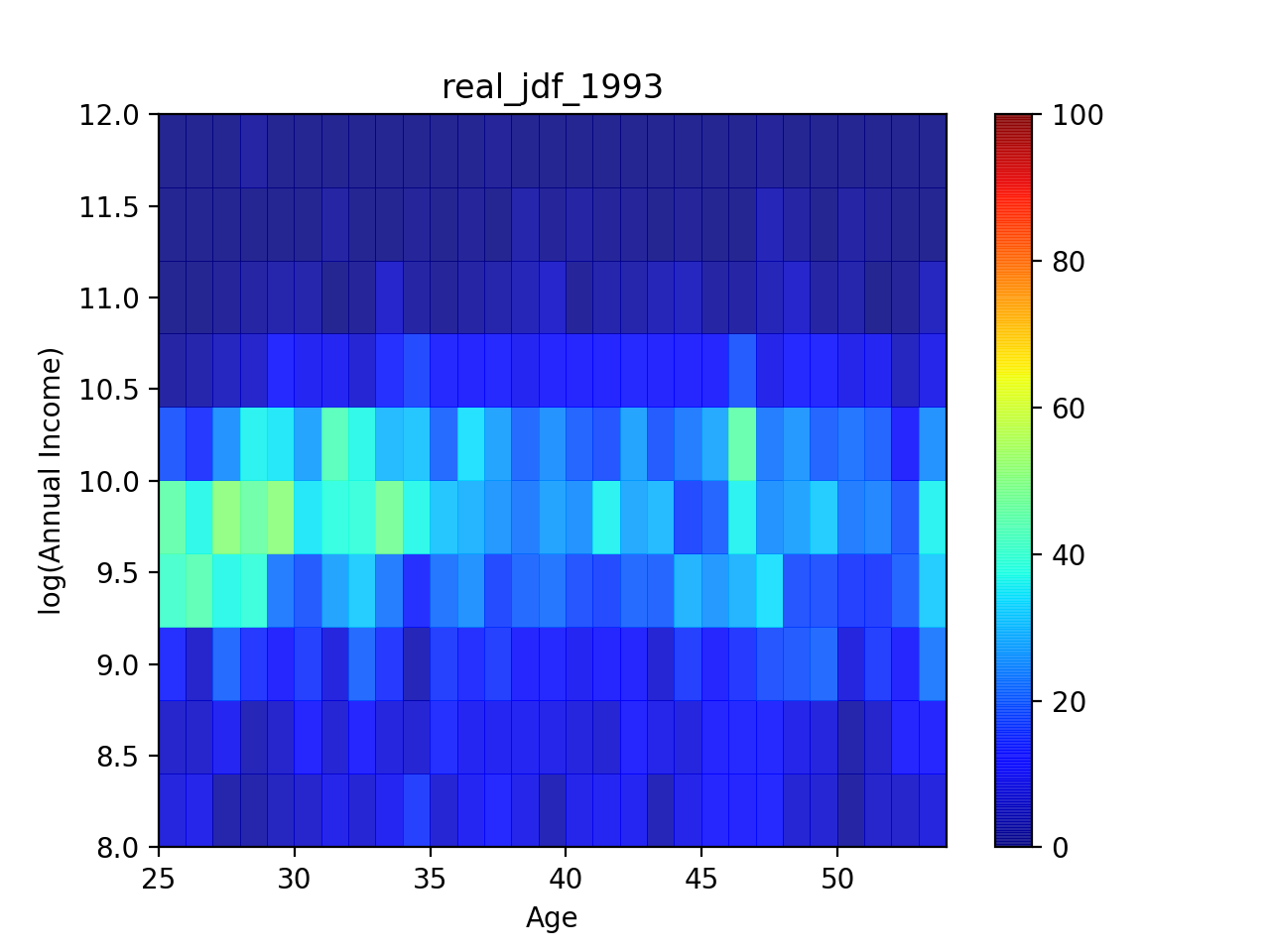}}\qquad
         \subfloat[Wave 1993 JDF of Sim Data]{\includegraphics[width=0.33\linewidth]{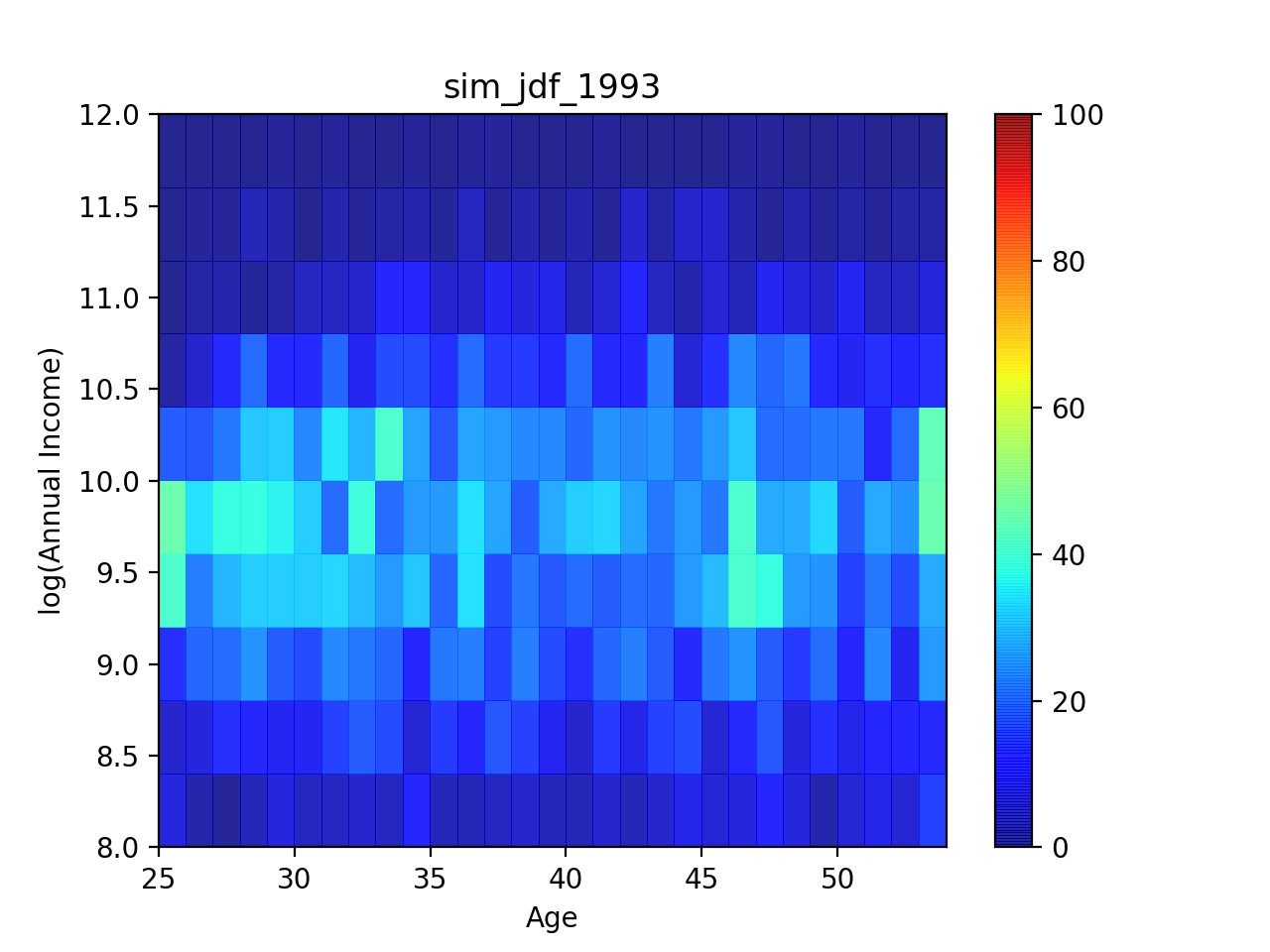}}\\
         \subfloat[Wave 1994 JDF of Observed Data.]{\includegraphics[width=0.33\linewidth]{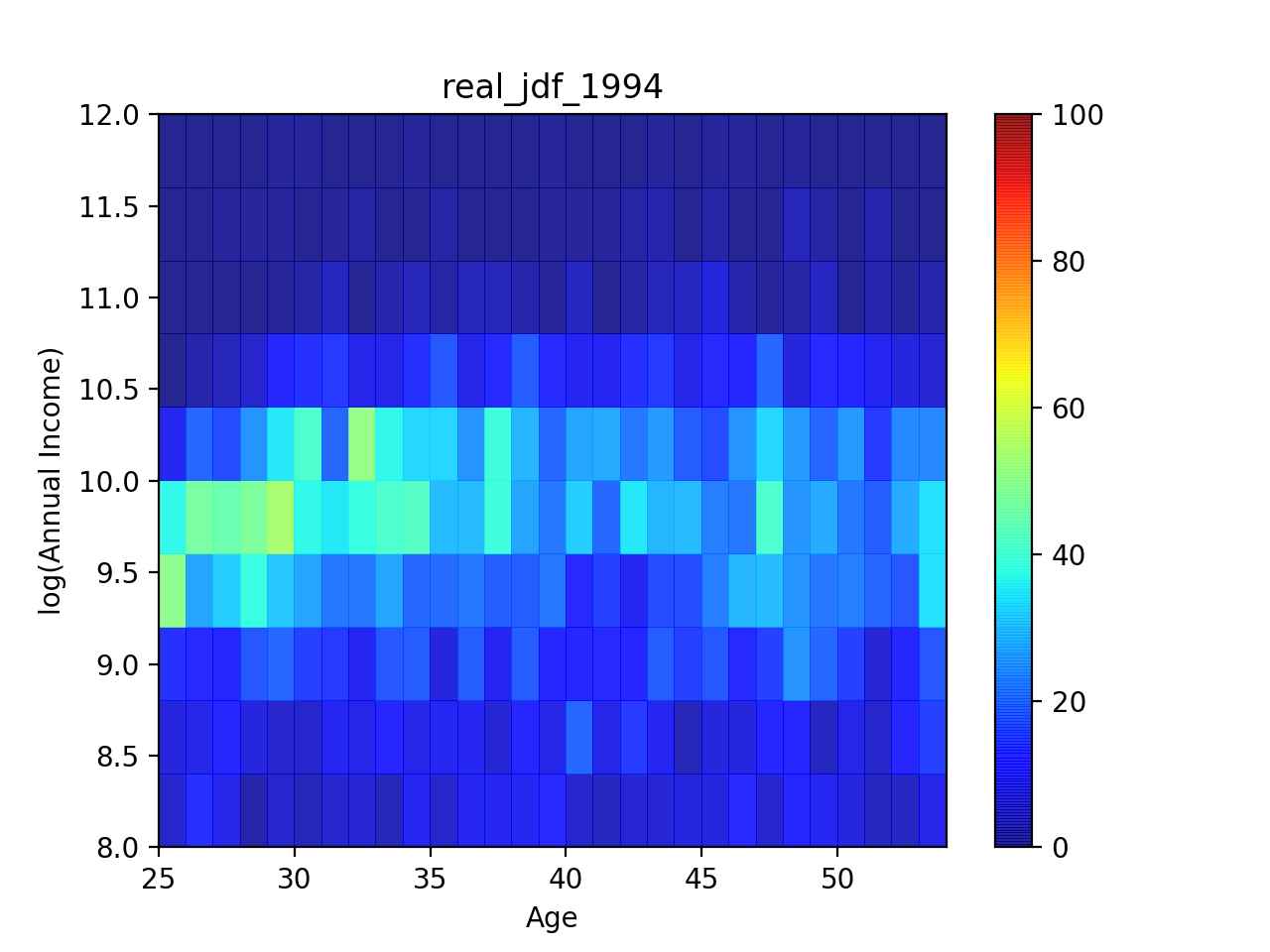}}\qquad
         \subfloat[Wave 1994 JDF of Sim Data]{\includegraphics[width=0.33\linewidth]{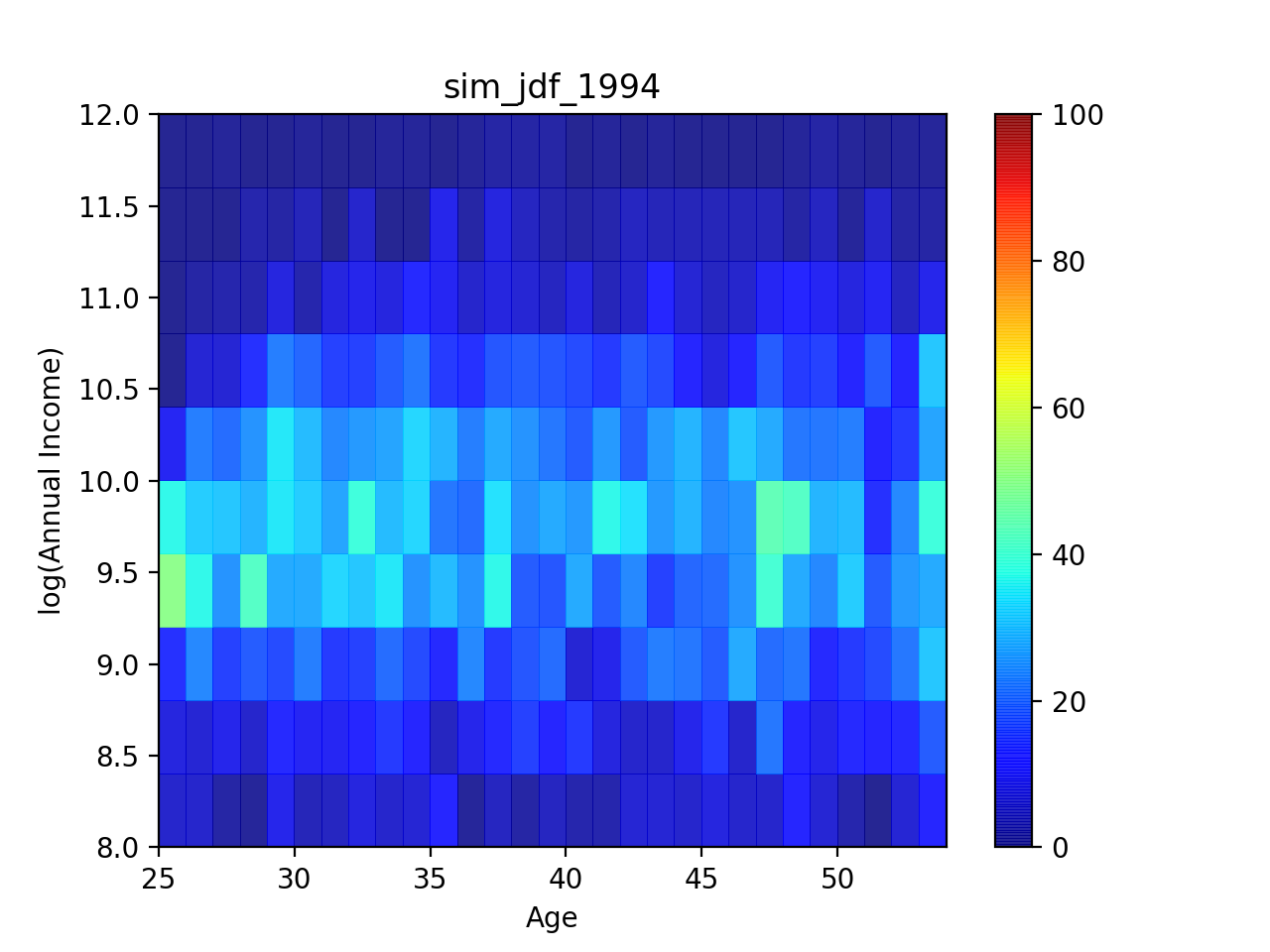}}\\
         \subfloat[Wave 1995 JDF of Observed Data.]{\includegraphics[width=0.33\linewidth]{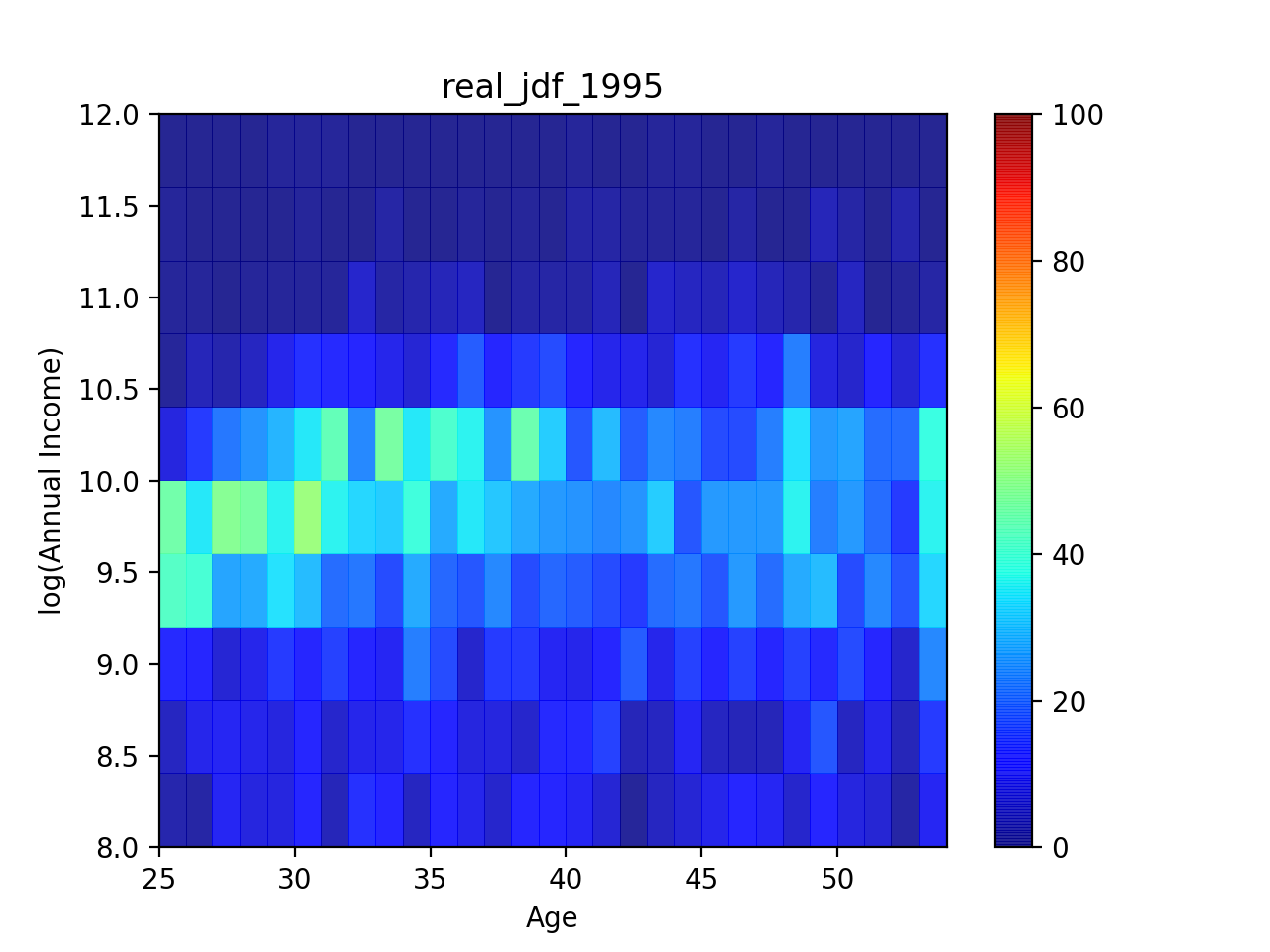}}\qquad
         \subfloat[Wave 1995 JDF of Sim Data]{\includegraphics[width=0.33\linewidth]{figs/uk_lsm_labour_bootstrap/sim_jdf_1995.png}}\\
         \caption{JDF for Waves 1991-1995}
         \label{fig:1991_1995_waves_jdf}
\end{figure}

\begin{figure}%
         \centering
         \subfloat[Wave 1996 JDF of Observed Data.]{\includegraphics[width=0.33\linewidth]{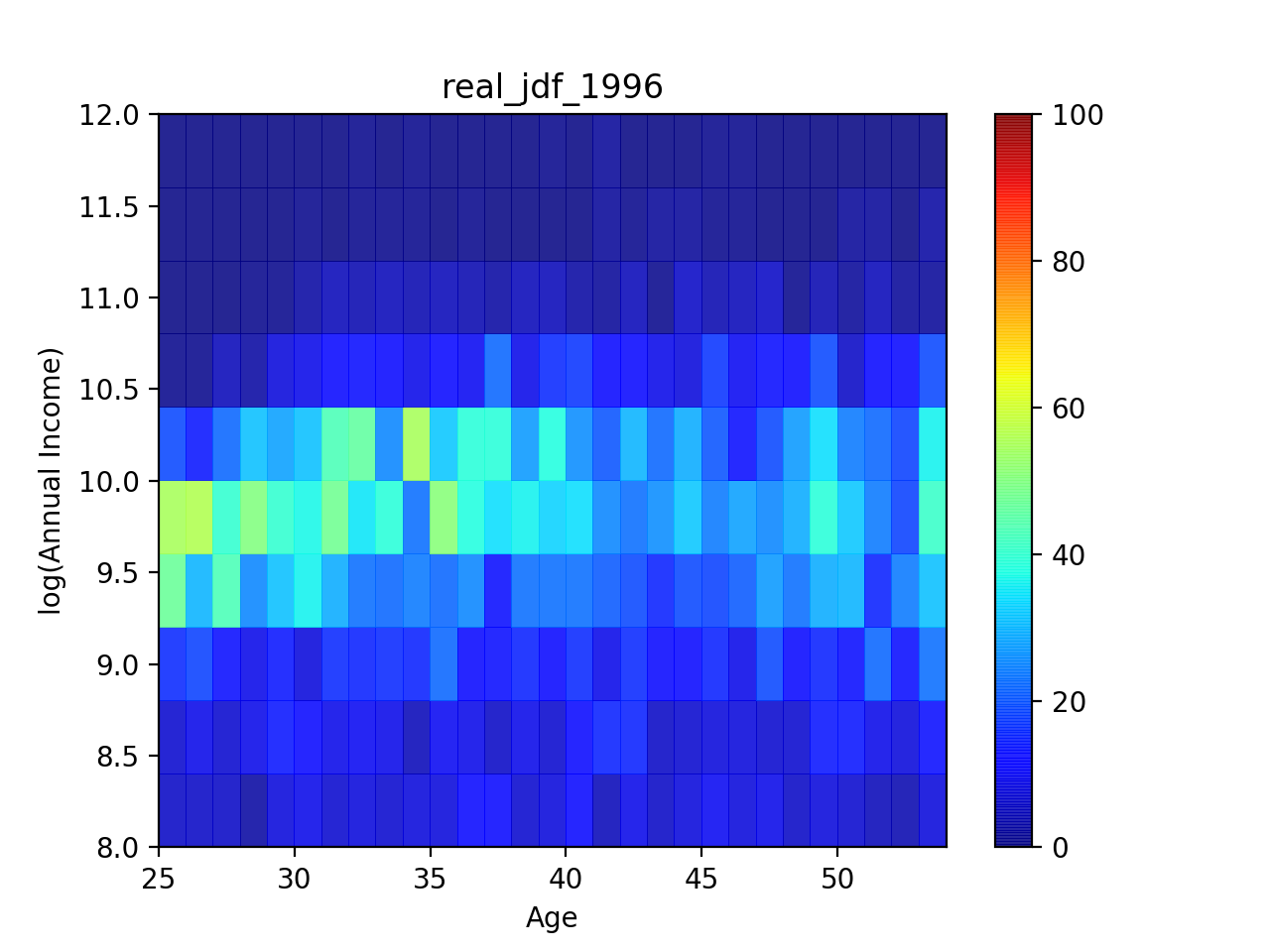}}\qquad
         \subfloat[Wave 1996 JDF of Sim Data]{\includegraphics[width=0.33\linewidth]{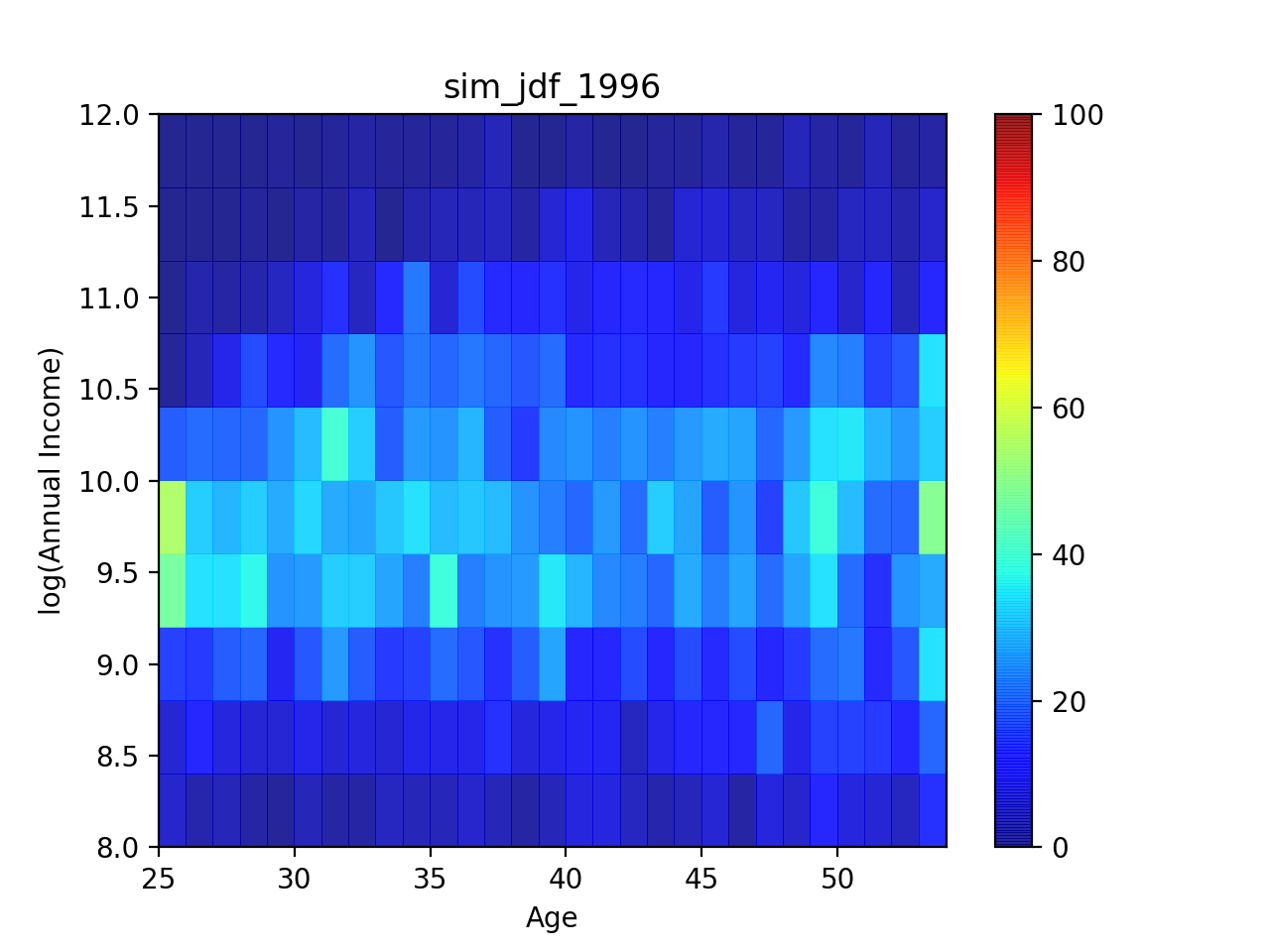}}\\
         \subfloat[Wave 1997 JDF of Observed Data.]{\includegraphics[width=0.33\linewidth]{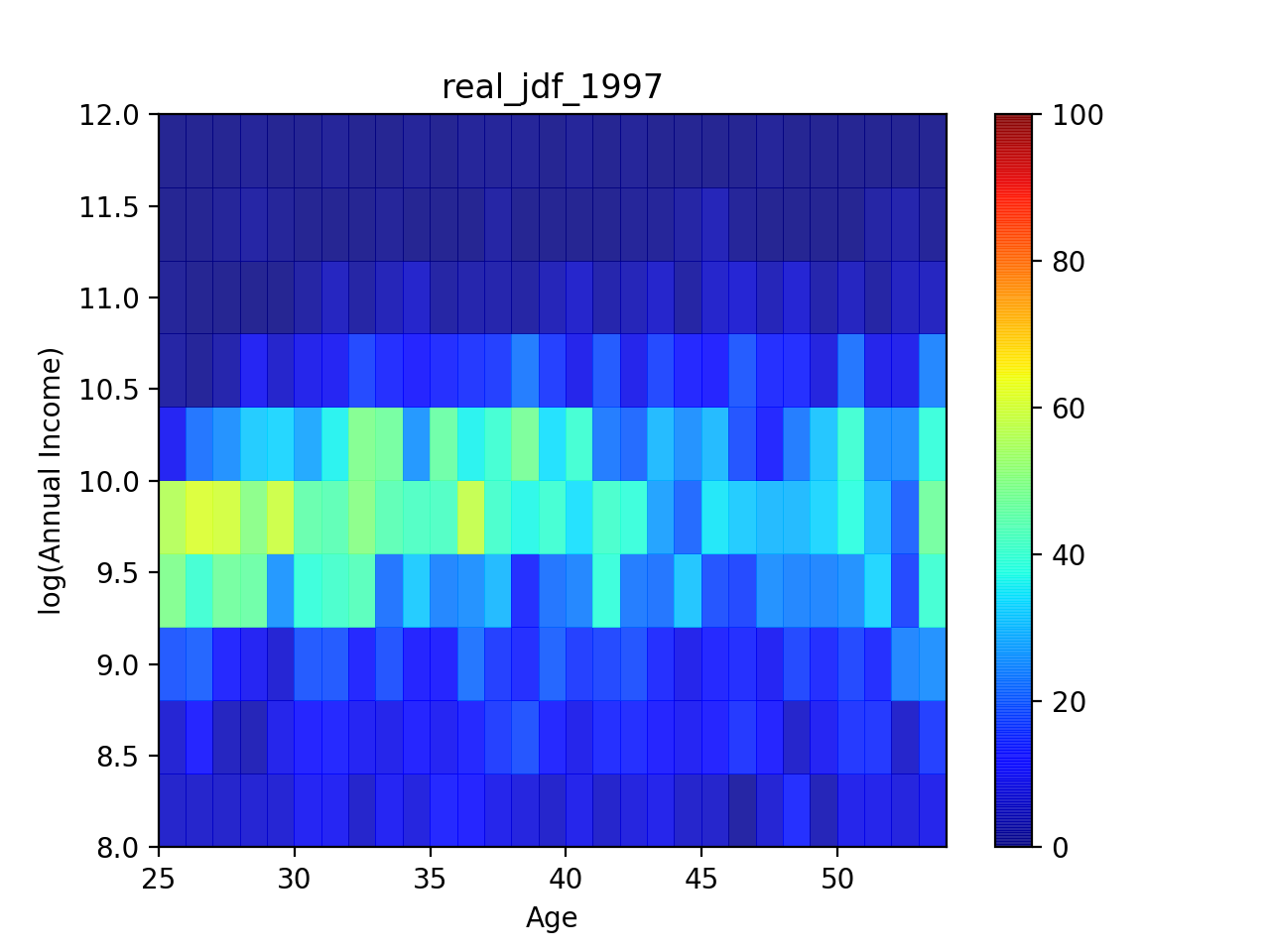}}\qquad
         \subfloat[Wave 1997 JDF of Sim Data]{\includegraphics[width=0.33\linewidth]{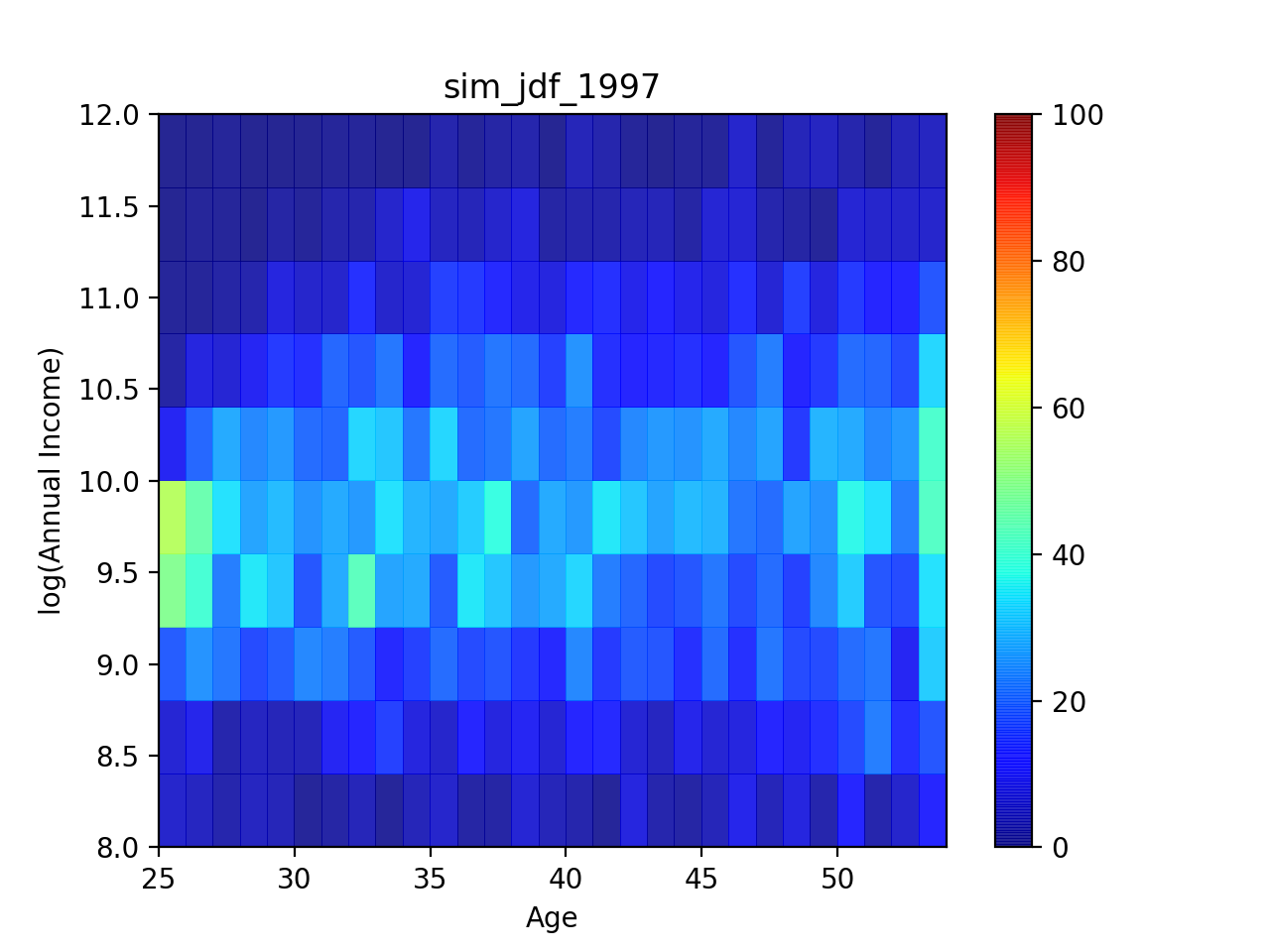}}\\
         \subfloat[Wave 1998 JDF of Observed Data.]{\includegraphics[width=0.33\linewidth]{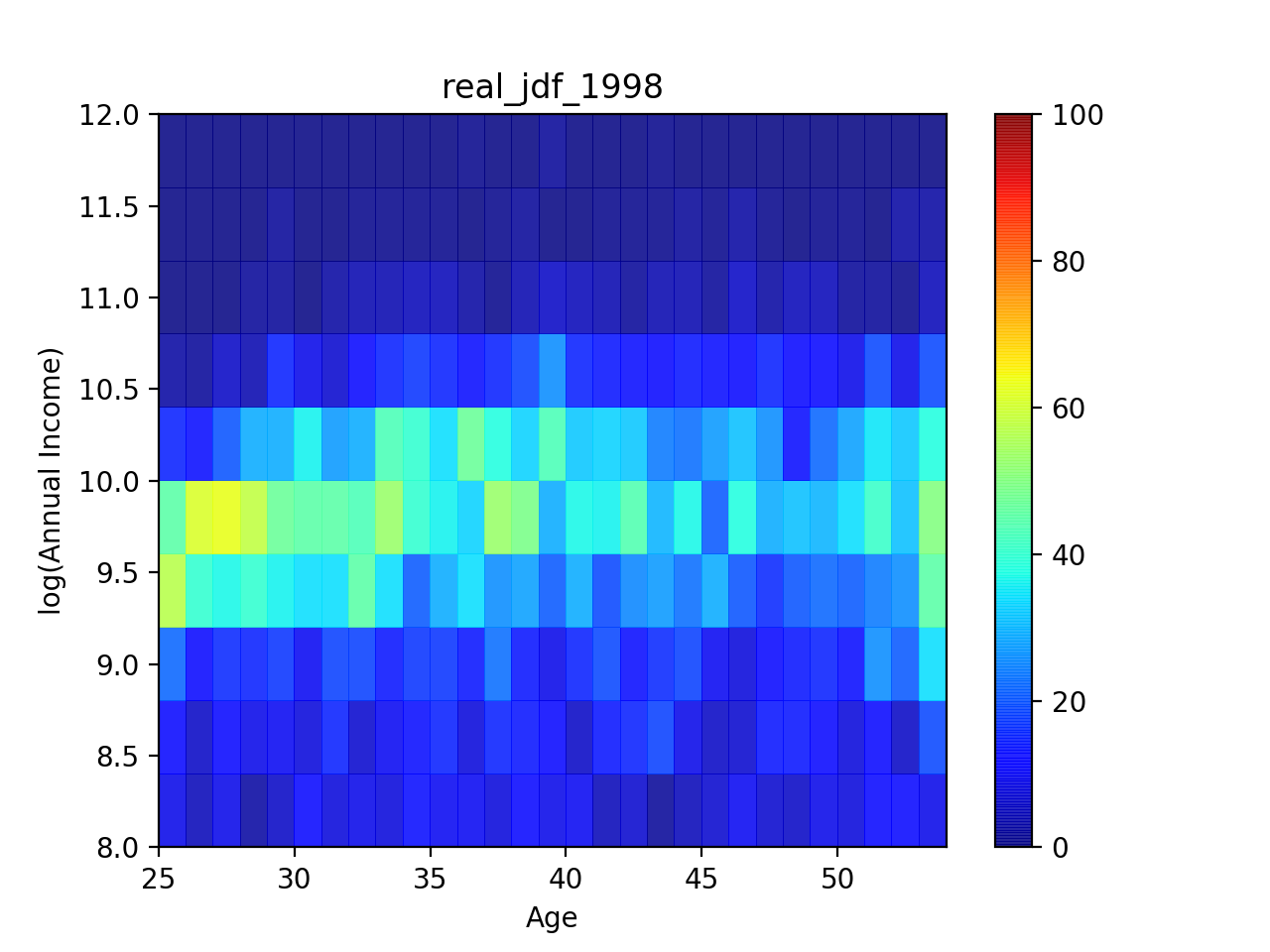}}\qquad
         \subfloat[Wave 1998 JDF of Sim Data]{\includegraphics[width=0.33\linewidth]{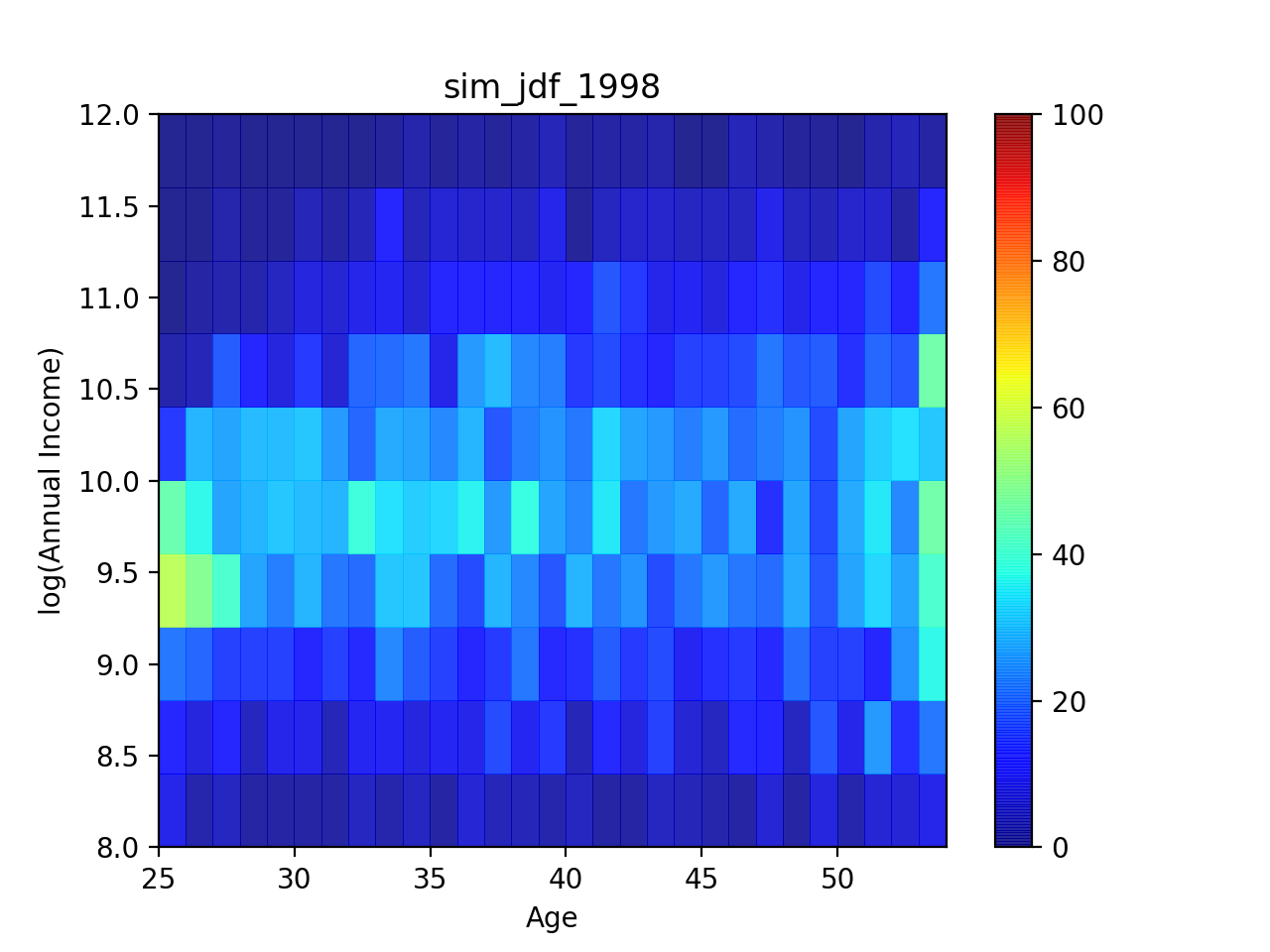}}\\
         \subfloat[Wave 1999 JDF of Observed Data.]{\includegraphics[width=0.33\linewidth]{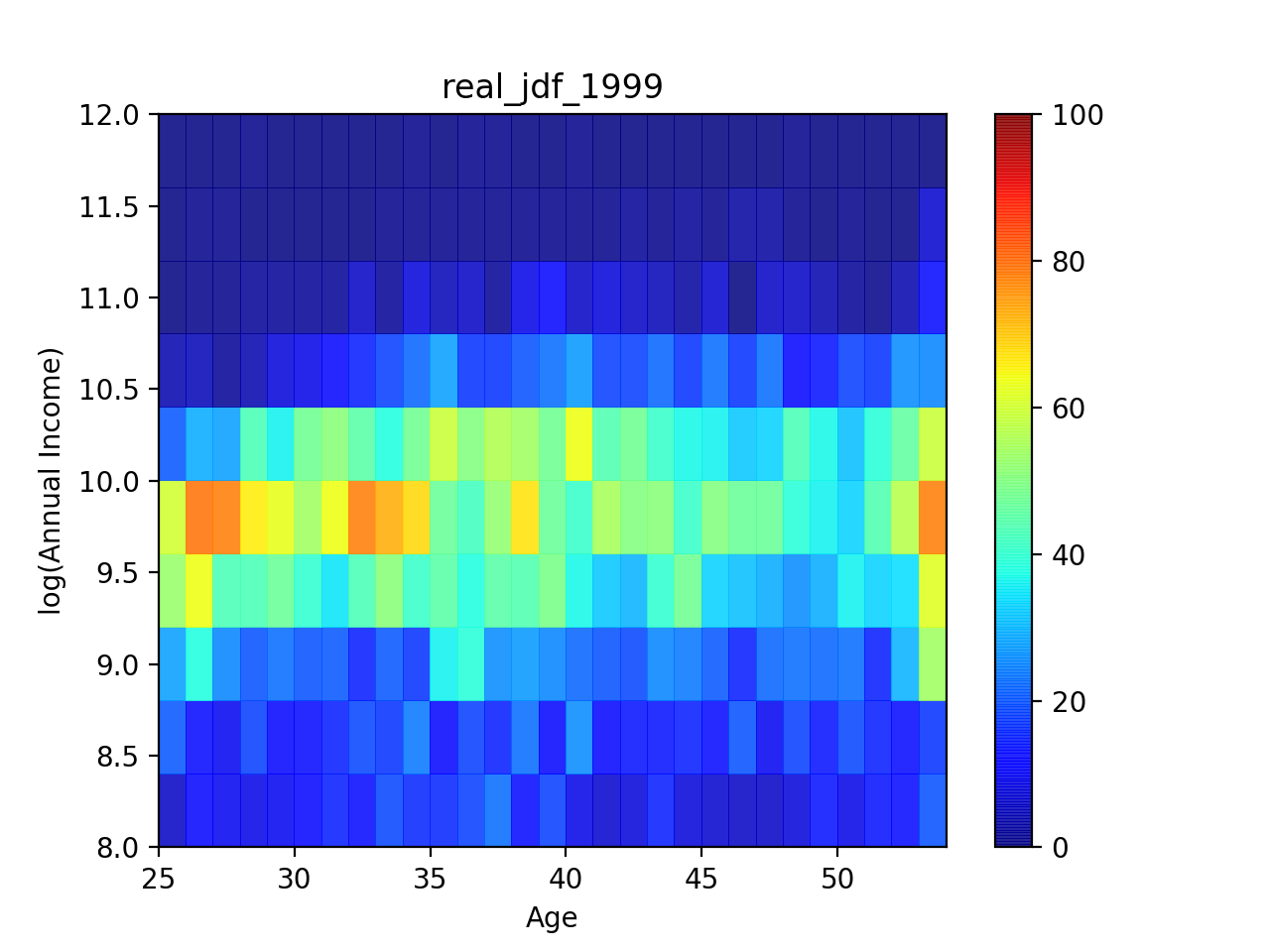}}\qquad
         \subfloat[Wave 1999 JDF of Sim Data]{\includegraphics[width=0.33\linewidth]{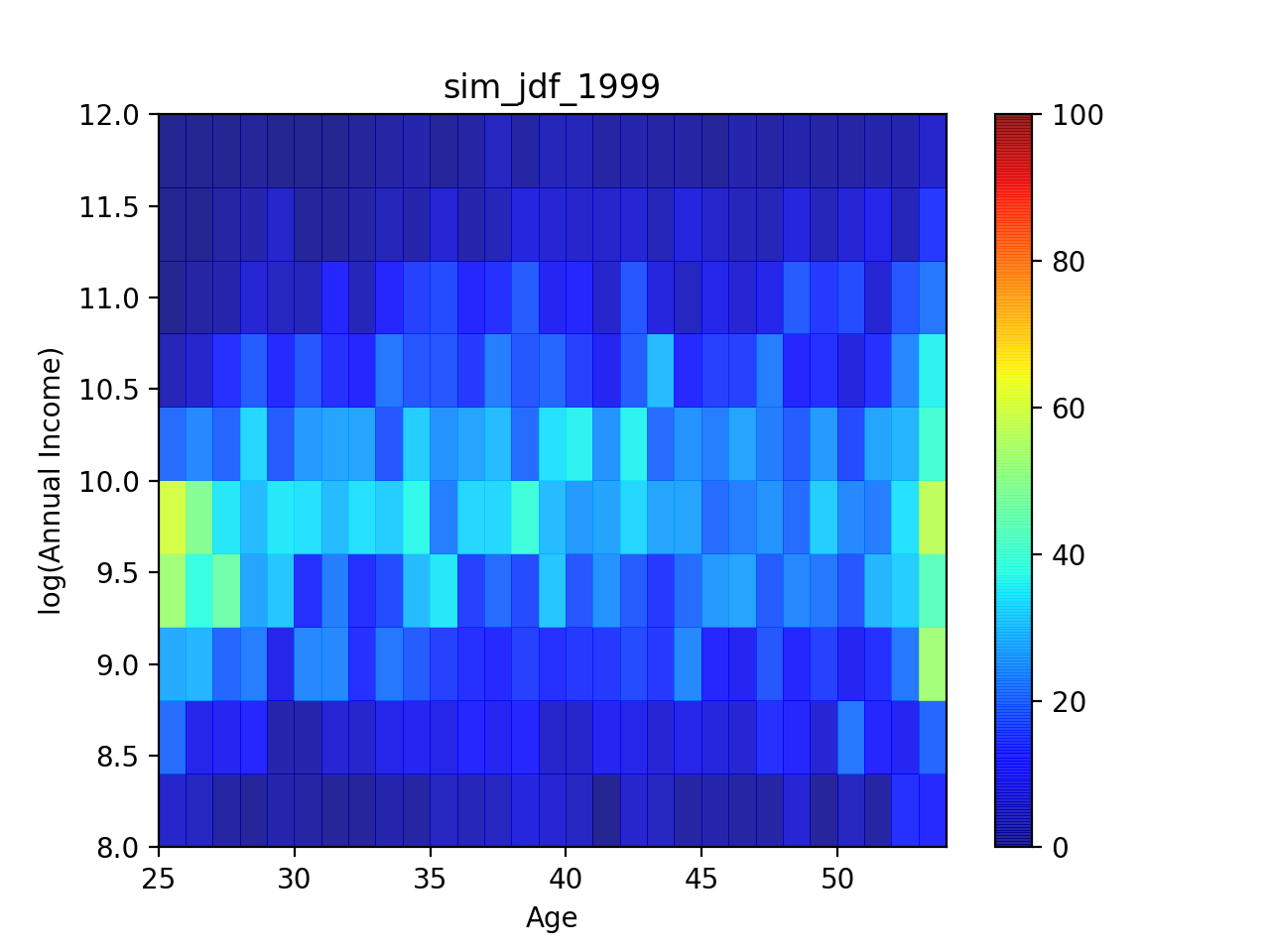}}\\
         \subfloat[Wave 2000 JDF of Observed Data.]{\includegraphics[width=0.33\linewidth]{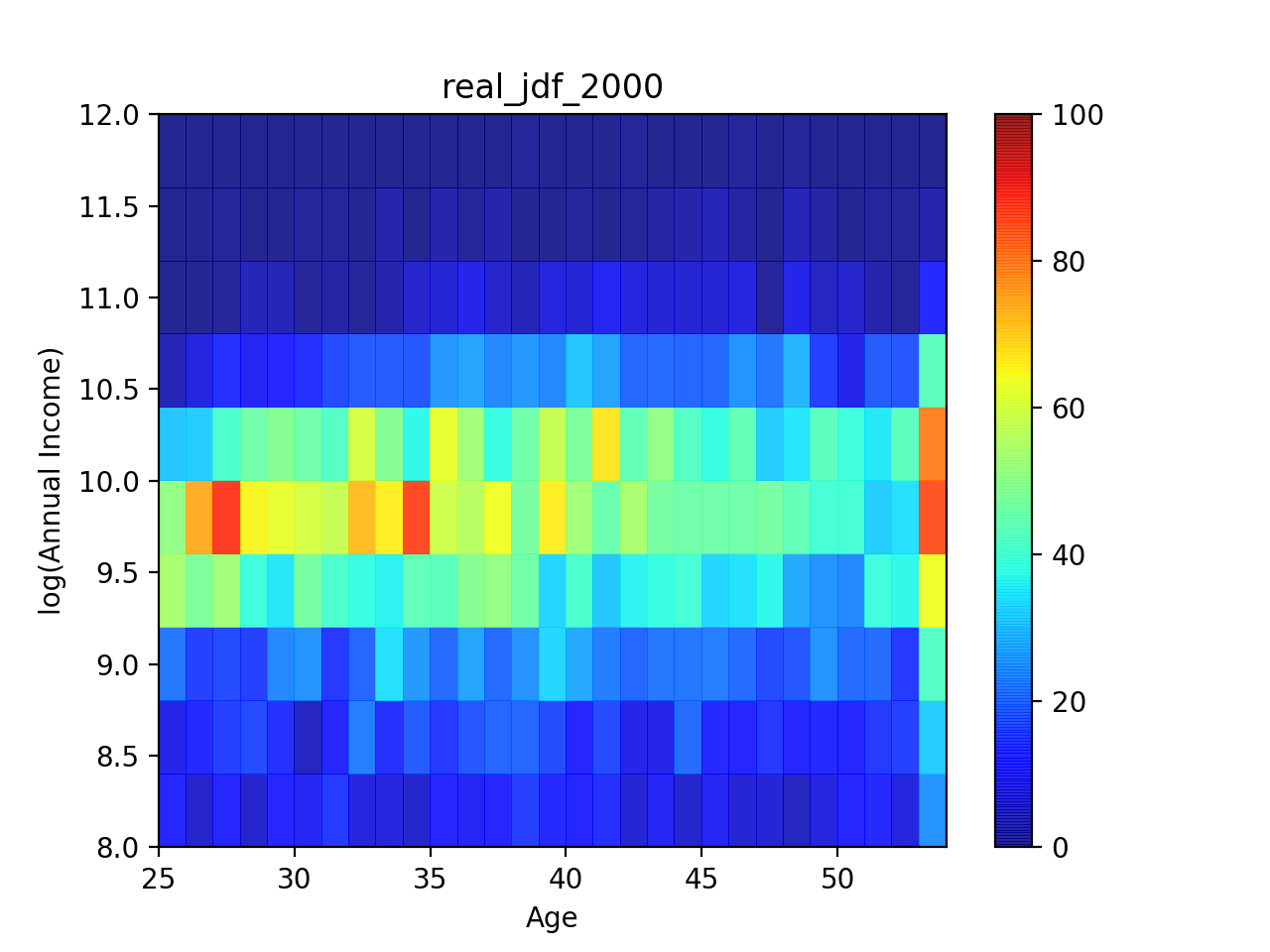}}\qquad
         \subfloat[Wave 2000 JDF of Sim Data]{\includegraphics[width=0.33\linewidth]{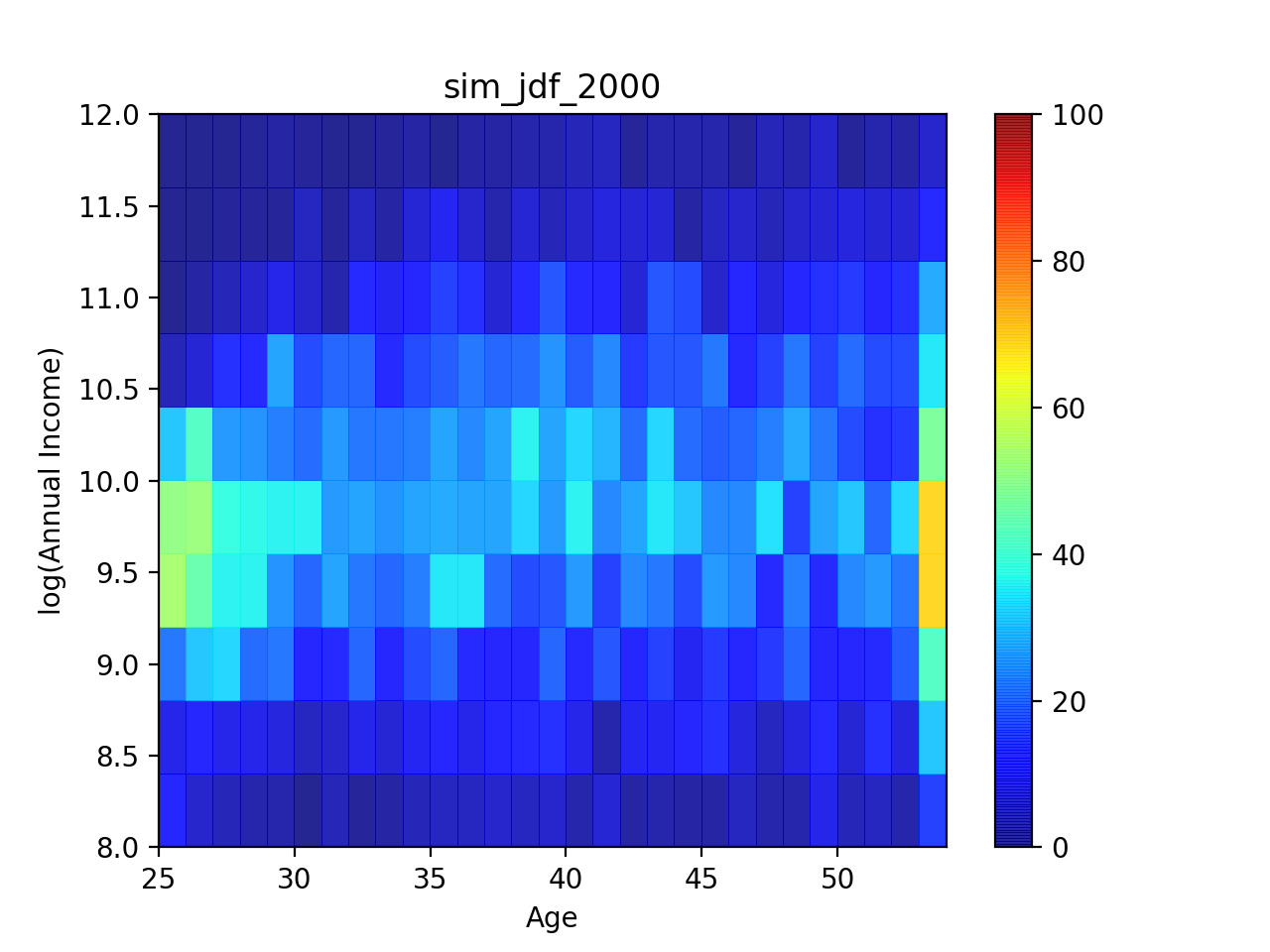}}\\
          \caption{JDF for Waves 1995-2000}
         \label{fig:1995_2000_waves_jdf}
\end{figure}
 \begin{figure}%
         \centering
         \subfloat[Wave 2001 JDF of Observed Data.]{\includegraphics[width=0.33\linewidth]{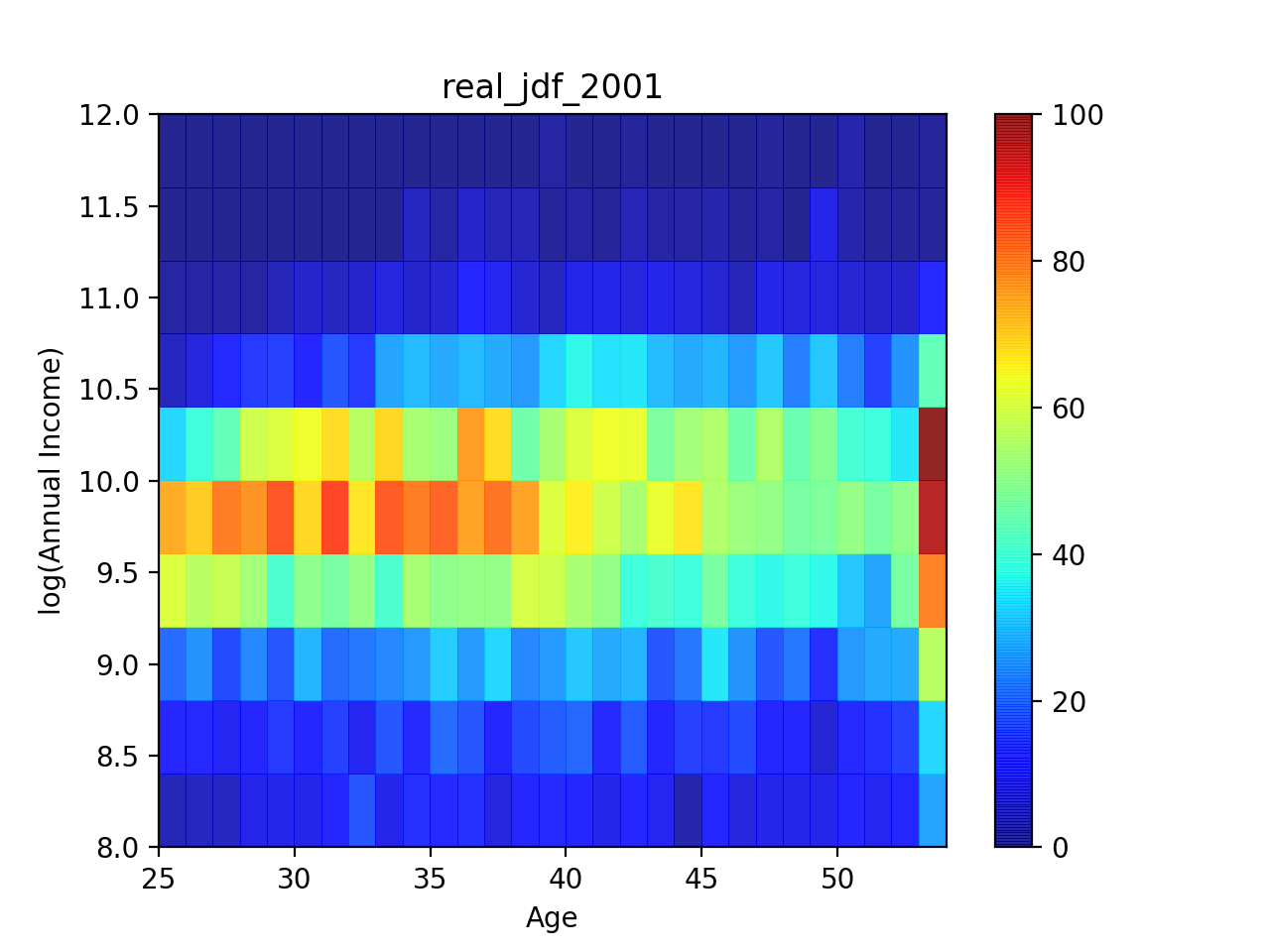}}\qquad
         \subfloat[Wave 2001 JDF of Sim Data]{\includegraphics[width=0.33\linewidth]{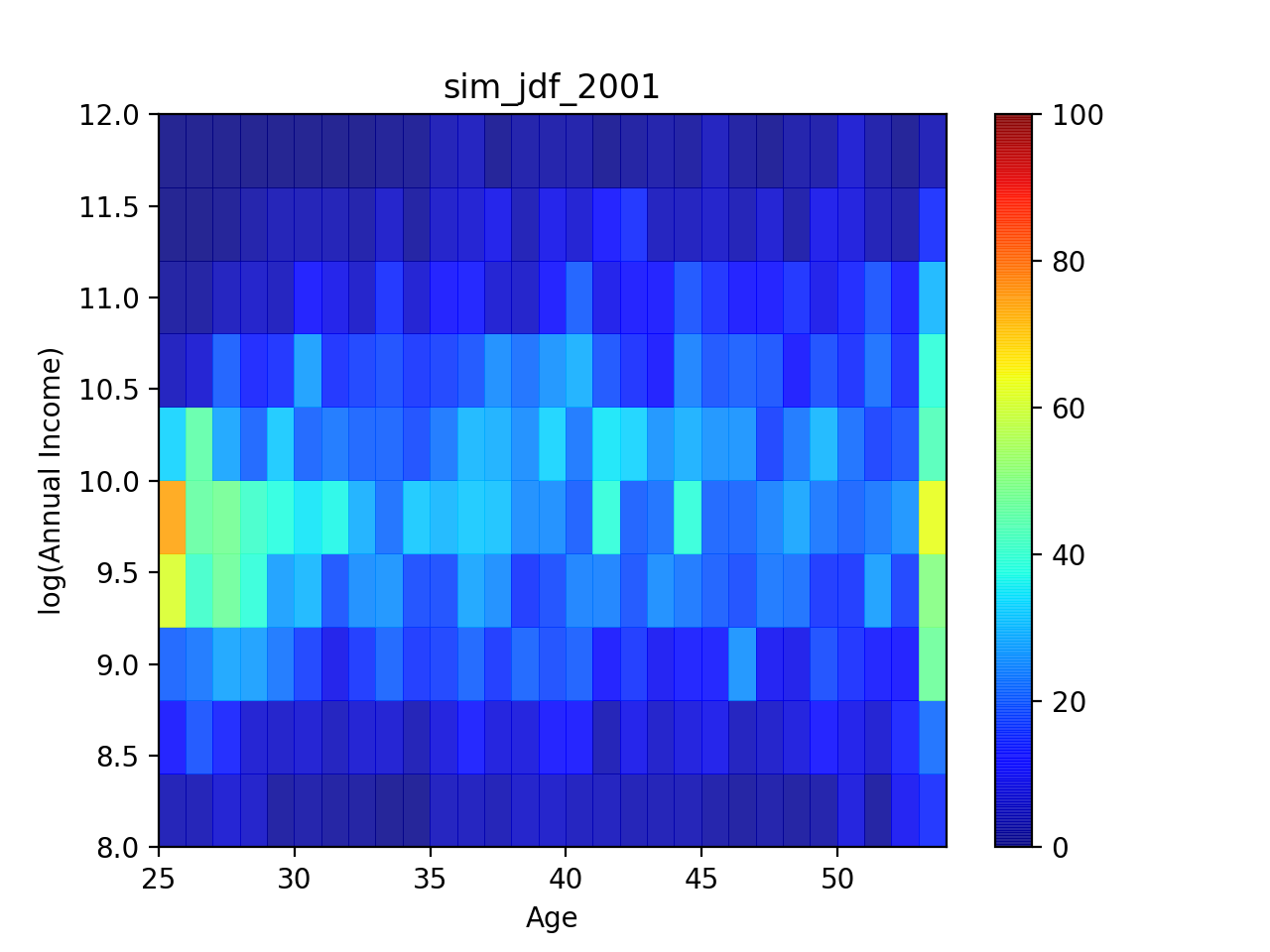}}\\
         \subfloat[Wave 2002 JDF of Observed Data.]{\includegraphics[width=0.33\linewidth]{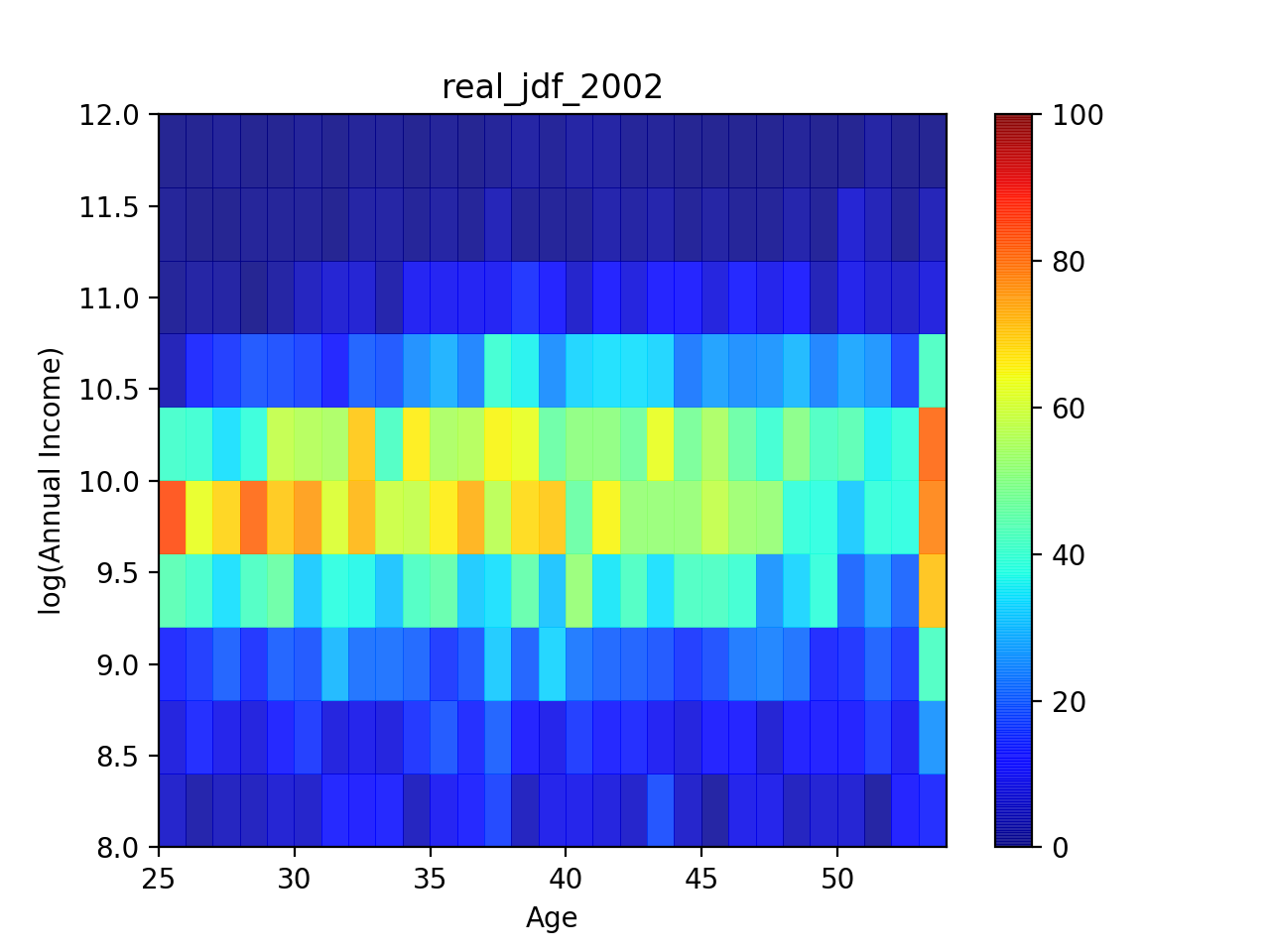}}\qquad
         \subfloat[Wave 2002 JDF of Sim Data]{\includegraphics[width=0.33\linewidth]{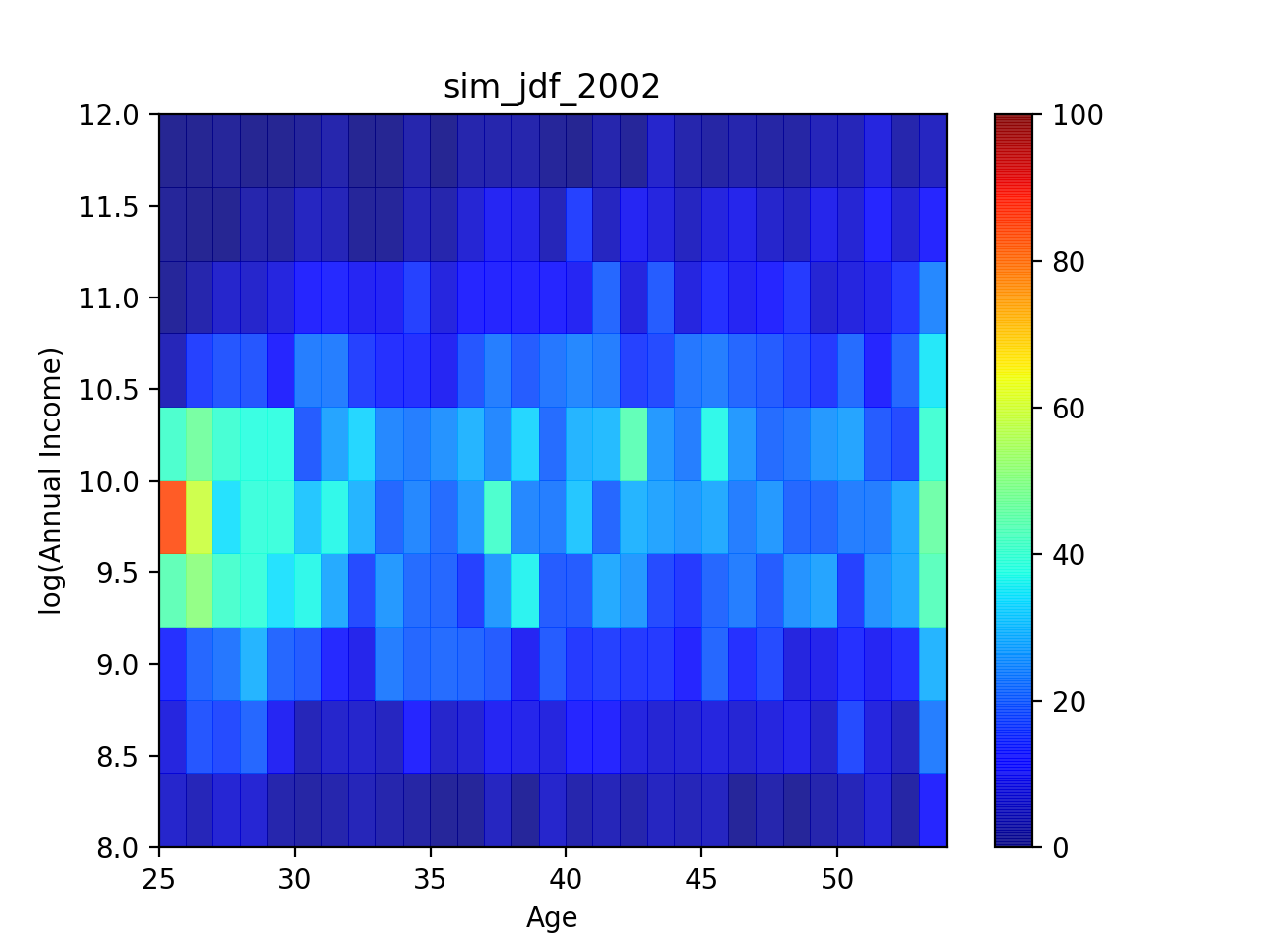}}\\
         \subfloat[Wave 2003 JDF of Observed Data.]{\includegraphics[width=0.33\linewidth]{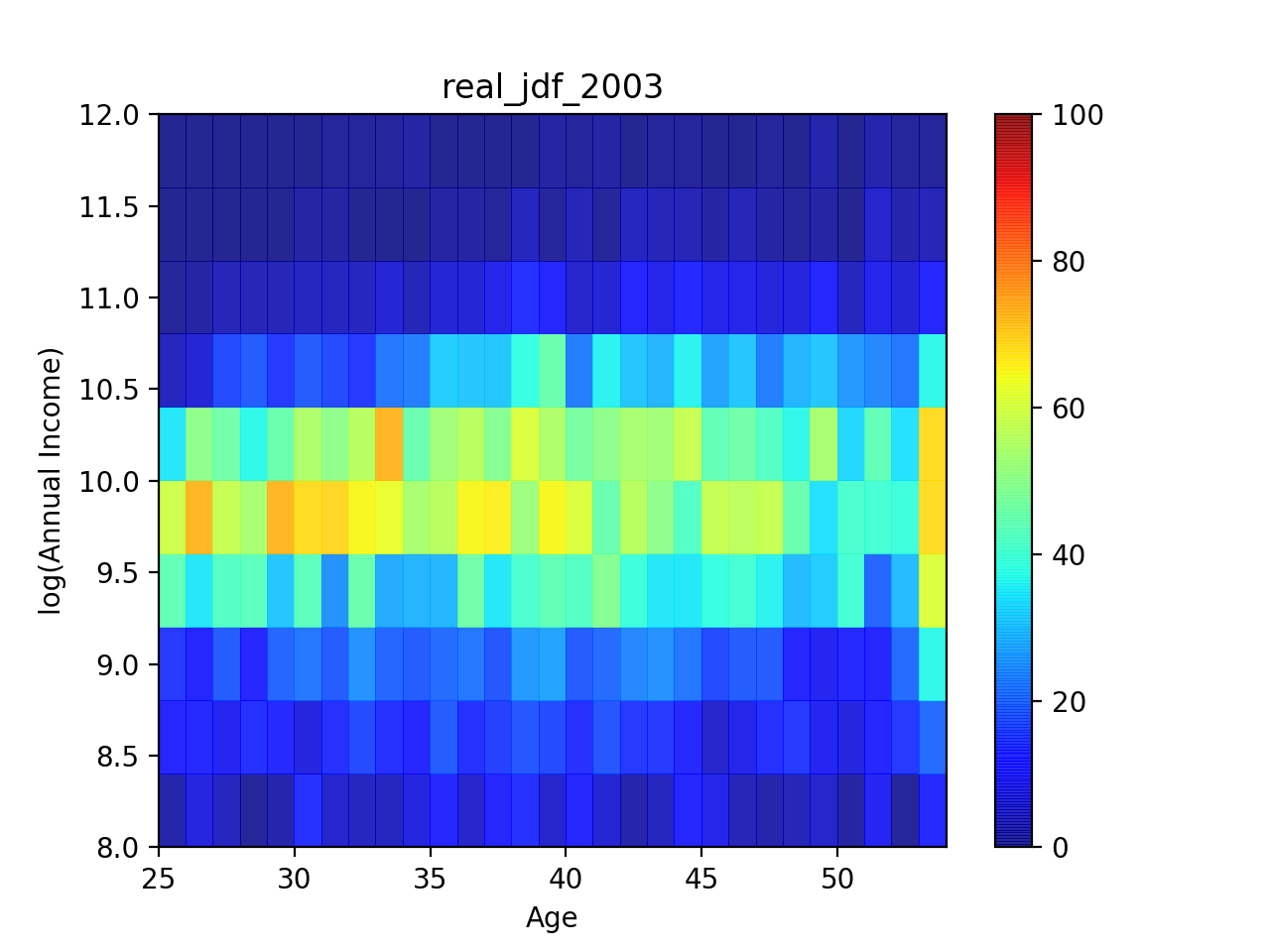}}\qquad
         \subfloat[Wave 2003 JDF of Sim Data]{\includegraphics[width=0.33\linewidth]{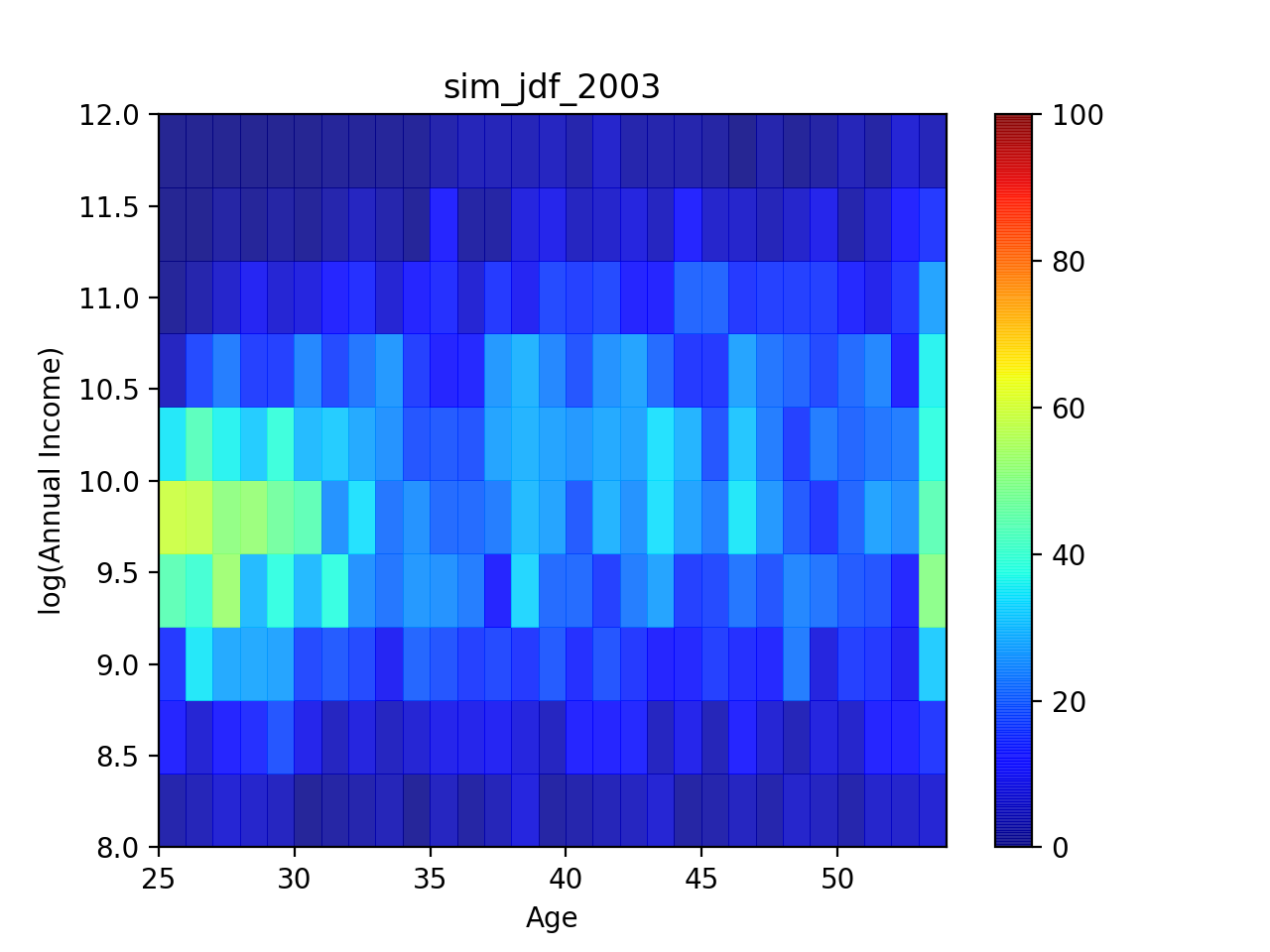}}\\
         \subfloat[Wave 2004 JDF of Observed Data.]{\includegraphics[width=0.33\linewidth]{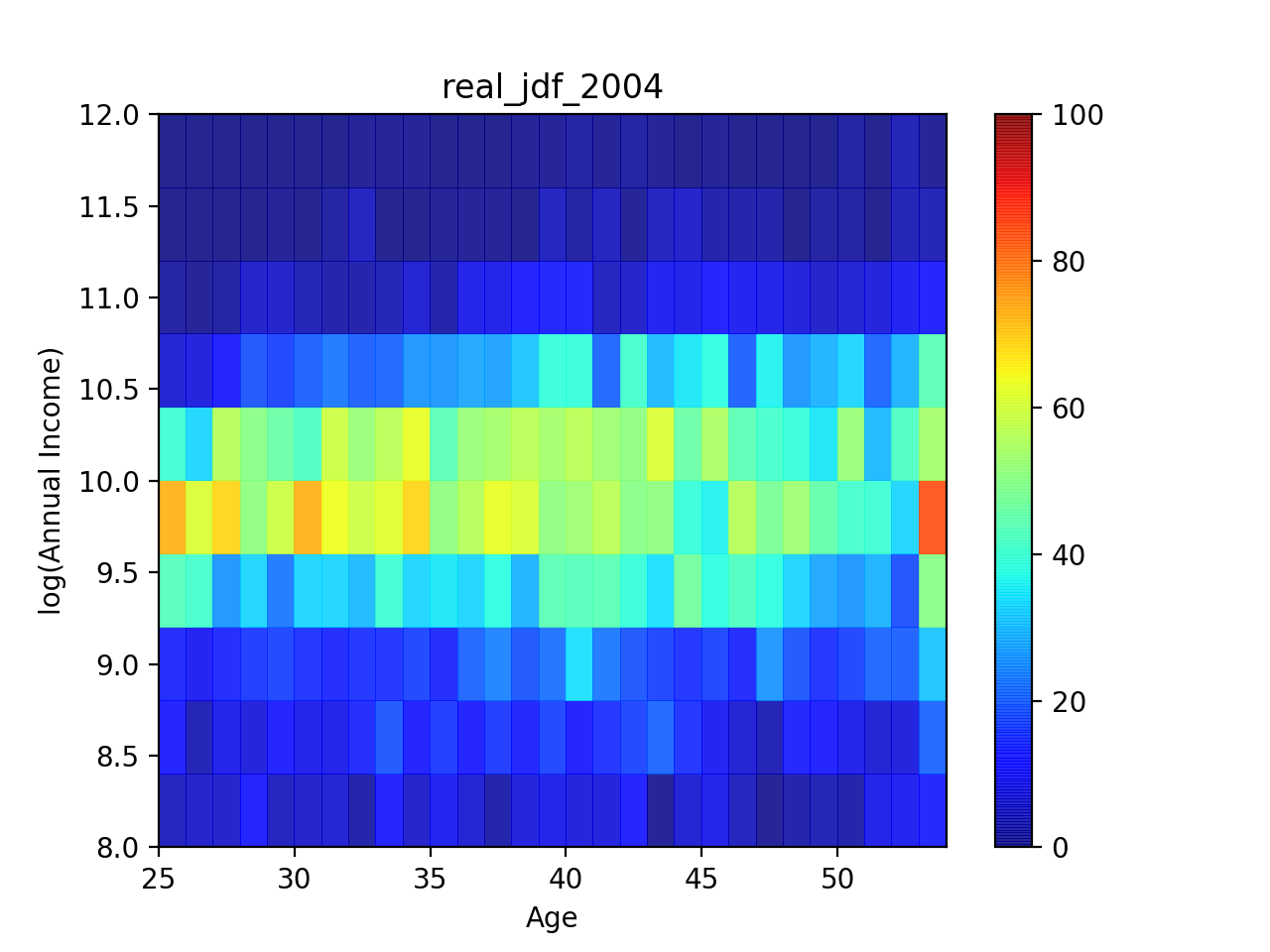}}\qquad
         \subfloat[Wave 2004 JDF of Sim Data]{\includegraphics[width=0.33\linewidth]{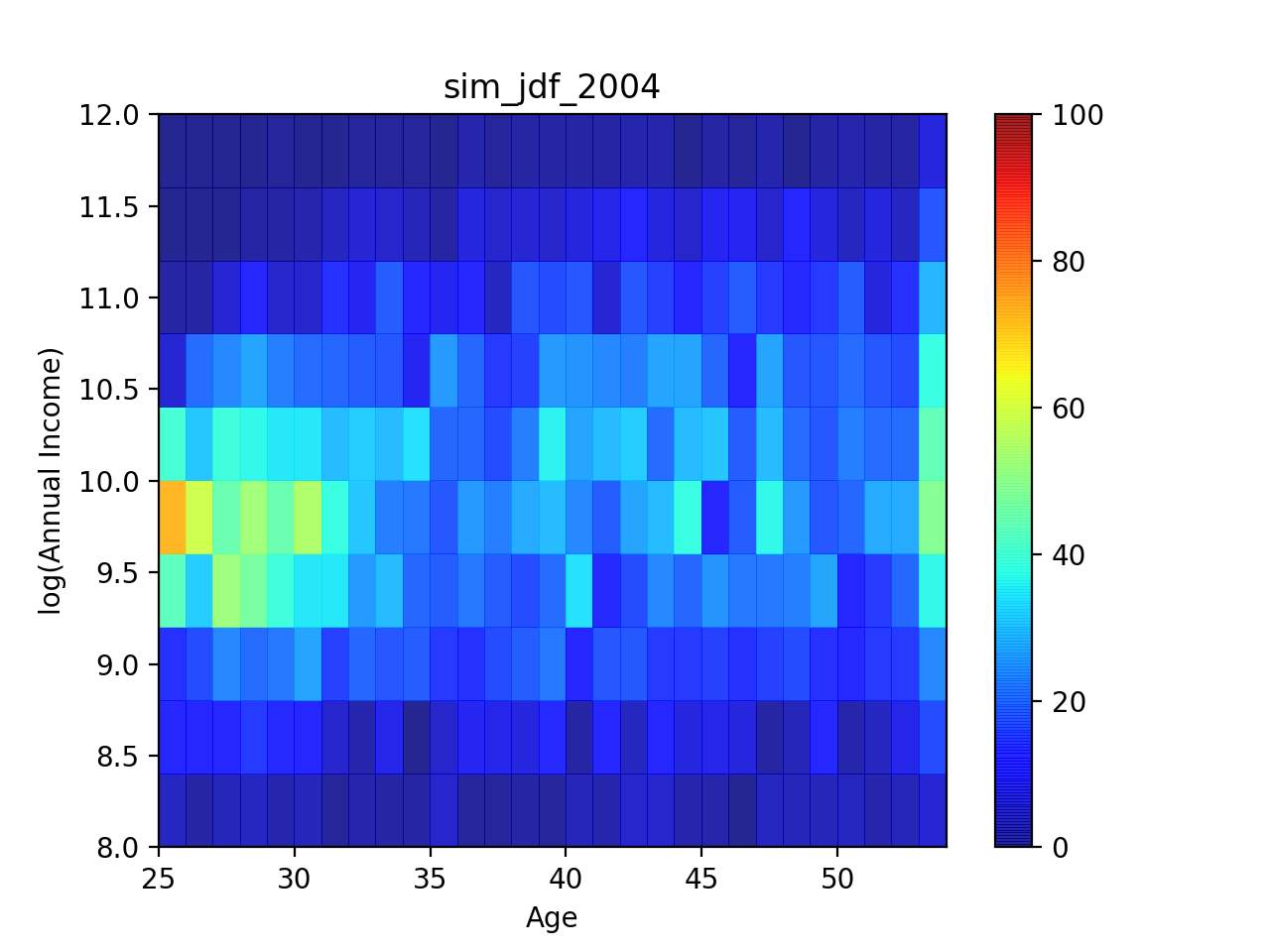}}\\
         \subfloat[Wave 2005 JDF of Observed Data.]{\includegraphics[width=0.33\linewidth]{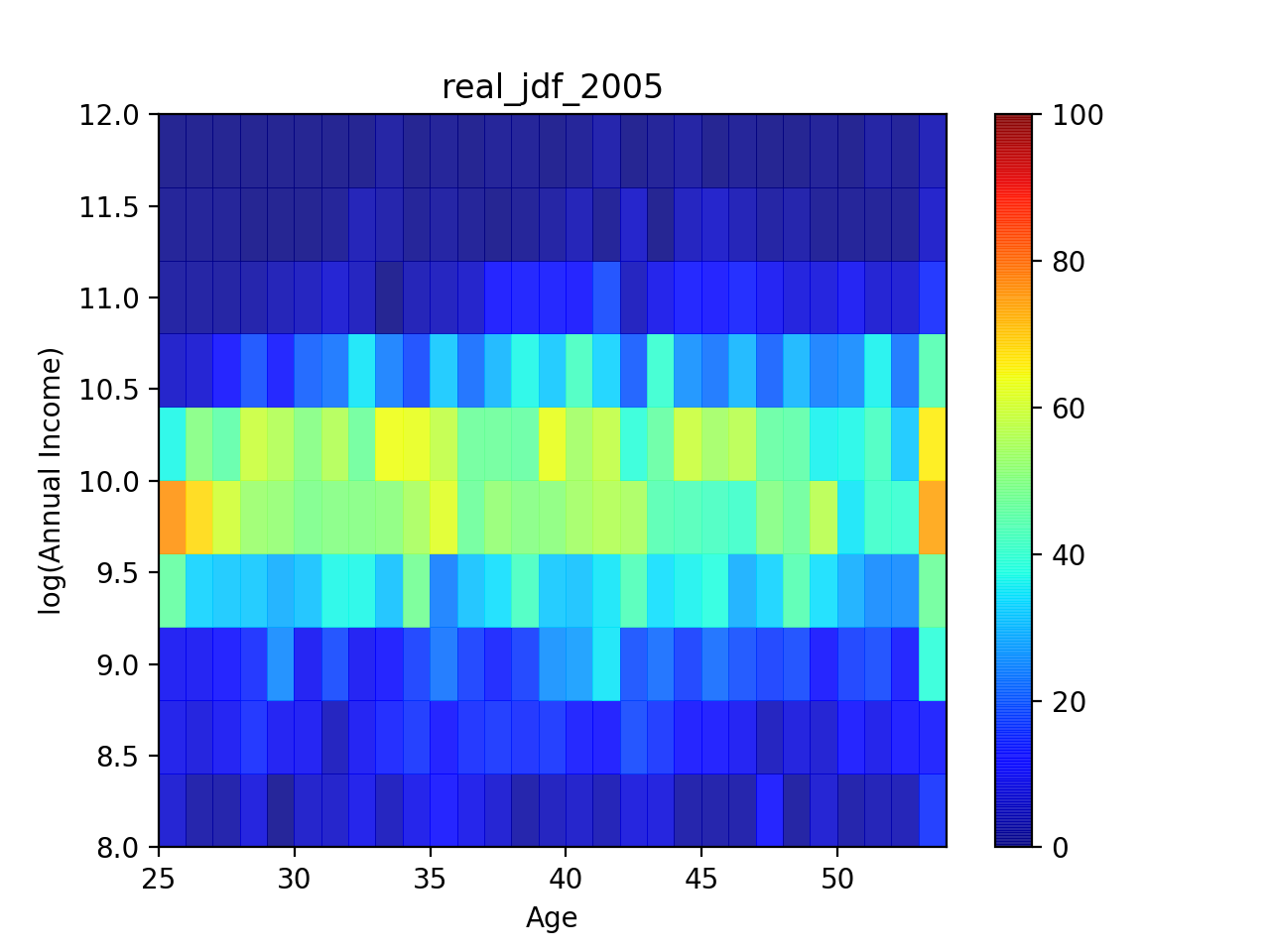}}\qquad
         \subfloat[Wave 2005 JDF of Sim Data]{\includegraphics[width=0.33\linewidth]{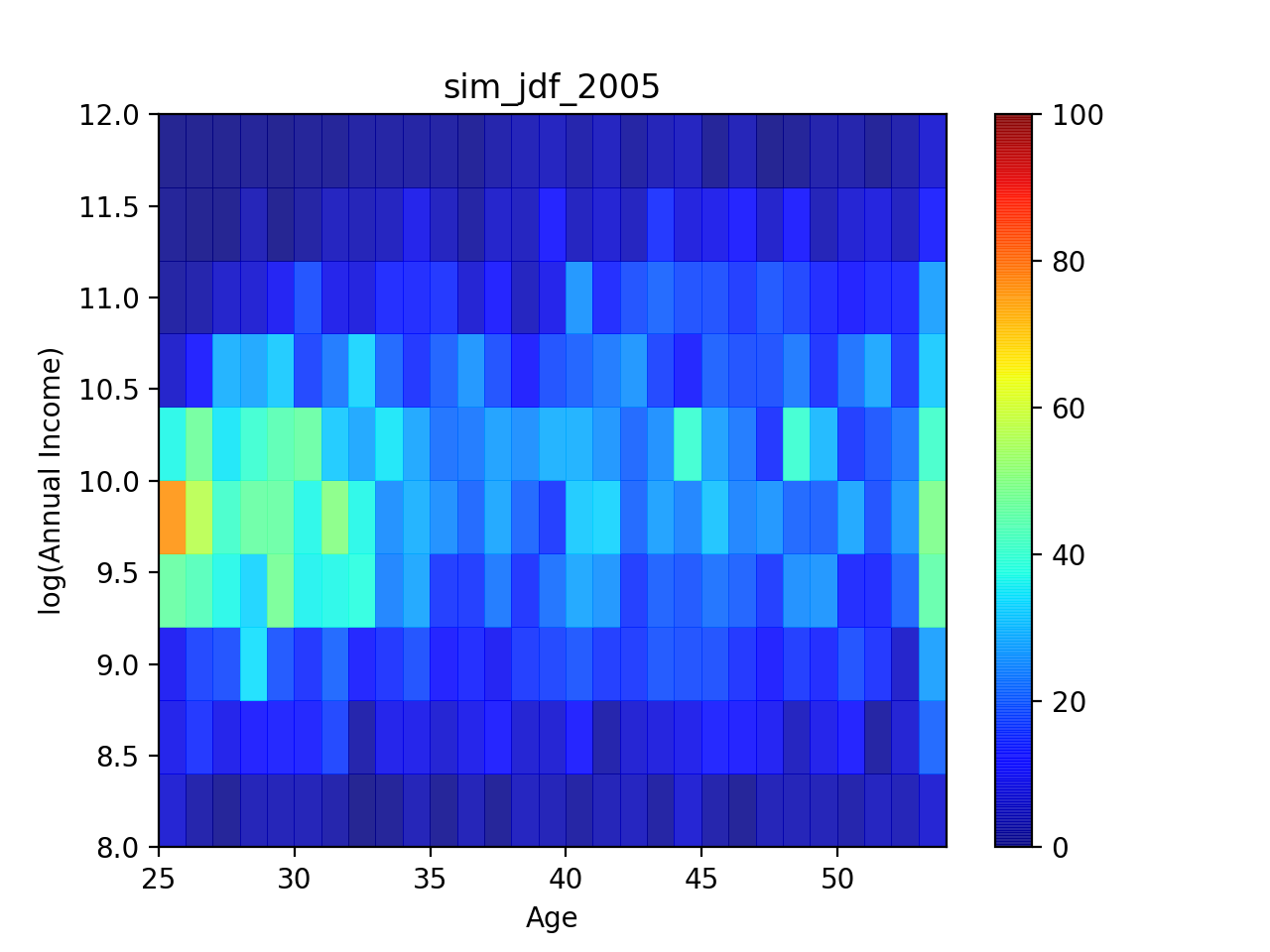}}\\
          \caption{JDF for Waves 2001-2005}
         \label{fig:2001_2005_waves_jdf}
\end{figure}

 \begin{figure}%
         \centering
         \subfloat[Wave 2006 JDF of Observed Data.]{\includegraphics[width=0.40\linewidth]{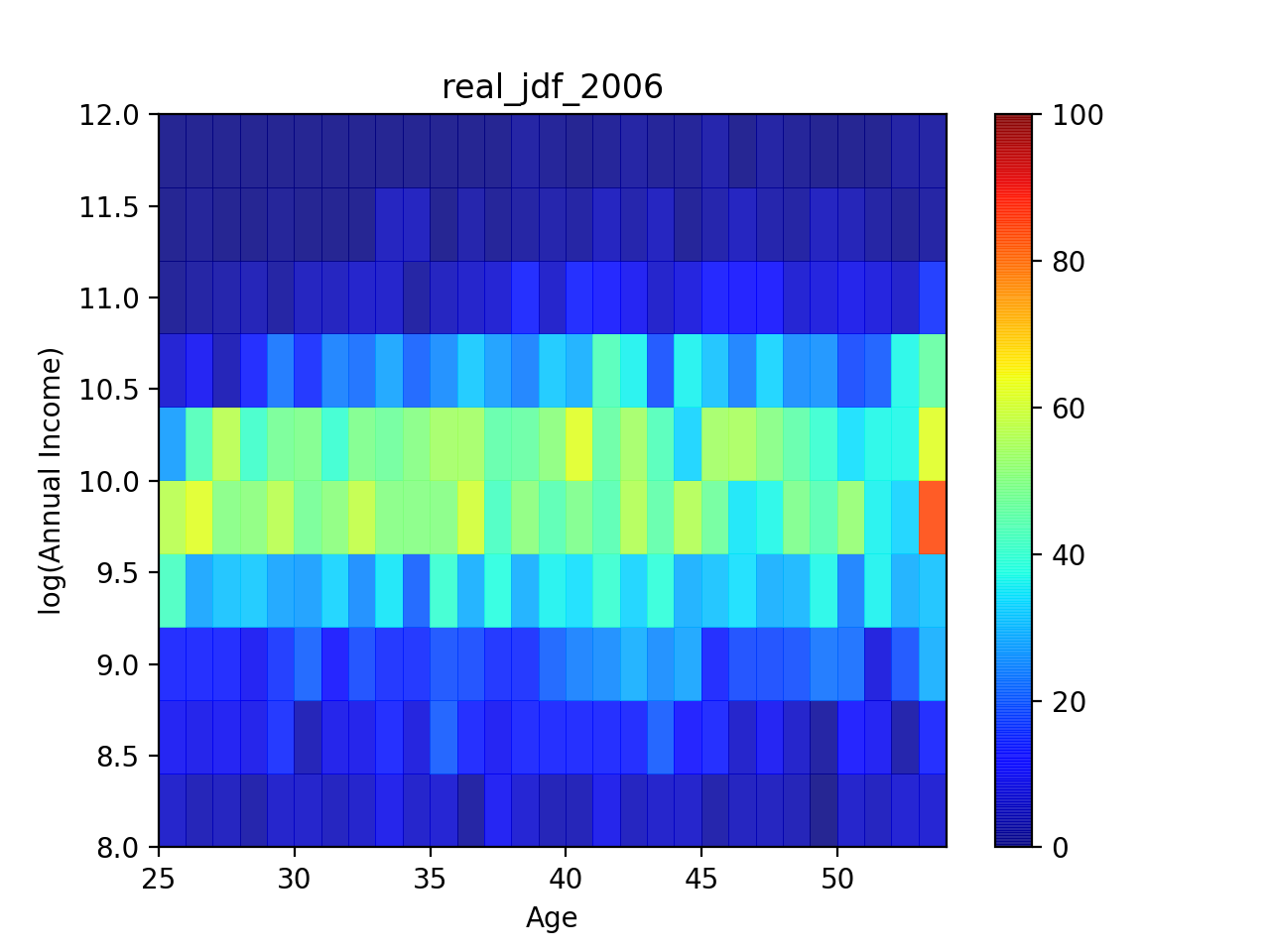}}\qquad
         \subfloat[Wave 2006 JDF of Sim Data]{\includegraphics[width=0.40\linewidth]{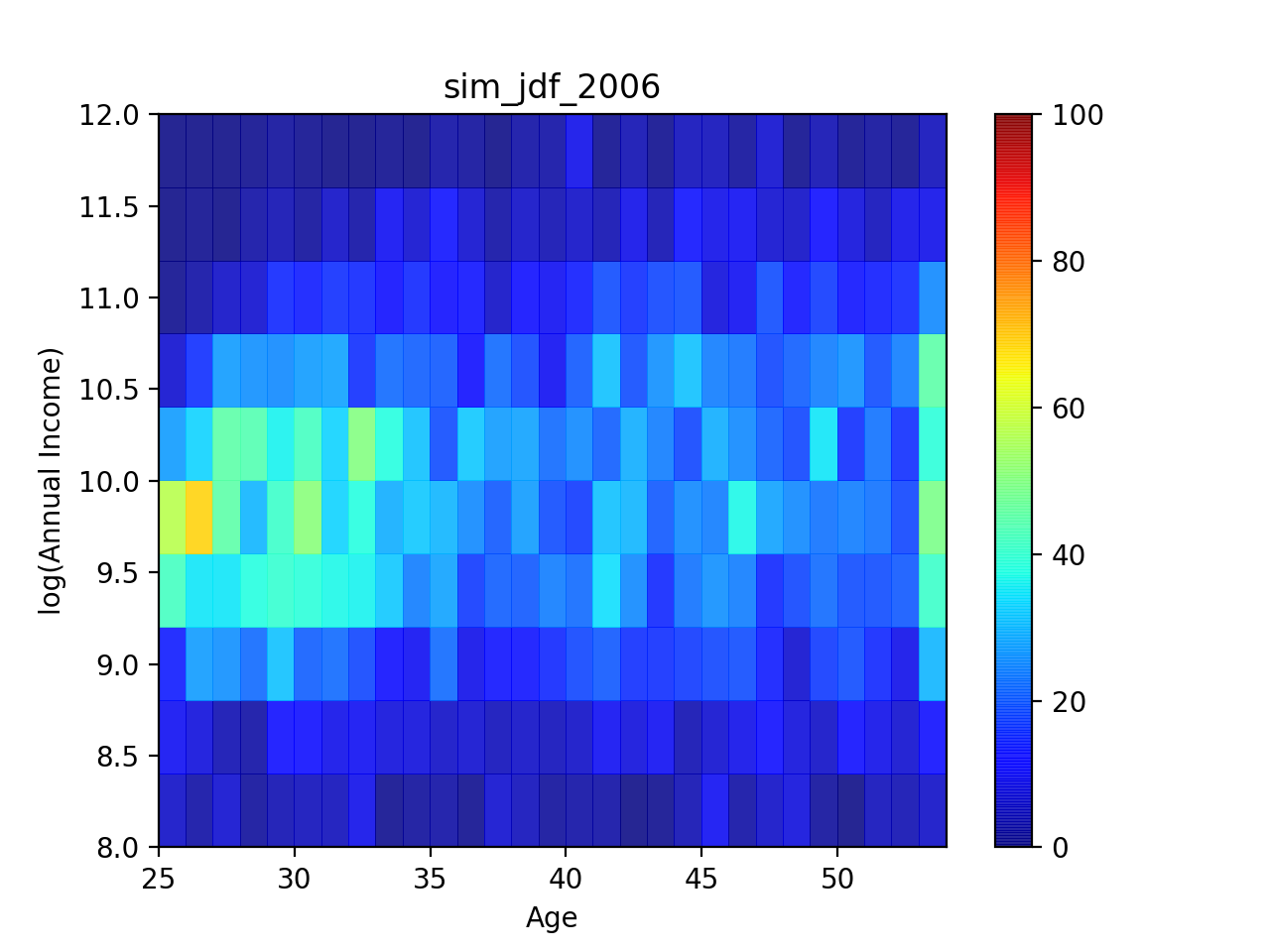}}\\
         \subfloat[Wave 2007 JDF of Observed Data.]{\includegraphics[width=0.40\linewidth]{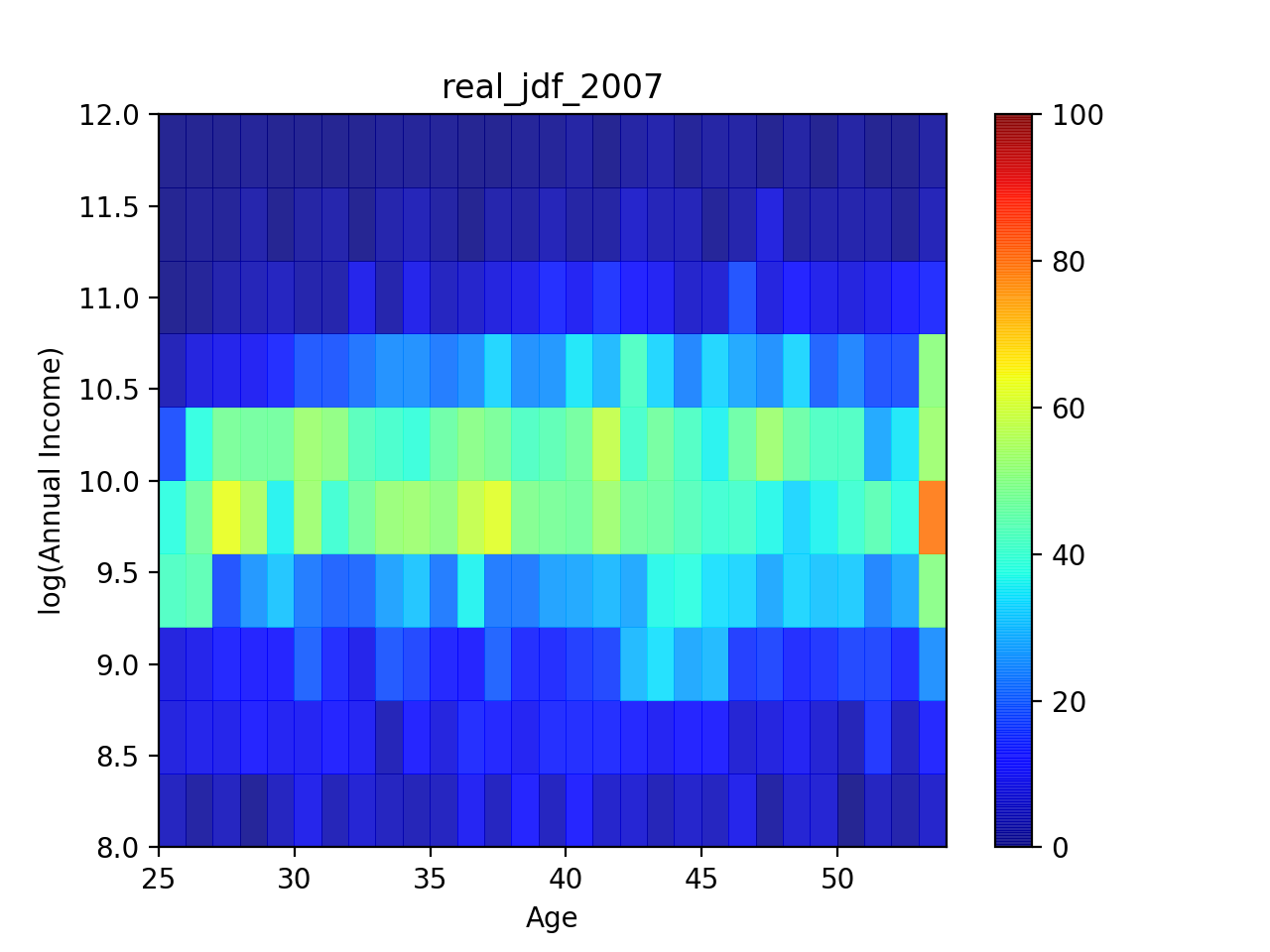}}\qquad
         \subfloat[Wave 2007 JDF of Sim Data]{\includegraphics[width=0.40\linewidth]{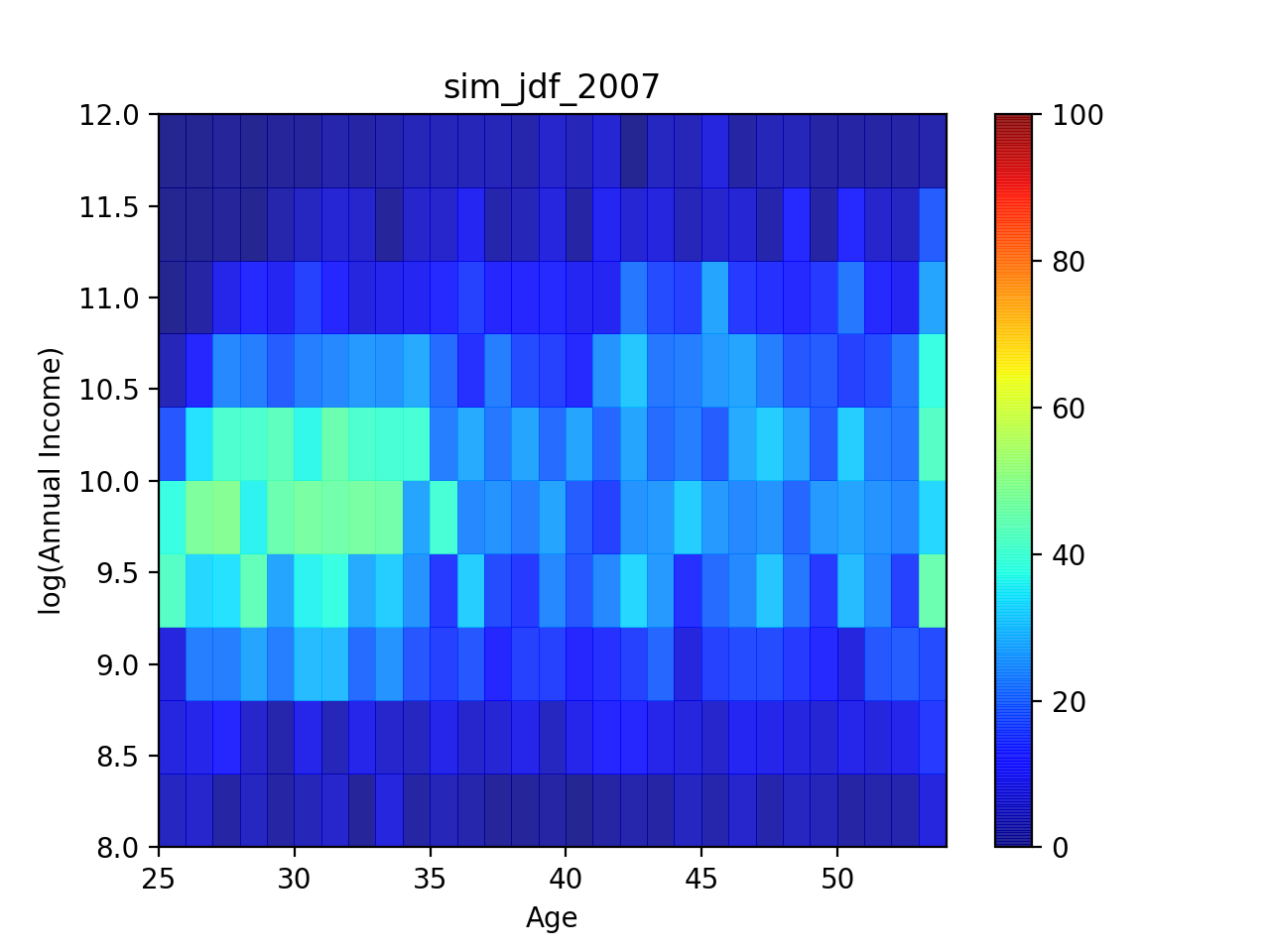}}\\
         \subfloat[Wave 2008 JDF of Observed Data.]{\includegraphics[width=0.40\linewidth]{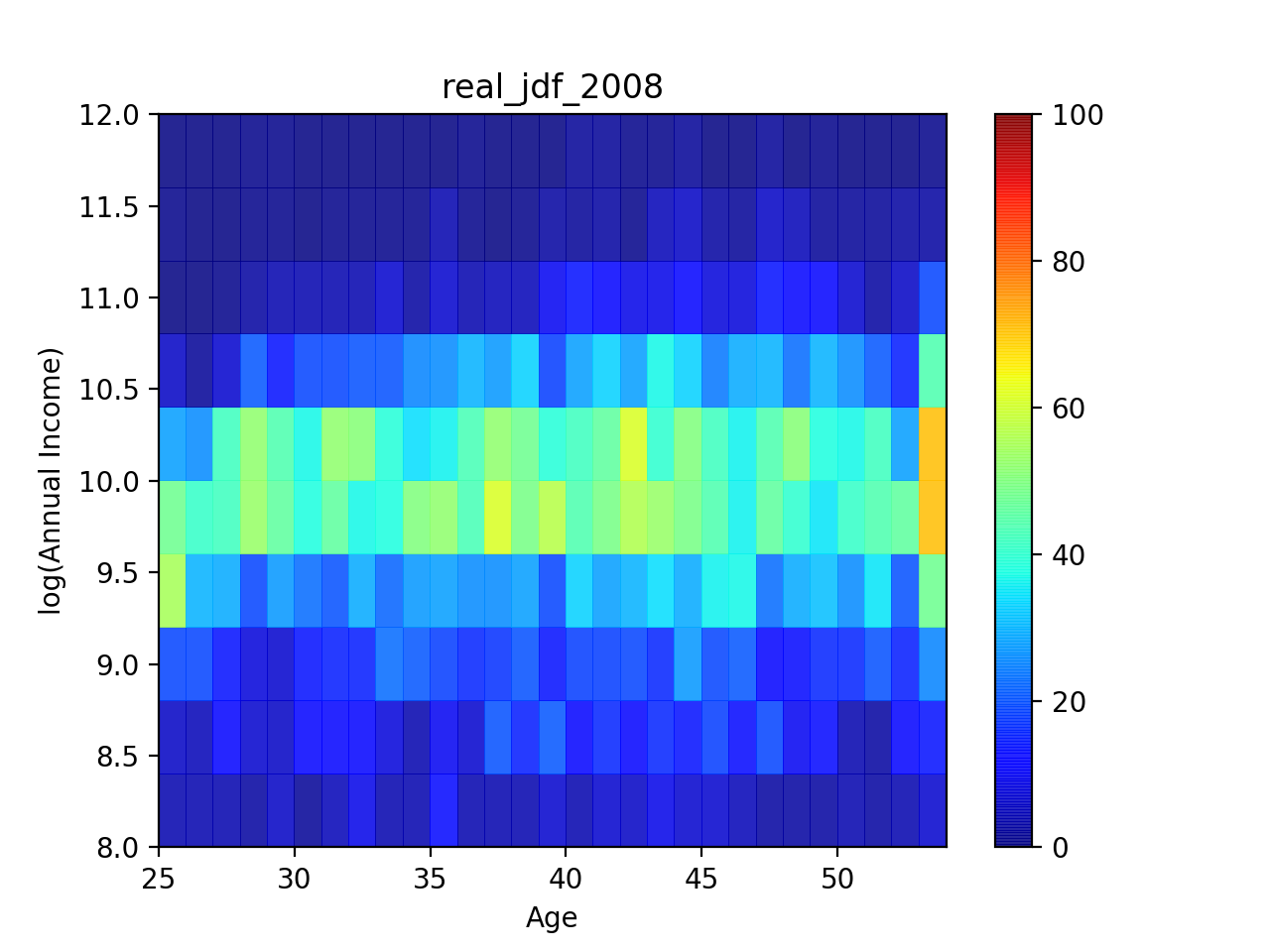}}\qquad
         \subfloat[Waves 2008 JDF of Sim Data]{\includegraphics[width=0.40\linewidth]{figs/uk_lsm_labour_bootstrap/sim_jdf_2008.png}}\\
         \caption{JDF for Waves 2005-2008}
         \label{fig:2005_2008_waves_jdf}
\end{figure}
\clearpage

\end{document}